\documentclass[twocolumn,aps,prc,superscriptaddress,floatfix,nofootinbib,10pt]{revtex4-2}
\usepackage{epsfig,graphics}
\usepackage{appendix}
\usepackage{graphicx}
\usepackage{dcolumn,multirow}
\usepackage{bm}
\usepackage{braket,overpic}
\usepackage{amsmath,amssymb}
\usepackage[usenames]{color}
\usepackage{pgffor}  
\usepackage{ulem} 
\newcommand{\er}{$\pm$}

\newcommand{\bc}           {\begin{center}}
\newcommand{\ec}           {\end{center}}
\newcommand{\bq}           {\begin{eqnarray}}
\newcommand{\eq}           {\end{eqnarray}}
\newcommand{\be}           {\begin{equation}}
\newcommand{\ee}           {\end{equation}}
\newcommand{\bi}           {\begin{itemize}}
\newcommand{\ei}           {\end{itemize}}
\newcommand{\rd}           {\color{red}}
\newcommand{\bl}           {\color{blue}}

\begin{document}
\title{\boldmath Photoproduction of two charged pions off protons in the resonance region}

\newcommand*{\JLAB}{Thomas Jefferson National Accelerator Facility, Newport News, Virginia 23606, United States of America}
\newcommand*{\JLABindex}{36}
\affiliation{\JLAB}

\newcommand*{\HISKP}{Helmholtz--Institut für Strahlen- und Kernphysik, Universit\"at Bonn, 53115 Bonn, Germany}
\newcommand*{\HISKPindex}{36}
\affiliation{\HISKP}

\newcommand*{\NRC}{NRC Kurchatov Institute, PNPI, Gatchina 188300, Russia}
\newcommand*{\NRCindex}{30}
\affiliation{\NRC}

\newcommand*{\FSU}{Florida State University, Tallahassee, Florida 32306, United States of America}
\newcommand*{\FSUindex}{14}
\affiliation{\FSU}
\newcommand*{\ANL}{Argonne National Laboratory, Argonne, Illinois 60439}
\newcommand*{\ANLindex}{1}
\affiliation{\ANL}
\newcommand*{\CSUDH}{California State University, Dominguez Hills, Carson, CA 90747}
\newcommand*{\CSUDHindex}{2}
\affiliation{\CSUDH}
\newcommand*{\CANISIUS}{Canisius College, Buffalo, NY}
\newcommand*{\CANISIUSindex}{3}
\affiliation{\CANISIUS}
\newcommand*{\CMU}{Carnegie Mellon University, Pittsburgh, Pennsylvania 15213}
\newcommand*{\CMUindex}{4}
\affiliation{\CMU}
\newcommand*{\CUA}{Catholic University of America, Washington, D.C. 20064}
\newcommand*{\CUAindex}{5}
\affiliation{\CUA}
\newcommand*{\SACLAY}{IRFU, CEA, Universit\'{e} Paris-Saclay, F-91191 Gif-sur-Yvette, France}
\newcommand*{\SACLAYindex}{6}
\affiliation{\SACLAY}
\newcommand*{\CNU}{Christopher Newport University, Newport News, Virginia 23606}
\newcommand*{\CNUindex}{7}
\affiliation{\CNU}
\newcommand*{\UCONN}{University of Connecticut, Storrs, Connecticut 06269}
\newcommand*{\UCONNindex}{8}
\affiliation{\UCONN}
\newcommand*{\DUKE}{Duke University, Durham, North Carolina 27708-0305}
\newcommand*{\DUKEindex}{9}
\affiliation{\DUKE}
\newcommand*{\DUQUESNE}{Duquesne University, 600 Forbes Avenue, Pittsburgh, PA 15282 }
\newcommand*{\DUQUESNEindex}{10}
\affiliation{\DUQUESNE}
\newcommand*{\FU}{Fairfield University, Fairfield CT 06824}
\newcommand*{\FUindex}{11}
\affiliation{\FU}
\newcommand*{\FERRARAU}{Università di Ferrara, 44121 Ferrara, Italy}
\newcommand*{\FERRARAUindex}{12}
\affiliation{\FERRARAU}
\newcommand*{\FIU}{Florida International University, Miami, Florida 33199}
\newcommand*{\FIUindex}{13}
\affiliation{\FIU}
\newcommand*{\GWUI}{The George Washington University, Washington, DC 20052}
\newcommand*{\GWUIindex}{15}
\affiliation{\GWUI}
\newcommand*{\GSIFFN}{GSI Helmholtzzentrum fur Schwerionenforschung GmbH, D-64291 Darmstadt, Germany}
\newcommand*{\GSIFFNindex}{16}
\affiliation{\GSIFFN}
\newcommand*{\INFNFE}{INFN, Sezione di Ferrara, 44100 Ferrara, Italy}
\newcommand*{\INFNFEindex}{17}
\affiliation{\INFNFE}
\newcommand*{\INFNGE}{INFN, Sezione di Genova, 16146 Genova, Italy}
\newcommand*{\INFNGEindex}{18}
\affiliation{\INFNGE}
\newcommand*{\INFNRO}{INFN, Sezione di Roma Tor Vergata, 00133 Rome, Italy}
\newcommand*{\INFNROindex}{19}
\affiliation{\INFNRO}
\newcommand*{\INFNTUR}{INFN, Sezione di Torino, 10125 Torino, Italy}
\newcommand*{\INFNTURindex}{20}
\affiliation{\INFNTUR}
\newcommand*{\INFNPAV}{INFN, Sezione di Pavia, 27100 Pavia, Italy}
\newcommand*{\INFNPAVindex}{21}
\affiliation{\INFNPAV}
\newcommand*{\ORSAY}{Université Paris-Saclay, CNRS/IN2P3, IJCLab, 91405 Orsay, France}
\newcommand*{\ORSAYindex}{22}
\affiliation{\ORSAY}
\newcommand*{\JMU}{James Madison University, Harrisonburg, Virginia 22807}
\newcommand*{\JMUindex}{23}
\affiliation{\JMU}
\newcommand*{\KNU}{Kyungpook National University, Daegu 41566, Republic of Korea}
\newcommand*{\KNUindex}{24}
\affiliation{\KNU}
\newcommand*{\MIT}{Massachusetts Institute of Technology, Cambridge, Massachusetts  02139-4307}
\newcommand*{\MITindex}{25}
\affiliation{\MIT}
\newcommand*{\MISS}{Mississippi State University, Mississippi State, MS 39762-5167}
\newcommand*{\MISSindex}{26}
\affiliation{\MISS}
\newcommand*{\UNH}{University of New Hampshire, Durham, New Hampshire 03824-3568}
\newcommand*{\UNHindex}{27}
\affiliation{\UNH}
\newcommand*{\NMSU}{New Mexico State University, PO Box 30001, Las Cruces, NM 88003, USA}
\newcommand*{\NMSUindex}{28}
\affiliation{\NMSU}
\newcommand*{\NSU}{Norfolk State University, Norfolk, Virginia 23504}
\newcommand*{\NSUindex}{29}
\affiliation{\NSU}
\newcommand*{\OHIOU}{Ohio University, Athens, Ohio  45701}
\newcommand*{\OHIOUindex}{30}
\affiliation{\OHIOU}
\newcommand*{\ODU}{Old Dominion University, Norfolk, Virginia 23529}
\newcommand*{\ODUindex}{31}
\affiliation{\ODU}
\newcommand*{\JLUGiessen}{II Physikalisches Institut der Universität Giessen, 35392 Giessen, Germany}
\newcommand*{\JLUGiessenindex}{32}
\affiliation{\JLUGiessen}
\newcommand*{\RPI}{Rensselaer Polytechnic Institute, Troy, New York 12180-3590}
\newcommand*{\RPIindex}{33}
\affiliation{\RPI}
\newcommand*{\ROMAII}{Università di Roma Tor Vergata, 00133 Rome Italy}
\newcommand*{\ROMAIIindex}{34}
\affiliation{\ROMAII}
\newcommand*{\MSU}{Skobeltsyn Institute of Nuclear Physics, Lomonosov Moscow State University, 119234 Moscow, Russia}
\newcommand*{\MSUindex}{35}
\affiliation{\MSU}
\newcommand*{\SCAROLINA}{University of South Carolina, Columbia, South Carolina 29208}
\newcommand*{\SCAROLINAindex}{36}
\affiliation{\SCAROLINA}
\newcommand*{\TEMPLE}{Temple University,  Philadelphia, PA 19122 }
\newcommand*{\TEMPLEindex}{37}
\affiliation{\TEMPLE}
\newcommand*{\UTFSM}{Universidad T\'{e}cnica Federico Santa Mar\'{i}a, Casilla 110-V Valpara\'{i}so, Chile}
\newcommand*{\UTFSMindex}{39}
\affiliation{\UTFSM}
\newcommand*{\BRESCIA}{Università degli Studi di Brescia, 25123 Brescia, Italy}
\newcommand*{\BRESCIAindex}{40}
\affiliation{\BRESCIA}
\newcommand*{\UCR}{University of California Riverside, 900 University Avenue, Riverside, CA 92521, USA}
\newcommand*{\UCRindex}{41}
\affiliation{\UCR}
\newcommand*{\GLASGOW}{University of Glasgow, Glasgow G12 8QQ, United Kingdom}
\newcommand*{\GLASGOWindex}{42}
\affiliation{\GLASGOW}
\newcommand*{\YORK}{University of York, York YO10 5DD, United Kingdom}
\newcommand*{\YORKindex}{43}
\affiliation{\YORK}
\newcommand*{\VIRGINIA}{University of Virginia, Charlottesville, Virginia 22901}
\newcommand*{\VIRGINIAindex}{44}
\affiliation{\VIRGINIA}
\newcommand*{\YEREVAN}{Yerevan Physics Institute, 375036 Yerevan, Armenia}
\newcommand*{\YEREVANindex}{45}
\affiliation{\YEREVAN}

\newcommand*{\NOWJLAB}{Thomas Jefferson National Accelerator Facility, Newport News, Virginia 23606}
\newcommand*{\NOWISU}{Idaho State University, Pocatello, Idaho 83209}
\newcommand*{\NOWSCAROLINA}{University of South Carolina, Columbia, South Carolina 29208}

\author{A.V.~Sarantsev} \affiliation{\NRC}
\author{E.~Klempt}\thanks{Corresponding author: \texttt{klempt@hiskp.uni-bonn.de}} \affiliation{\JLAB} \affiliation{\HISKP}
\author{K.V.~Nikonov} \affiliation{\HISKP}
\author{P.~Achenbach}\affiliation{\JLAB}
\author {V.D.~Burkert} \affiliation{\JLAB}
\author {V.~Crede} \affiliation{\FSU}
\author {V.~Mokeev}\affiliation{\JLAB}
\author {J. S. Alvarado} 
\affiliation{\ORSAY}
\author {M.J.~Amaryan} 
\affiliation{\ODU}
\author {H.~Atac} 
\affiliation{\TEMPLE}
% ...
\vspace{2mm}
\collaboration{The CLAS Collaboration}
\noaffiliation
\date{\today}

\begin{abstract}
Photoproduction of charged pions pairs off protons is studied within the invariant masses of the final state hadrons from 1.6 to 2.4\,GeV  
at the Thomas Jefferson National Accelerator Facility with the CLAS detector. 
The total and differential cross sections and spin-density matrix elements
are presented for the isobars $p\rho^0(770)$, $\Delta(1232)^{++}\pi^-$, and  $\Delta(1232)^{0}\pi^+$.
The data are included in the Bonn--Gatchina coupled-channel analysis and provide the information
necessary to determine the branching fractions of $N\rho(770)$ decays for most known $N^*$ and $\Delta^*$ resonances.
For the first time, the $N\rho$ branching ratios are obtained here from an event-based likelihood to
$\gamma p\to \pi^+\pi^-p$.  
\end{abstract}

\maketitle

\section{Introduction}
Photoproduction of charged pion pairs off protons is born out of rich dynamics.
At high energies, neutral $\rho(770)$ mesons provide a large contribution to the cross section:
the incident photon can convert into a vector meson, e.g.\ into a $\rho^0(770)$ meson, and
the virtual vector meson may then scatter off the proton by Pomeron
or by Reggeon exchange~\cite{H1:2009cml}. Understanding these processes in terms of
quark and gluon interactions is a major challenge for Quantum Chromodynamics (see
Ref.~\cite{Gross:2022hyw} for a review).

At lower energies, further production modes are important~\cite{Lueke:1971cx}:
The initial photon can dissociate the proton into a $\Delta(1232)^{++}$ and a $\pi^-$ meson 
in a contact
interaction (Kroll--Ruderman mechanism). 
The $\Delta(1232)$ and higher-mass $N$ and $\Delta$ excitations can
be formed as intermediate isobars decaying into the $N\pi\pi$ final state in a cascade process.
The two final-state pions can interact forming scalar, vector or tensor mesons.

In this paper, we report a detailed study of the reaction
\be\label{reaction}
  \gamma p\to \pi^+\pi^-p.
\ee
The mass and angular distributions for invariant masses up to 2.15\,GeV have been presented in a letter publication \cite{CLAS:2018drk} in which
the total cross section and photocouplings of four-star $N$ and $\Delta$ resonances below 2\,GeV
were determined. These distributions and data on reaction~(\ref{reaction}) but with linearly polarized photons
and transversely polarized protons~\cite{Crede:2024tbd}
are included in a large database used in a coupled-channel analysis. Here, part of the data on reaction \ref{reaction} are
used event-by-event in an event-based likelihood fit. 
Results on the properties of $N^*$ and $\Delta^*$ resonances will be presented elsewhere~\cite{Sarantsev:2024tbd}.

The two-pion production process has been studied before in pion, photo-, and electroproduction (for
electroproduction, see Refs.~\cite{Mokeev:2023zhq} and \cite{Burkert:2022ioj}, and references therein). \\[-2ex]

\paragraph{Pion-induced experiments:} 
Early analyses of $\pi N$ scattering into three-body final states exploited bubble chambers.
The low-energy region covering the  $N(1440)1/2^+$ resonance was studied in
Refs.~\cite{Kirz:1963zz,Saxon:1970ac}.
The resonance region was investigated in different charge combinations of the $N\pi\pi$ final state covering
the 1390--1535\,MeV~\cite{DeBeer:1969ak,DeBeer:1969aj}, 1400--2000\,MeV~\cite{Brody:1971rb}, 1500--1740\,MeV~\cite{Dolbeau:1974an},
and 1400--1700\,MeV~\cite{Barnham:1980za} mass ranges.

First counter experiments were
devoted to the threshold region to study the breaking of chiral symmetry \cite{Kravtsov:1976sy,Belkov:1979bp,Aaron:1979hc,Bjork:1980vb,%
OMICRON:1988ocn,OMICRON:1989yyd,OMICRON:1990yei,Sevior:1990yp,Lowe:1991gd}.
The energy range was increased when the Crystal Ball detector was moved to BNL where the reaction
$\pi^-p\to\pi^0\pi^0 n$ was studied up to $\sqrt s=1520$\,MeV \cite{CrystalBall:2004qln}.
Recently, the HADES Collaboration reported results on the reaction $\pi^-p\to \pi^+\pi^-p$ at four beam momenta
from 650 to 786\,MeV and derived excitation functions of different partial waves
and of the $\Delta(1232)\pi$, $N f_0(500)$, and $N\rho(770)$ isobar configurations  \cite{HADES:2020kce}.\\[-2ex]

\paragraph{Photoproduction experiments with bubble chambers:} The first photoproduction experiments
were carried out at photon energies in the $3<E_\gamma <8$\,GeV range exploiting bubble chambers
with correspondingly low statistics
\cite{CambridgeBubbleChamberGroup:1967zz,Aachen-Berlin-Bonn-Hamburg-Heidelberg-Munich:1968rzt,%
Aachen-Berlin-Bonn-Hamburg-Heidelberg-Muenchen:1969pjo,Ballam:1971yd,Ballam:1971wq}.\\[-2ex]

\paragraph{Photoproduction experiments at MAMI:} Photoproduction using a counter experiment  was first studied at the electron accelerator MAMI for energies up to 0.8\,GeV~\cite{Braghieri:1994rf};
three  different two-pion states ($p\pi^+\pi^-$, $p\pi^0\pi^0$, $n\pi^+\pi^0$)  were studied and
cascade decays via $\Delta^0(1232)\pi^0$ were observed~\cite{Harter:1997jq}. The decays were
tentatively assigned to $N(1440)1/2^+$ or $N(1520)3/2^-\to \Delta^0(1232)\pi^0$
decays~\cite{Wolf:2000qt}. The helicity dependence of the $\gamma p\to p\;2\pi^0$ \cite{GDH:2005jgl,Zehr:2012tj}
and $\gamma p\to \pi^+\pi^-p$ \cite{GDH:2007nkn,Zehr:2012tj} cross sections
indicated a preference for the decay sequence $N(1520)3/2^-\to \Delta^0(1232)\pi^0$. 
The beam helicity asymmetry was compared to models, in the case of
$\pi^0\pi^0$ production with reasonable agreement, in the case of $\pi^+\pi^0$ production with
large discrepancies~\cite{GDH:2007nkn,Zehr:2012tj} for all models used.
In Refs.~\cite{Kotulla:2003cx,GDH:2005jgl,Zehr:2012tj}, the $\pi^0\pi^0$ production
threshold was investigated and predictions of the chiral perturbation theory for the threshold
behavior were confirmed. 

In Ref.~\cite{Kashevarov:2012wy}, the differential cross sections for $2\pi^0$ photoproduction were
expanded into spherical harmonics and moments. Large contributions from spin-parity $J^P=3/2^-$-waves
were required below the $N(1520)3/2^-$ resonance. This effect was tentatively ascribed to contributions
of the wide $\Delta(1700)3/2^-$ or to rescattering effects.
The A2 Collaboration compared $\pi^0\pi^0$ photoproduction off free protons with photoproduction
off quasi-free protons and neutrons bound in deuterons \cite{A2:2015pgk}. The results on bound and
free protons agree in shape reasonably well, although the cross section for the quasi-free
protons was found to be smaller than the cross section measured with an H$_2$ target. The cross section
for the reaction $\gamma n\to \pi^0\pi^0 n$ showed a much less pronounced valley between 
the peaks due to the second and third resonance region at about 1.5 and 1.7\,GeV. 
The peak at about 1.7\,GeV from the $\gamma p$ initial state was mainly assigned to the
$N f_0(500)$ decay mode while for $\gamma n$, this structure was assigned to the $\Delta(1232)^0\pi^0$ decay mode.
Apparently, different resonances in the third resonance region are predominantly produced off protons
or off neutrons. 

The A2 Collaboration determined the double-polarization observable $E$, and helicity-dependent cross 
sections $\sigma_{1/2}$ and $\sigma_{3/2}$ were reported for photoproduction of $\pi^0$ pairs off 
quasifree protons and neutrons, and determined the helicity- and isospin-dependent structure of the 
$\gamma N\to N\pi^0\pi^0$ reaction \cite{Dieterle:2020vug}. The polarization observables
$P_x, P_y$ (unpolarized beam, target polarized in the $x,y$ direction),  $P_x ^\odot, P_y ^\odot$ (circularly polarized beam, 
target polarized in the $x,y$ direction) were measured~\cite{A2:2022ipx}. A partial wave analysis was performed that included
Born terms and a sum of $s$-channel Breit--Wigner resonances.\\[-2ex]

\paragraph{Photoproduction experiments at LEPS:} 
The Graal Collaboration studied the energy range up to $E_\gamma=1.5$\,GeV;
the total cross section showed strong peaks at $W = 1.5$ and 1.7\,GeV
for $\gamma p\to p\;2\pi^0$~\cite{Assafiri:2003mv} and for $\gamma n\to n\;2\pi^0$~\cite{Ajaka:2007zz}.
The LEPS Collaboration studied the differential cross section and photon-beam asymmetry for
the reaction $\gamma p \to \pi^- \Delta(1232)^{++}$ at forward $\pi^-$ angles~\cite{LEPS:2018pbi}. \\[-2ex]

\paragraph{Photoproduction experiments at ELSA:} The reaction~(\ref{reaction}) was studied by the SAPHIR Collaboration~\cite{Wu:2005wf}
throughout the resonance region. The total cross section up to $E_\gamma = 2.6$~GeV was determined as well
as the $\rho^0(770)$, $\Delta(1232)^{++}$, and $\Delta(1232)^{0}$ excitation functions and the $t$-dependence
of the production of the isobars.

The results of the CBELSA / TAPS Collaboration on $\gamma p\to p\;2\pi^0$ were included in a coupled-channel analysis
of a larger data set and provided masses, widths, helicity amplitudes, and decay frequencies for
resonances up to 1700\,MeV~\cite{Thoma:2007bm,Sarantsev:2007aa}. Later, the photon energy range was extended
and photon polarization data were included. The coupled-channel analysis yielded a large 
number of cascade decays \cite{CBELSATAPS:2015kka,CBELSATAPS:2015tyg,CBELSATAPS:2015taz}.
It was shown that baryon resonances with a component in the wave function in which both oscillators
are simultaneously excited have a significant fraction of decays into orbitally or radially
excited mesons or baryons; resonances with only one oscillator de-excite mostly into mesons
and baryons with no orbital or radial excitation. The findings were confirmed in a recent analysis of the reaction $\vec\gamma \vec p\to p\;2\pi^0$ with linearly polarized photons and transversely polarized protons~\cite{CBELSATAPS:2022uad}.\\[-2ex]

\paragraph{Photoproduction experiments at Jefferson Lab:} 
Photoproduction of $\rho(770)$ mesons off protons has been studied by the CLAS
Collaboration for $E_\gamma$ between 3.19 and 3.91\,GeV and squared momentum
transfers $-t$ from 0.1 to 5\,GeV$^2$ \cite{CLAS:2001zxv}. The beam helicity asymmetry was studied in the energy range $1.35<E_\gamma < 2.3$~GeV, but
without firm conclusions on the contributing resonances~\cite{CLAS:2005oqk}.
In the energy range from 3.0 to 3.8~GeV, moments in the $\pi\pi$ angular distributions were extracted in Ref.~\cite{CLAS:2009ngd}.
As discussed above, contributions from the excitation spectrum of the 
nucleon were identified in \cite{CLAS:2018drk}. A comparison of the properties of the resonances produced in photo- and electroproduction showed
inconsistent results for the $N(1720)3/2^+$ decays to $\Delta\pi$ and 
$N\rho(770)$, which were only resolved by introducing an additional resonance
$N'(1720)3/2^+$ \cite{Mokeev:2020hhu}. Recently, the spin-density matrix elements were measured 
for photon energies in the 8.2 to 8.8\,GeV range by the GlueX Collaboration for $\rho(770)$ photoproduction with natural and unnatural parity exchange~\cite{GlueX:2023fcq} and for the $\Delta(1232)^{++}\pi^-$ final state~\cite{Afzal:2024yaq}.

In this paper, we present details of the extraction of the $\gamma p\to p\pi^+\pi^-$ reaction reported in a Letter \cite{CLAS:2018drk}. 
We give a short description of the CLAS detector and discuss the four event topologies,
events with three particles detected or with $p$, $\pi^+$, or $\pi^-$ missing. The data  are included in the BnGa coupled-channel analysis. The analysis
yields Dalitz plots, mass and angular distributions, the total cross section and excitation functions for $\gamma p \to$
$\Delta(1232)^{++}\pi^-$, $\Delta(1232)^{0}\pi^+$, $p \rho^0(770)$, and $pf_2(1270)$,
differential cross section as a function of squared momentum transfer $t$, 
spin-density matrix elements for $p\rho^0(770)$, and
branching ratios of $N^*$ and $\Delta^*$ resonances for their decays into $N\rho(770)$. Analysis details and
branching ratios for $N^*$ and $\Delta^*$ decays into $\Delta(1232)\pi$ and $N^*\pi$ will be presented elsewhere~\cite{Sarantsev:2024tbd}.
The new results profit from the large number of events from reaction~(\ref{reaction}) which is used here in
a event-by-event likelihood fit that fully exploits the correlations between the variables in their 5-dimensional
phase space.

\section{The CLAS experiment and event selection}
\label{SectionNewData}

The results presented here were derived using the CEBAF Large Acceptance Spectrometer (CLAS)~\cite{CLAS:2003umf}
in Hall~B at the Thomas Jefferson National Accelerator Facility during the ``g11a'' data taking period in 2004.
The photon beam was produced by an unpolarized electron beam of 4.019\,GeV energy. The beam current
ranged from 60 to 75\,nA. The electron beam was impinged on a gold foil radiator with a thickness of $10^{-4}$ radiation
lengths. The beam delivered tagged photons with energies from 1.6 to 2.6\,GeV. The photon energies were determined by
detecting scattered electrons in the tagging spectrometer~\cite{tagger,Stepanyan:2006vv}.

\subsection{Target and photon beam}
The tagged photon flux on the target within the data acquisition live time was obtained by the standard CLAS {\it gflux}
method~\cite{flux}. The number of photons for each tagger counter was calculated independently as $N_\gamma =
\epsilon \cdot N_{e^-}$, where $N_{e^-}$ is the number of electrons detected by a tagger counter and $\epsilon$ is the
tagging ratio. The tagging ratio was determined by placing a total absorption counter directly in the photon beam at
low intensity and determining the ratio of the number of beam photons and the number of electrons detected in coincidence
in the tagger. The global normalization uncertainty derived from the run-to-run variance and the estimated normalization
variance with the electron beam current together were found to be 1\%, employing the method described in
Ref.~\cite{willthesis}.

The photons hit a 40-cm-long liquid H$_2$ target.
Temperature and pressure of this cryogenic target were monitored throughout
the run period. The mean calculated density of H$_2$, 0.0718\,g$/$cm$^3$, was nearly constant with relative
fluctuations of about 0.1\% \cite{willthesis,bradnote}.\vspace{-1mm}

\begin{figure}[thb]
\includegraphics[width=0.45\textwidth,height=0.39\textwidth]{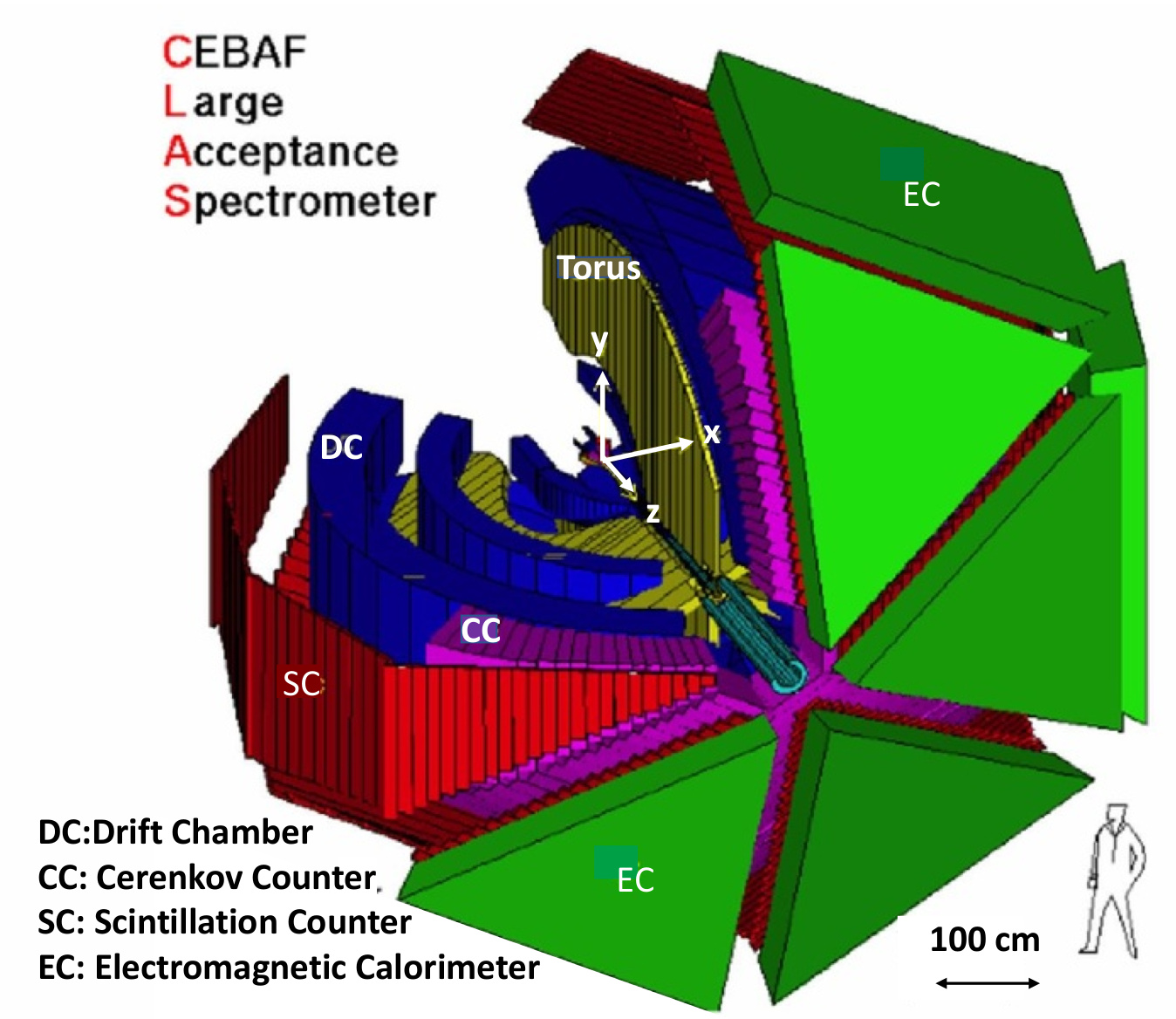}
\vspace{-1mm}
\caption{\label{fig:The-CLAS-detector} Cut-away view of the CLAS
    detector~\cite{CLAS:2003umf} illustrating the torus magnet,
    three regions of drift chambers (R1 -- R3), Cherenkov counters (CC),
    time-of-flight (TOF) scintillators, and electromagnetic calorimeters
    (EC). The CLAS detector was approximately 10\,m in diameter. \vspace{-6mm}}
\end{figure}
\subsection{The CLAS detector}
The CLAS detector, shown in Fig.~\ref{fig:The-CLAS-detector}, was designed
around six superconducting coils arranged in a hexagonal configuration that
produced an approximately toroidal magnetic field surrounding the beamline.
The magnetic field bent charged particles through the regions of
multi-layer drift chambers for momentum measurements. Drift chambers were
positioned between the superconducting coils within six sectors in azimuthal
$\phi$, each spanning approximately $60^\circ$. Charged particles reconstructed
in the fiducial volume and produced at a momentum
of 1\,GeV were measured with average angular resolutions of
$\sigma(\Theta)$, $\sigma(\phi)$ $\sim$2\,mrad and with a momentum
 resolution of $\sigma(p)/p$ $\leq$ 0.5\% ~\cite{Mestayer:2000we}.

The start counter scintillators~\cite{Sharabian:2005kq} (not shown) with a time resolution of 260\,ps
 surrounded the target cell and were used to determine the start time of the event. A set of 342 time-of-flight (TOF) scintillators was used for particle identification~\cite{Smith:1999ii}.
Their timing resolution ranged between 150--250\,ps, depending on
the length of the paddle. The Cherenkov counters and electromagnetic calorimeters were used for the detection of electrons and neutral particles~\cite{Amarian:2001zs}.
A coincidence between the start counter and TOF scintillators in at least two of the
six CLAS sectors was required to trigger the data acquisition. 
Some TOF paddles were dead or showed an excessive rate.
Hits in 22 paddles were later removed from further analysis.
 During the g11a run period, the integrated luminosity was 70\,pb$^{-1}$, and $\sim2\times10^{10}$ triggers were recorded.

\subsection{Event selection}
\label{evsel}

The selection of $\pi^+\pi^-p$ photoproduction events was carried out within the framework developed by the CLAS Collaboration \cite{CLAS:2018drk,Golovatch:2015not}. Here we will highlight the specific features of this approach, which are relevant for physics analysis on an event-by-event basis.\\[-2ex]

\paragraph{Fiducial cuts:}

The CLAS detector contained insensitive regions for particle detection. These insensitive regions were located in the regions blocked by
   the torus coils, as well as at very forward and very backward angles in the laboratory
frame~\cite{CLAS:2003umf}. The final-state particles were selected \cite{Golovatch:2015not} to be within the ``fiducial'' regions with reliable particle
detection efficiency, away from the insensitive regions. In addition, kinematic regions were excluded where the particle detection
acceptance was less than 5\%.

\begin{figure*}[pt]
\includegraphics[width=0.9\textwidth]{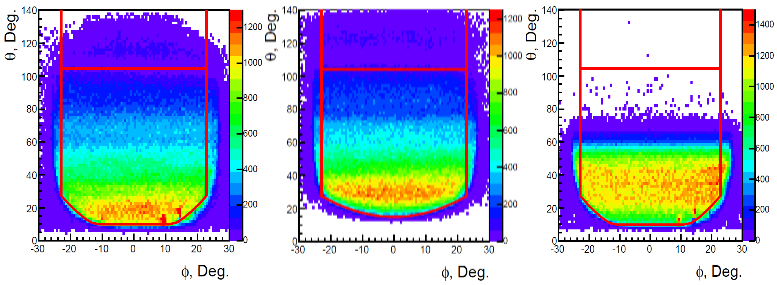}\vspace{-2mm}
\caption{\label{fiducial cuts}Polar ($\Theta$) versus azimuthal ($\phi$) angular distributions for identified $\pi^+$ (left), 
$\pi^-$ (middle) and protons (right) in
one of the six sectors of the CLAS spectrometer.  Fiducial cuts are shown in red.
}
\end{figure*}

In a first step, fiducial cuts based on the polar ($\Theta$) and azimuthal  ($\phi$) angle of a particle
were applied. Tracks with a polar angle $\Theta <27^\circ$ in the forward direction were removed.
A fiducial cut on the azimuthal ($\phi$) angle removed tracks close to
insensitive regions in the six sectors, with the cut on $\Theta$ being $\phi$-dependent.
Figure~\ref{fiducial cuts} shows the distribution of $\Theta$ versus $\phi$
for charged pions and protons for the first sector. The event distributions for the other sectors look very similar.
These fiducial cuts were defined iteratively, with preliminary and final particle identification.

Timing information was used to clean the event sample.
 A cut of 1.5\,ns was applied for the time difference between tagger time and the start time derived from the CLAS
Start Counters. The kinematic fit probed all photons within this time window.\\[-2ex]

\paragraph{Monte Carlo:}
The CLAS detector is simulated with the standard CLAS GSIM package~\cite{gsim} and an event generator based on the JM05 version of the $\pi^+\pi^-p$ reaction
model \cite{Aznauryan:2005tp, Ripani:2000va}. The Monte Carlo events
undergo the same reconstruction and selection chain as for data events.\\[-2ex]

\paragraph{Kinematic fit:}


The sample of events with two or three tracks consisted of four mutually
exclusive topologies,  one with all three final-state hadrons detected and three others in which one out of the three final-state hadrons was missing. In events with
all three particles detected, the energy given by the tagger ($E_t$) was compared with the energy
calculated from the reconstructed final-state particles ($E_p$). The momenta of the charged particles were corrected for
the energy loss in materials of the target assembly~\cite{eloss}. The tagged-photon energies were also corrected
taking into account all known tagger focal-plane mechanical deformations.
After corrections, the $E_t/E_p$ distribution was centered at {\it one} with a width of 0.52\%, which quantifies the relative
energy uncertainty $\Delta E/E$ related to the reconstructed photon energy and the invariant mass of the event. A
kinematic fit with four constraints (energy and momentum conservation) adjusts the total energy within this range.

A kinematic fit was used to select events due to the $\gamma p \to \pi^+ \pi^- p$ reaction~\cite{Williams}.
Events with the three particles detected were constrained by energy and momentum conservation
(4C fit); for events with a missing $\pi^-$, $\pi^+$ or $p$, its momentum had to be determined in the kinematic
fit, leaving one constraint (1C fit). We refer to these topologies as $C_0$ (4C) and $C_i, i=1,2,3$ (1C).

In kinematic fits, the observables $p$, $\lambda$, and $\phi$ ($p$ = modulus of the particle momentum,
$\lambda=\cos\Theta$, $\phi$ = azimuthal angle) are combined into a vector $y$. The measured quantities $\eta$ deviate
from $y$ by $\epsilon$,
\be
  \eta=y+\epsilon\,.
\ee
The improved values of $y$ are determined by the kinematic fit. The fit determines the value of 
the {\it confidence level} ($CL$) defined as
\be
  CL = \int_{\chi^2} ^\infty f(x;n)dx\,,
\ee
where $f(x;n)$ is the probability density function of the $\chi^2$ distribution with
$n$ degrees of freedom. Under ideal conditions, a flat $CL$ distribution is expected.
Given the increase in the distribution towards zero, small values of $CL$ may indicate a significant fraction of background events. Here, events with $CL < 0.1$ are rejected; 72\% of the events have $CL > 10$\%.

\begin{figure}[pb]
\includegraphics[width=0.4\textwidth]{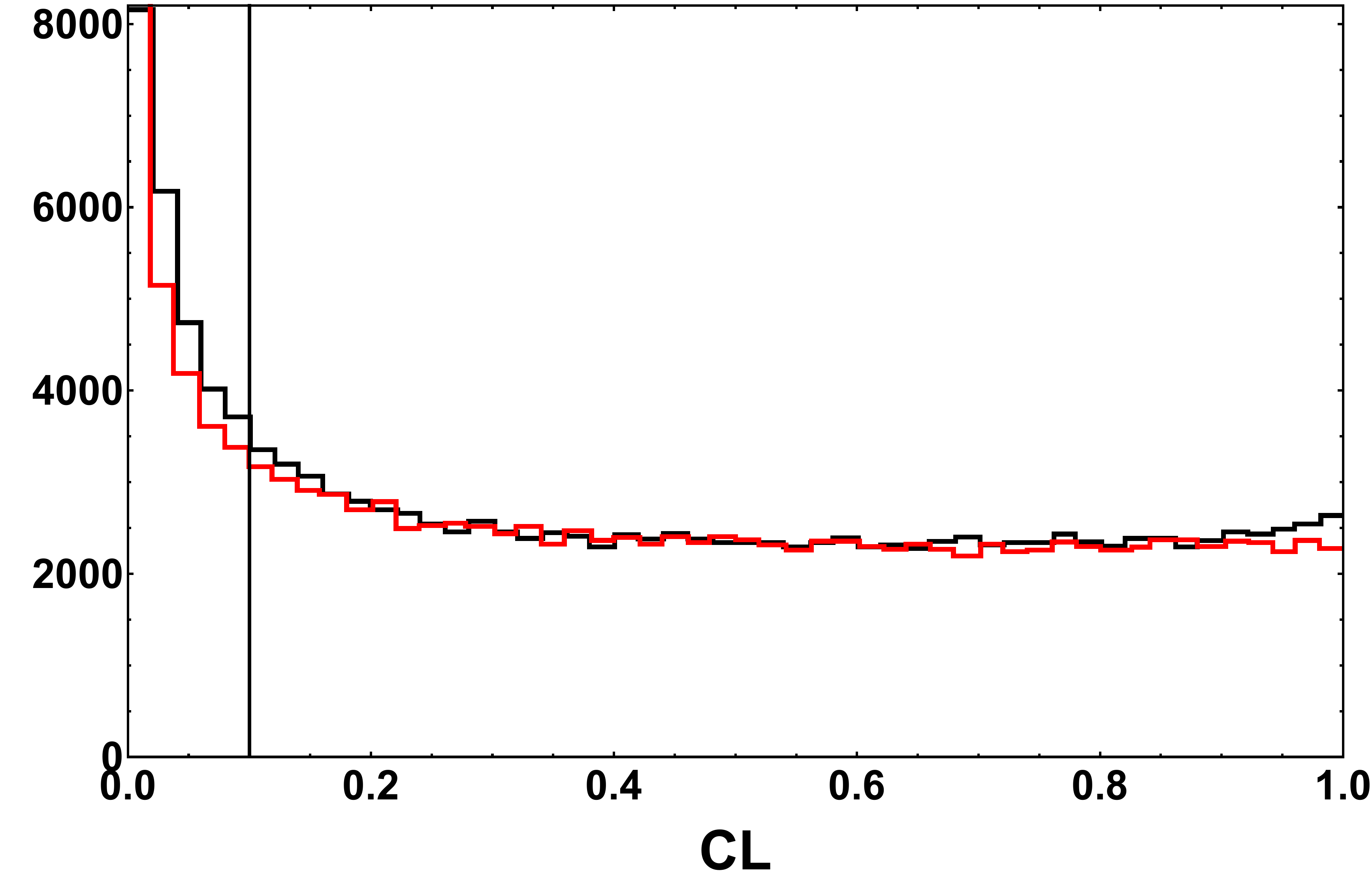}\vspace{-2mm}
\caption{\label{conflev}Confidence level distributions for data (black) and Monte Carlo events (red).
The number of Monte Carlo events is normalized to the number of data events.
}
\end{figure}
Figure~\ref{conflev} shows the confidence level distribution for events passing the 4C fit. The distribution for
measured events shows a reasonably flat distribution and a rise at low CL. This might
signal strong background contributions. These contributions are discussed in the next subsection.

\begin{figure}[htbp]
\includegraphics[width=0.48\textwidth]{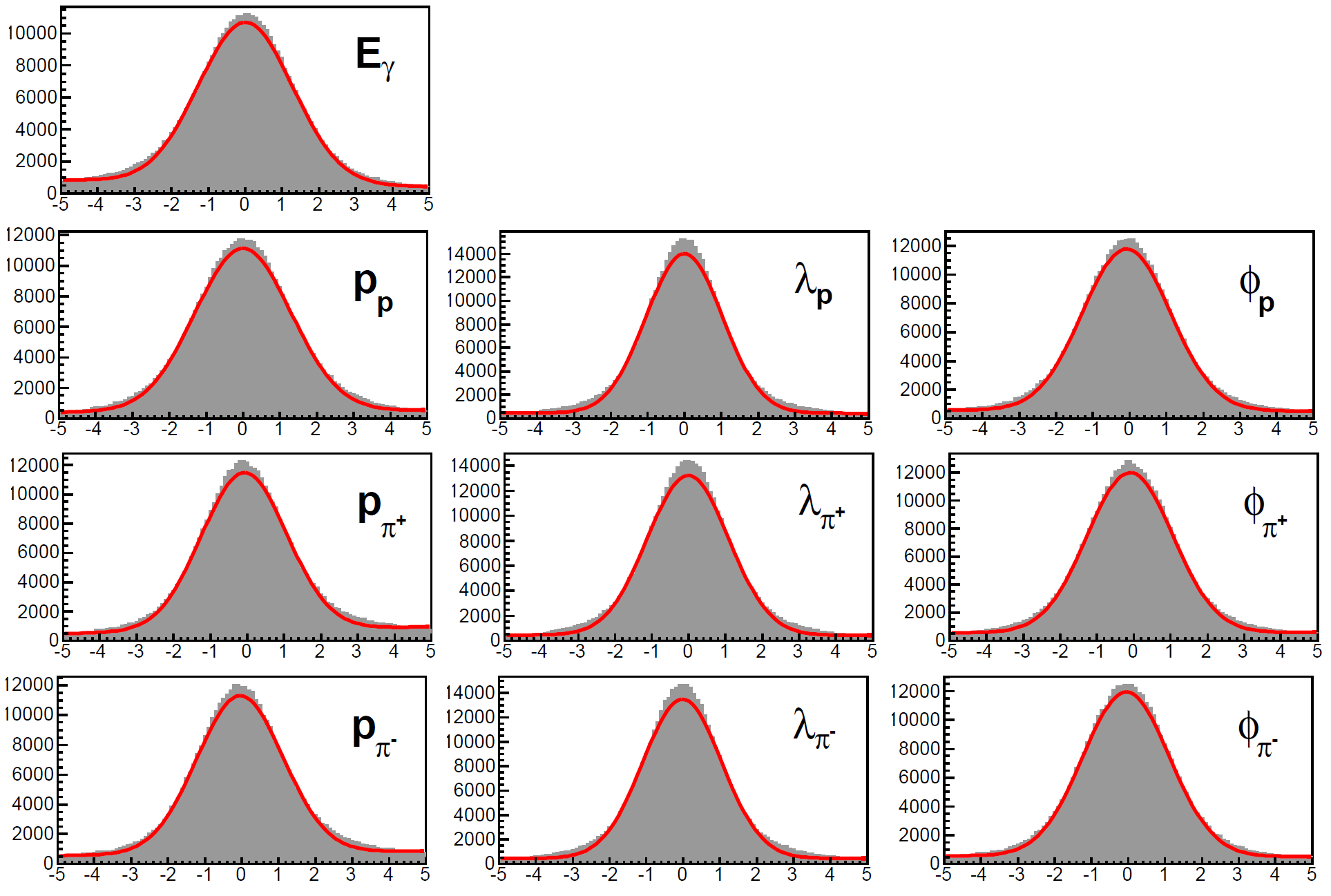}
\caption{\label{pulls} Pull distributions for 4C events: Number of events as a function 
of the pull (see eq.~\ref{pull}). The pull distributions for 1C events are nearly identical.}
\end{figure}

The quality of the error definition can be estimated from the pull distributions. The pull is defined as
\be
\label{pull}
  \frac{\epsilon_i}{\sigma(\epsilon_i)} = \frac{\eta_i - y_i}{\sigma(\eta_i) - \sigma(y_i)}\,
\ee
where $\sigma(y_i)$ is the standard deviation of $y_i$ found in the kinematic fit. When the errors
of the measured quantities are correctly determined, the pulls are
normally distributed around zero with a standard deviation equal to unity.
As an example, in Fig.~\ref{pulls} we show the pull distributions for events with the three measured particles (4C fit).
All pull distributions can be described by Gaussian distributions centered at $0.00 \pm 0.05$ with $\sigma = 1.0 \pm 0.1$.

\subsection{The background}
The selection of two-pion events due to reaction~(\ref{reaction}) relies upon kinematic fitting. 
The confidence level distribution indicates the presence of large background contributions.
A contamination may come from three-pion events $\gamma p\to p\pi^+\pi^-\pi^0$ which could be  
reconstructed as a two-pion event or from badly measured two-pion events.  

Both, two- and three-pion events were simulated with their known cross sections. 
A realistic event generator based on the results reported in Refs.~\cite{Aznauryan:2005tp,Ripani:2000va} was 
used for two-pion events. Three-pion events were generated using a phase space approximation.

\begin{figure}[htbp]
  \includegraphics[width=0.30\textwidth]{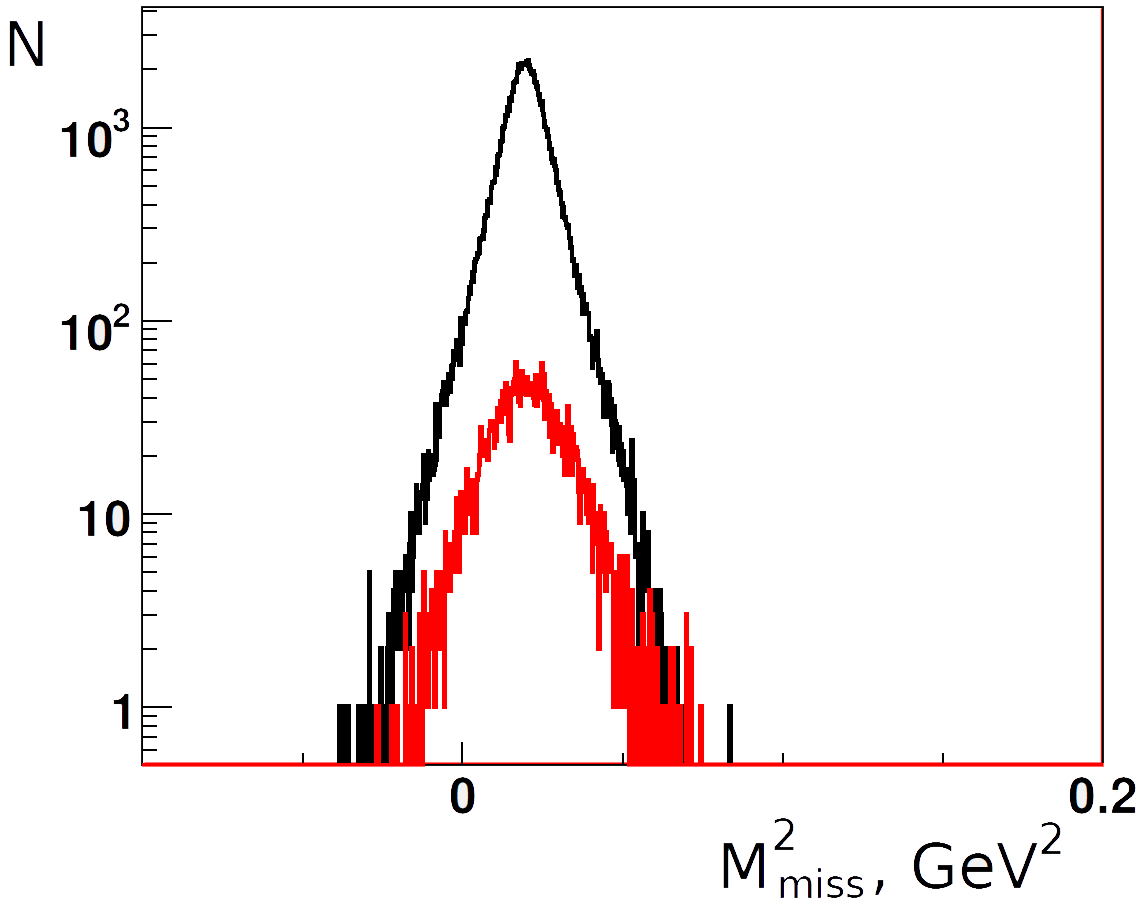}
  \caption{\label{3pi}Missing mass distribution for Monte Carlo events from $\gamma p\to p\pi^+\pi^-$ reactions with a missing $\pi^-$ (black) and $\gamma p\to p\pi^+\pi^-\pi^0$ (red), reconstructed as two-pion events.}
\end{figure}

Figure~\ref{3pi} shows events due to the two reactions with reconstructed $p$ and $\pi^+$, with the missing mass computed
under the assumption that they belong to reaction~(\ref{reaction}). The mean 
three-pion background for these events with a missing $\pi^-$, averaged over all kinematic bins, is 3.2\%. Similar distributions
show that the background for events with a missing $p$ or $\pi^+$ is
2.5\% or 1.3\%. This fraction rises about linearly from zero at $E_\gamma=0.9$\,GeV to (6\er 2)\% at 
$E_\gamma=1.7$\,GeV. The three-pion background vanishes for events in which all three particles
were reconstructed. Obviously, the three-pion background cannot be responsible for the large contribution with $CL<10$\%.

Indeed, the mass and angular distributions of the events with $CL < 10$\% are very similar to those with $CL > 10$\%. 
We conclude that events with $CL < 10$\% are mostly true $p\pi^+\pi^-$ events that are not so well measured,
or events with wrong particle identification. True two-pion events 
that form a spike at small values of $CL$ have similar distributions over all kinematic variables to the
events that pass the $CL$ cut. A cut of $CL < 0.1$ reduces the acceptance by a factor of 0.72. 
Simulated $p\pi^+\pi^-$ events passed the $CL$ cut at 10\% with a probability of 75\%.  
Of course, all simulated events are true $p\pi^+\pi^-$ events. The difference between
the 75\% acceptance for data and 72\% for Monte Carlo events is compatible with the detailed studies in
Ref.~\cite{Williams} that give a 3\% systematic uncertainty on yield extractions due to mismatches of the kinematic fit 
between the data and Monte Carlo events. Hence a correction factor of 1/(0.72\er 0.03) was applied to the obtained cross sections.

\section{Analysis}
\subsection{The coordinate frames}
\begin{figure}
\begin{tabular}{cc}
\includegraphics[width=0.24\textwidth]{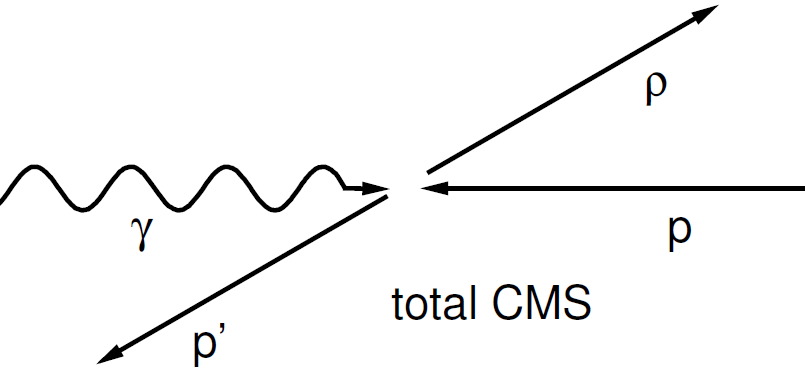}&
\includegraphics[width=0.24\textwidth]{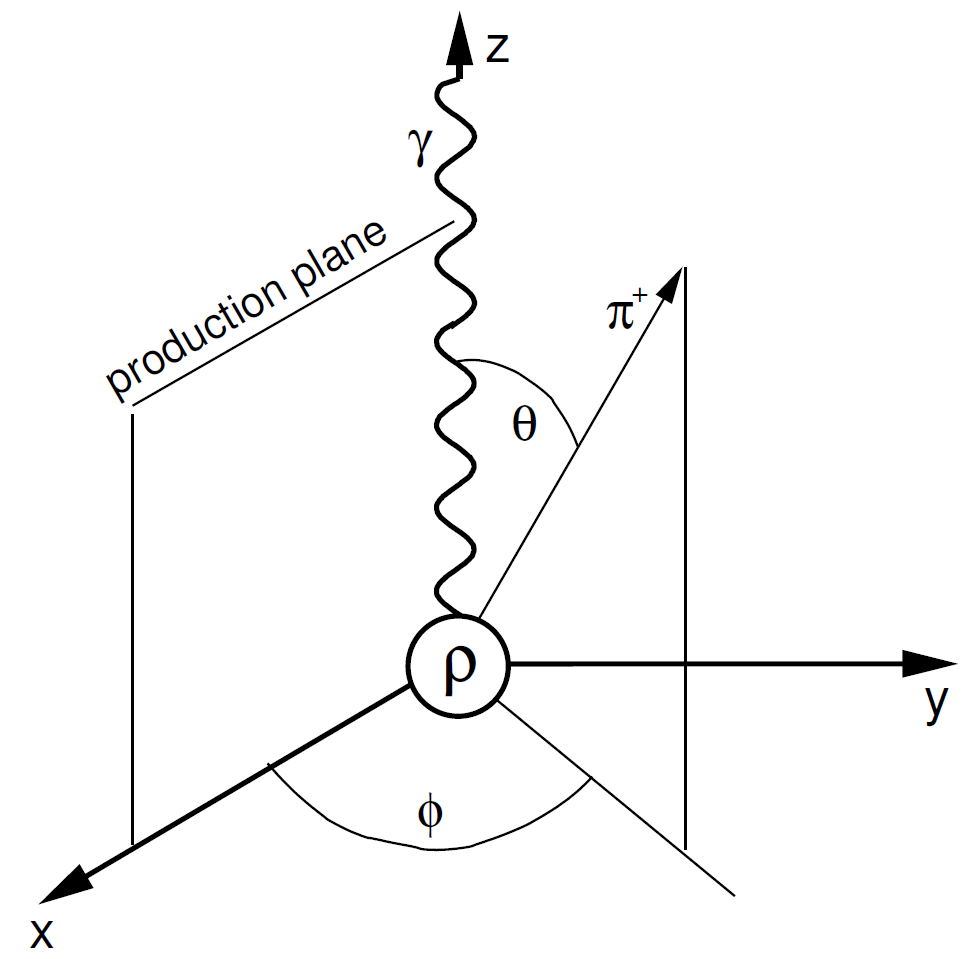}
\end{tabular}
\includegraphics[width=0.48\textwidth]{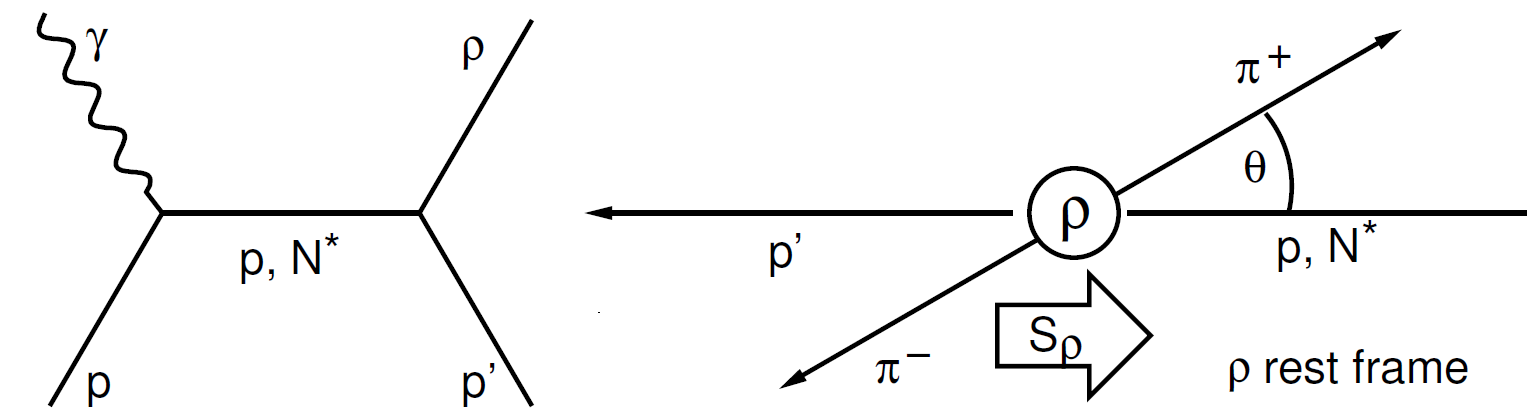}\\[1ex]
\includegraphics[width=0.48\textwidth]{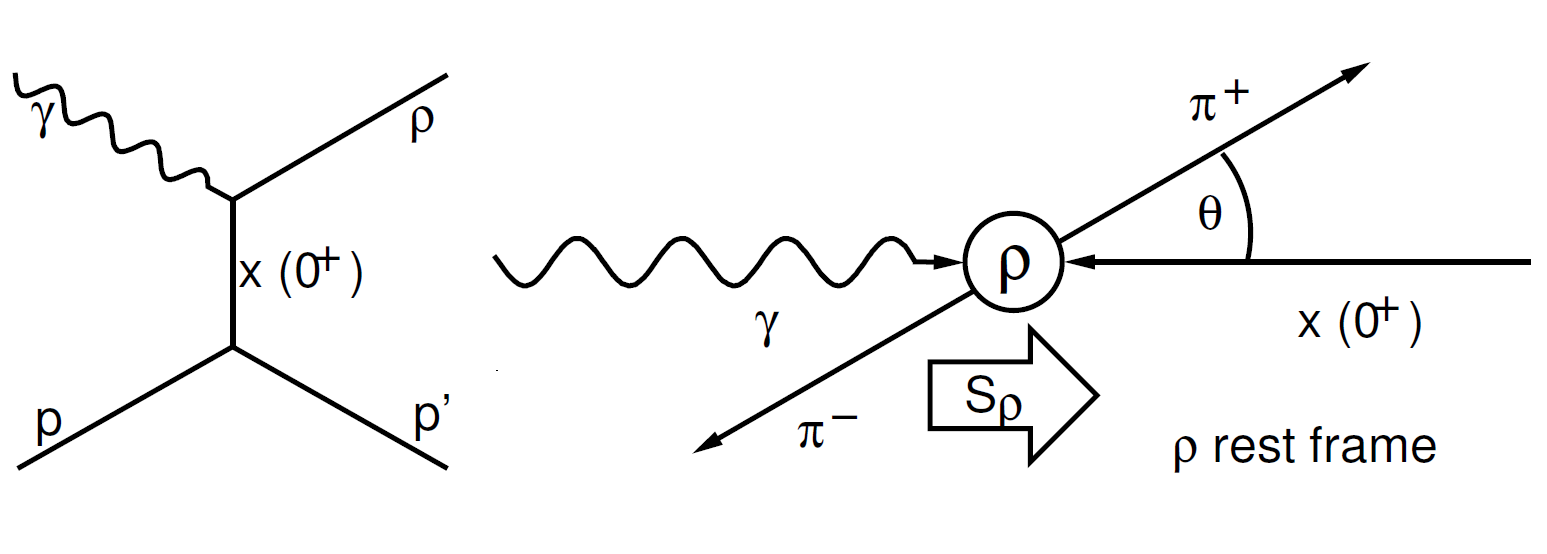}
\caption{\label{cms}In the center-of-mass system, the photon and proton collide having
momenta of the same magnitude. One particle, here the $\pi^+$, and a properly
chosen $z$-axis are used to define the production plane and the angles $\Theta$ and $\phi$
in the $\rho^0(770)$ rest frame (top). 
For the helicity frame the photon direction (center) and in the Gottfried--Jackson
frame the $\rho^0(770)$ direction (bottom) is chosen as the $z$-axis. The helicity 
and Gottfried--Jackson frames are used to characterize $s$- and $t$-channel exchanges.  
S$_\rho$ denotes the $\rho$ spin,  {\footnotesize X}($0^+$) stands for Pomeron exchange.  
(Adapted from \cite{Wu:2005wf}.)}
\end{figure}

The three four-momenta of events in reaction~(\ref{reaction}) are defined by 12 variables.
Energy-momentum conservation (4 constraints) and requirement for the final state hadrons to be on-shell (3 constraints) reduce 
this number to
5 variables, which we choose as invariant masses  $M_{\pi^+\pi^-}$, $M_{p\pi^+}$, $M_{p\pi^-}$, and two angles
$\Theta$ and $\phi$. The angles can be defined in different frames (see Fig.~\ref{cms}).

In the center-of-mass system (cms), the total momentum vanishes; it is related to the laboratory frame by
a Lorentz boost along the $z$-axis. Two further frames require a second Lorentz boost in
the rest frame of one of the three two-particle systems, e.g. in
the $\rho^0(770)$ rest frame; see Fig.~\ref{cms}. The $z$-axis can then be chosen as the direction of the 
photon (Gottfried--Jackson frame) or in the direction of one
of the three two-particle systems (helicity frame). The
polar angle $\Theta$ is then defined as the angle between the decay momentum in the two-particle
subsystem and the $z$-axis.  Figure~\ref{cms} shows these angles in the $\rho^0(770)$ meson rest frame.
For resonance production, the $s$-channel helicity conservation can be tested in the helicity frame.
For particle (or Reggeized particle) exchange in the $t$-channel, the photon spin can be transferred to
the $\rho^0(770)$ meson, and the $t$-channel helicity may be conserved.
 \\[-2ex]

\subsection{The  coupled-channel analysis}
\paragraph{The BnGa database:}
The BnGa database covers all of the main results on pion- and photo-induced reactions. In particular, we
used differential cross sections and polarization observables on the reactions shown in Table~\ref{data}.
The {\it full data set} and  the {\it  event-based data sample}, as well as
new CLAS data on $\vec\gamma \vec p\to p\pi^+\pi^-$~\cite{Crede:2024tbd}, are included in the BnGa database.

\begin{table}[htbp]
\caption{\label{data}The Bonn--Gatchina database used in the  coupled-channel analysis in addition
to the new data on $\gamma p\to p\pi^+\pi^-$. $\pi N\to \pi N$ stands for the real and imaginary parts
of the partial-wave amplitudes for $\pi N$ elastic and charge-exchange scattering derived by the GWU~\cite{GWU} or 
Karlsruhe-Helsinki~\cite{Koch:1980ay} groups.
}
\centering
\renewcommand\arraystretch{1.4}
\begin{tabular}{llccccccc}
\hline\hline
$\gamma p$&$\to$& $\pi^0\pi^0 p$&$\pi^0\eta p$ \ &$ \pi^+\pi^- p$ &\ $\omega p$&\multicolumn{2}{|c|}{$\pi N\to \pi N$}\\\cline{7-8}
$\gamma p$&$\to$ &$\pi N$ &$\eta p$ &$\eta' p$&$K^+\Lambda$&$K^+\Sigma^0$ & \  $ K^0\Sigma^+$\\
$\gamma n$&$\to$&$ \pi N$&$\eta n$& $K^0 \Lambda$ &$K^+\Sigma^-$ &\\
$\pi^- p$&$\to $& &$\eta n $&$K^0\Lambda $ &$K^0\Sigma^0 $& $K^+\Sigma^-$\\
$\pi^- p$&$\to $& $\pi^+\pi^-n$ \ & $\pi^0\pi^0n$ \ & $\pi^0\pi^- p$ \ &\multicolumn{3}{c}{$\pi^+p\to K^+\Sigma^+$}\\
\hline\hline
\end{tabular}
\end{table}

Also included are the real and imaginary parts of the elastic $\pi N\to \pi N$
scattering amplitudes from the GWU~\cite{GWU} or
Karlsruhe-Helsinki~\cite{Koch:1980ay} analysis. The full list of the data used can be
found on the BnGa website (\url{https://pwa.hiskp.uni-bonn.de}).
Events with three particles in the final state are included in the fit event-by-event maximizing the log likelihood.
The difference of the $\chi^2$ values from fits to distributions like differential cross sections, polarization
asymmetries and the likelihood for three-particle final states is minimized in the fit. 
All data sets are given a weight to guarantee that they 
have an impact on the fit result but that they do not lead to a significantly worse fit to other data sets.
In first studies, the weight of a new data set is increased until the overall $\chi^2$ starts to increase.\\[-2ex]

\paragraph{The BnGa PWA approach:}
In the BnGa approach, the energy-dependent part of
the photoproduction amplitude is described by a $D$-matrix that is
based on dispersion relations and a one-step subtraction. The amplitude is
covariant and preserves analyticity and two-particle unitarity. The
photoproduction amplitude contains coupling
constants describing transitions from initial to final states
via resonances and via non-resonant transitions.
Resonances correspond to $K$-matrix poles. Further
contributions with at most left-hand singularities are approximated by non-resonant terms. 
The angular momentum barrier is taken into account by
Blatt--Weisskopf form factors. Rescattering within the
three-body final state is taken into account. The explicit formulae
are given elsewhere~\cite{Sarantsev:2024tbd}.

\begin{figure*}
\centering
\includegraphics[width=0.85\textwidth]{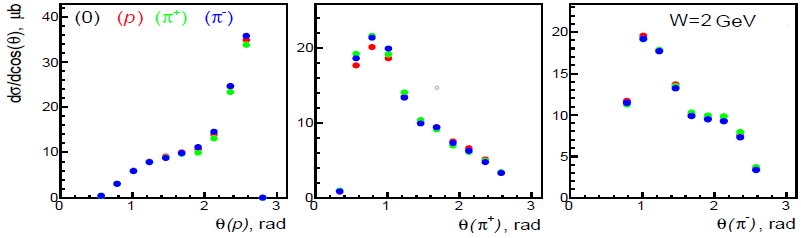}
\caption{\label{4top-full}Comparison of the differential cross sections $d\sigma/d\cos\Theta$ for different
missing particle topologies, where
$\Theta$ is the polar angle of the proton or pion. The color of the data points indicate the topology. The black points are not seen as they are covered by the other points.
}
\end{figure*}
\begin{figure}[tb]
\begin{overpic}[width=0.46\textwidth]{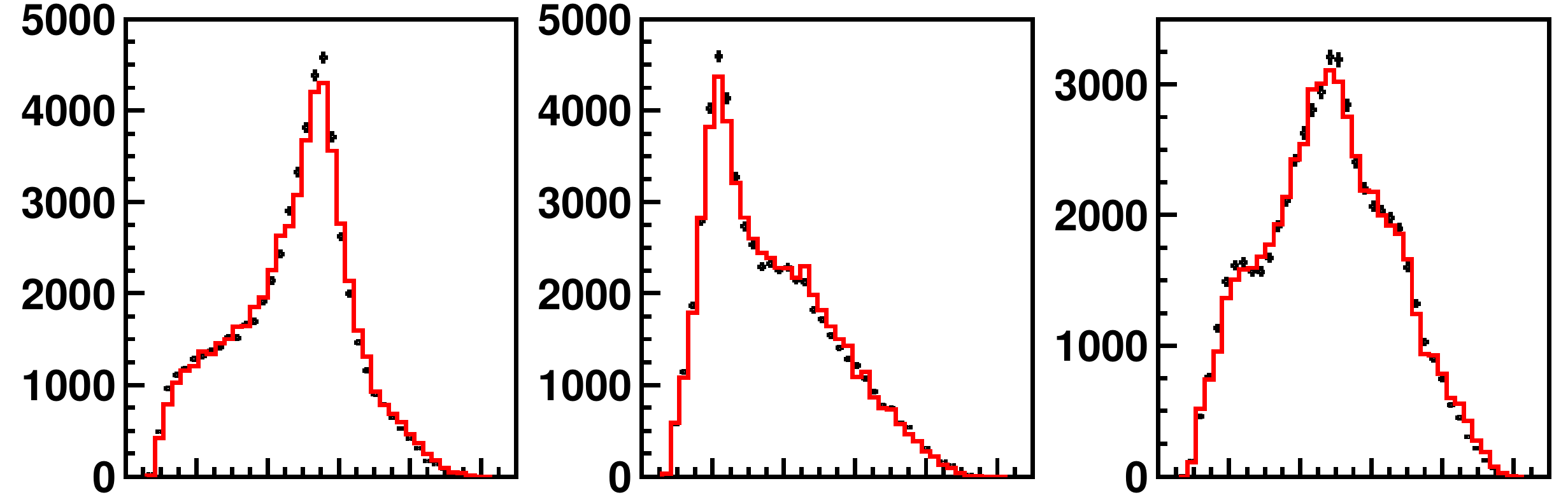}
\put(28,26){$a_0$}
\put(60,26){$b_0$}
\put(93,26){$c_0$}
\end{overpic}
\\
\begin{overpic}[width=0.46\textwidth]{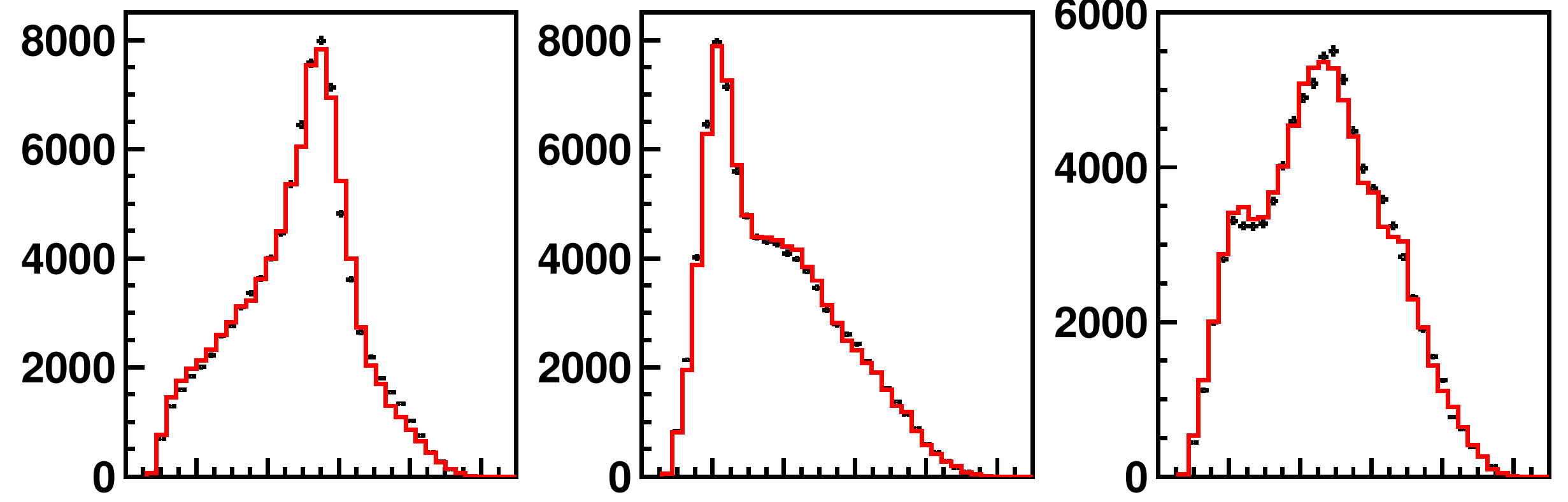}
\put(28,26){$a_1$}
\put(60,26){$b_1$}
\put(93,26){$c_1$}
\end{overpic}
\\
\begin{overpic}[width=0.46\textwidth]{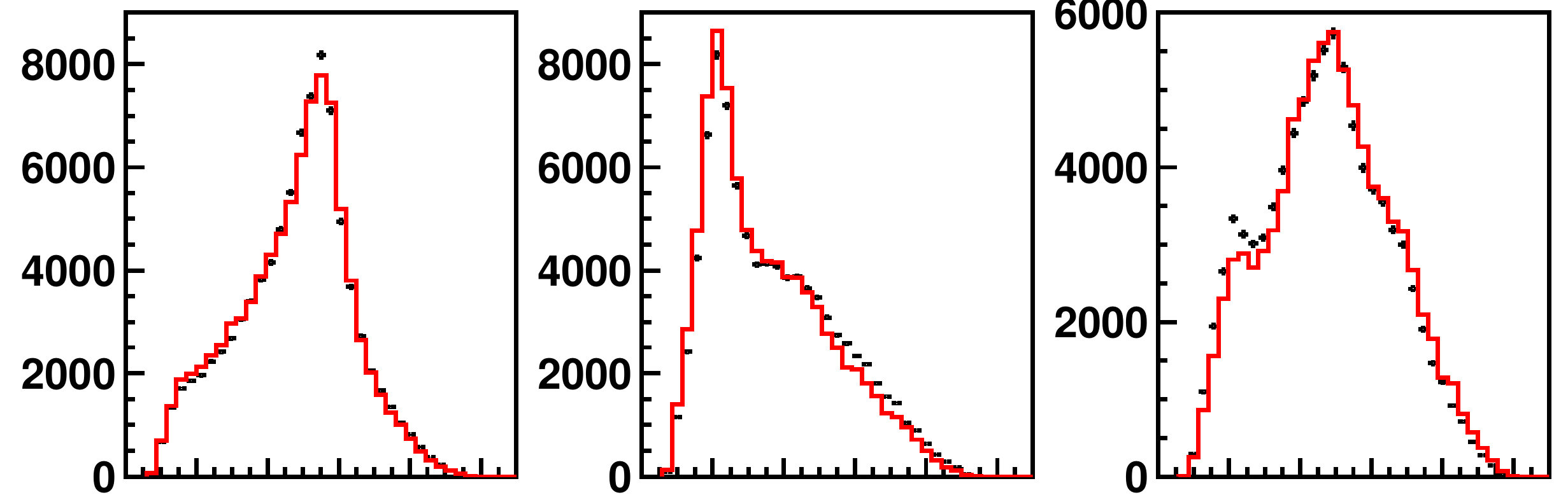}
\put(28,26){$a_2$}
\put(60,26){$b_2$}
\put(93,26){$c_2$}
\end{overpic}
\\
\begin{overpic}[width=0.46\textwidth]{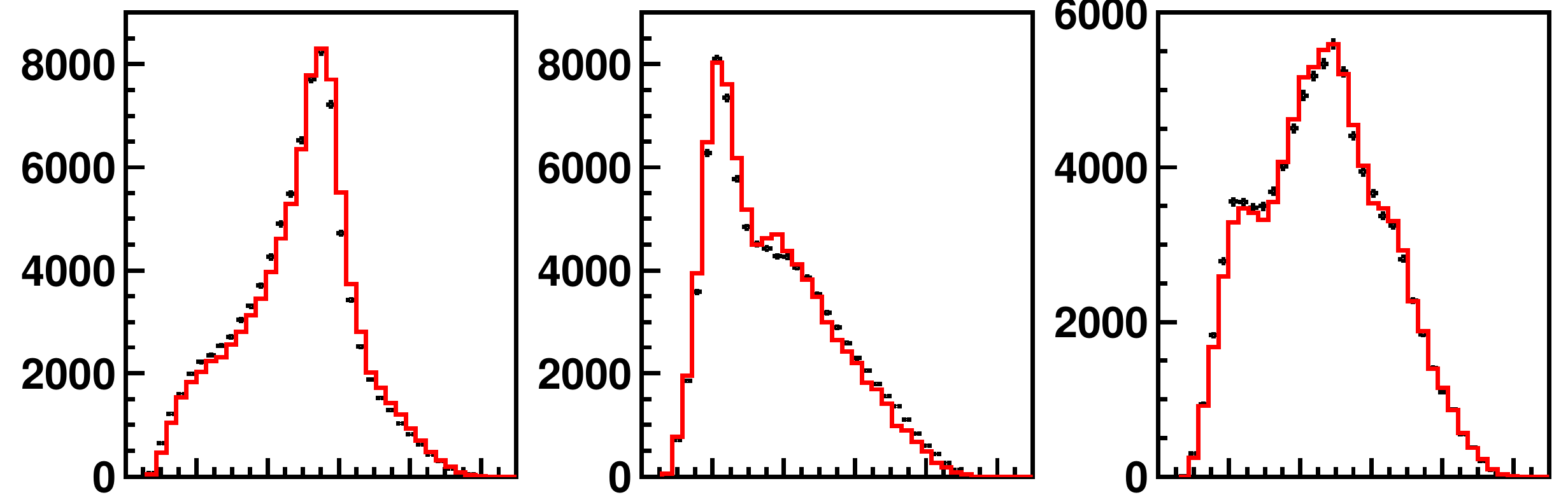}
\put(28,26){$a_3$}
\put(60,26){$b_3$}
\put(93,26){$c_3$}
\end{overpic}
\\
\includegraphics[width=0.48\textwidth]{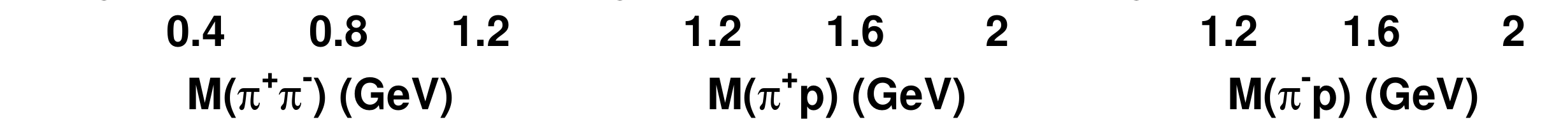}
\caption{\label{4top-mass}The $\pi^+\pi^-$ (a), $p\pi^+$ (b), and $p\pi^-$ (c) mass distributions for the mass range $1.9<W<2.1$\,GeV for the four topologies $C_i, i= 0, 1, 2, 3$. The distributions are not corrected for efficiency. The data are represented by crosses, our fit (in red)
by the histogram. 
}
\end{figure}

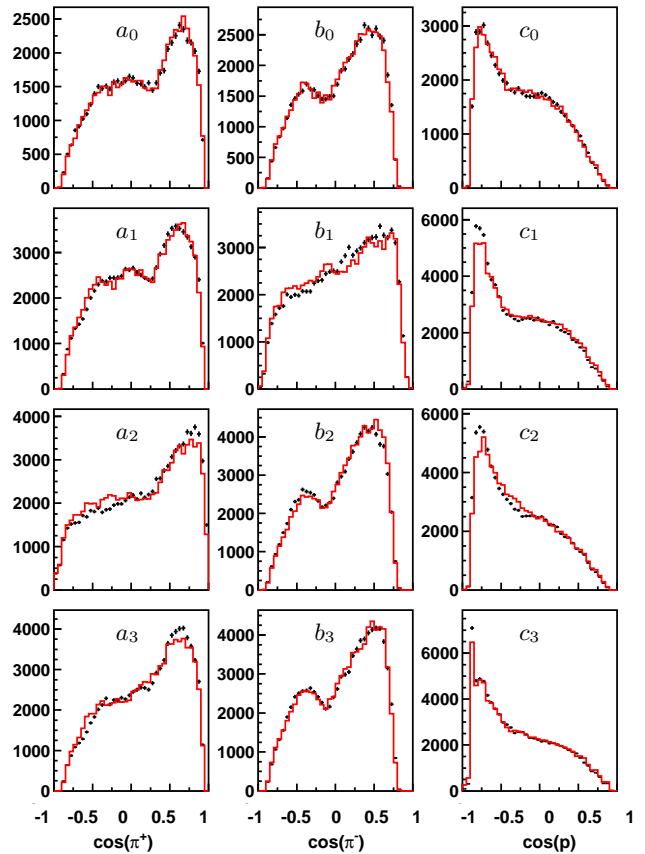
\begin{figure}[tb]
\begin{overpic}[width=0.46\textwidth]{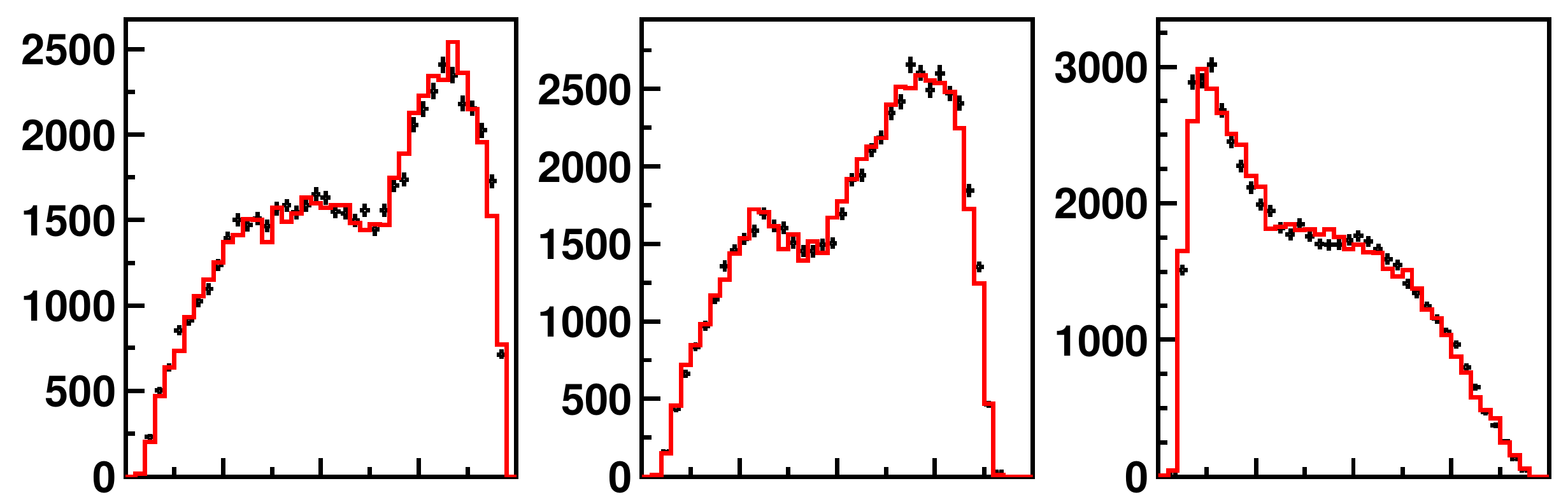}
\put(18,26){$a_0$}
\put(50,26){$b_0$}
\put(83,26){$c_0$}
\end{overpic}
\\
\begin{overpic}[width=0.46\textwidth]{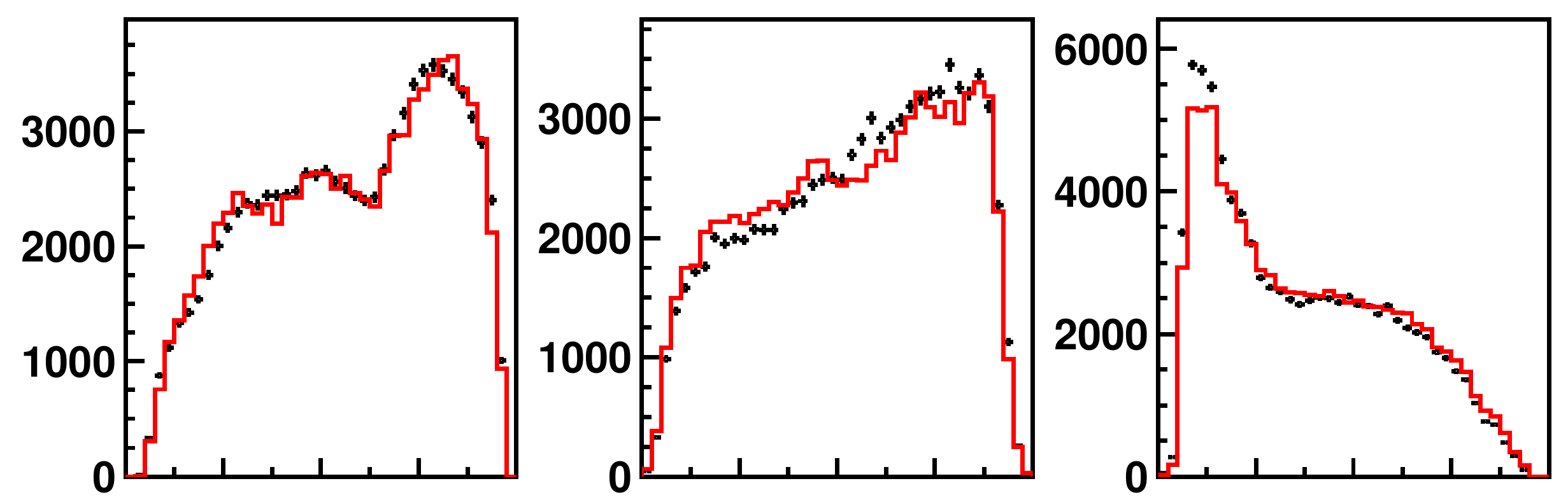}
\put(18,26){$a_1$}
\put(50,26){$b_1$}
\put(83,26){$c_1$}
\end{overpic}
\\
\begin{overpic}[width=0.46\textwidth]{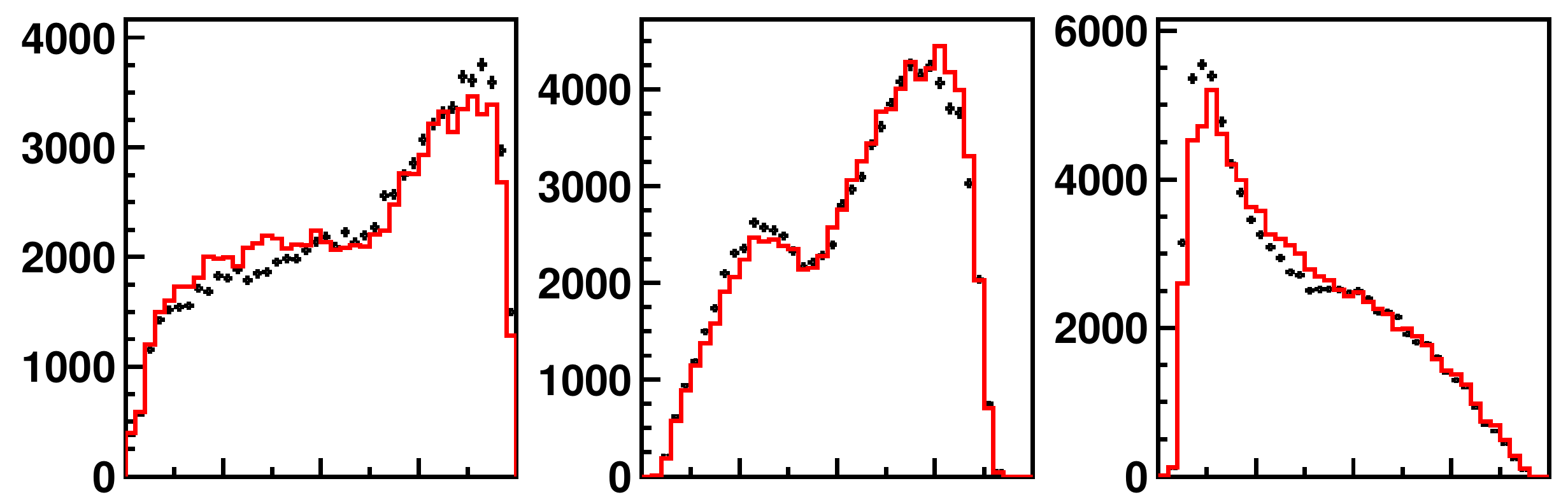}
\put(18,26){$a_2$}
\put(50,26){$b_2$}
\put(83,26){$c_2$}
\end{overpic}
\\
\begin{overpic}[width=0.46\textwidth]{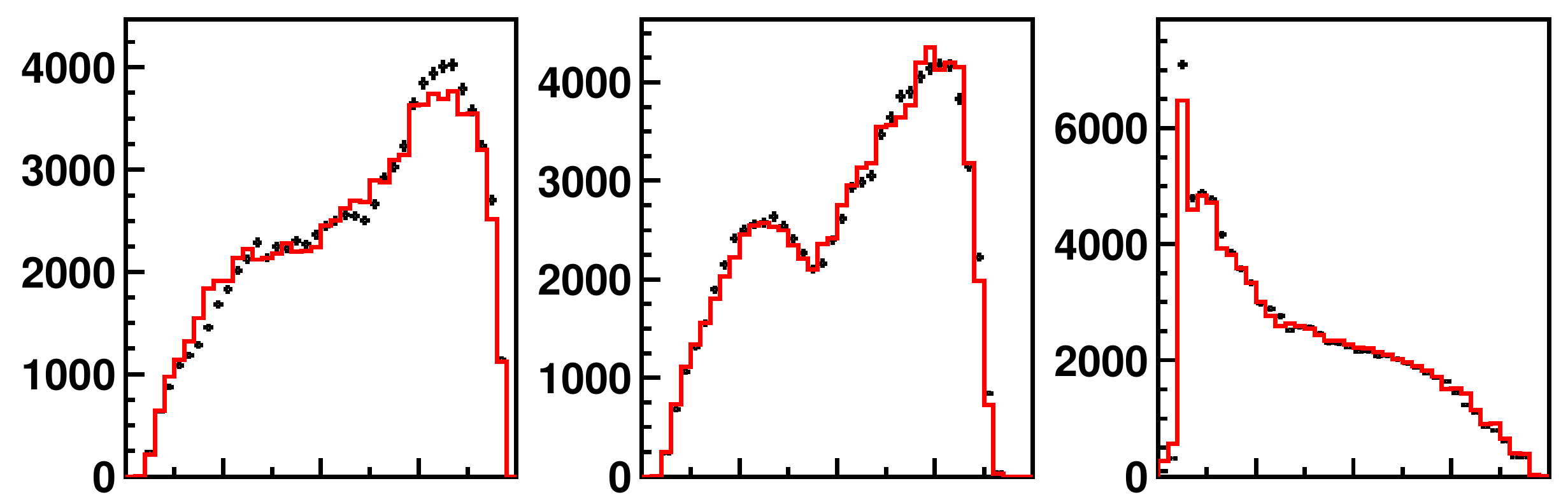}
\put(18,26){$a_3$}
\put(50,26){$b_3$}
\put(83,26){$c_3$}
\end{overpic}
\\
\includegraphics[width=0.48\textwidth]{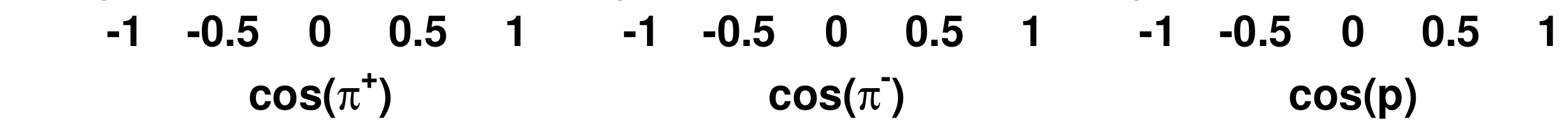}
\caption{\label{4top-angles}The $\cos\Theta_{\pi^+}$ (a), $\cos\Theta_{\pi^-}$ (b), and $\cos\Theta_p$ (c) angular
distributions in the c.m.s.\ for the mass range $1.9<W<2.1$\,GeV and for the
 four topologies $C_i, i= 0, 1, 2, 3$. The distributions are not corrected for efficiency.
 The data are represented by crosses, our fit (in red) by the histogram. 
}
\end{figure}

\subsection{Full data set and event-based data sample}
Overall, about 400 million $\pi^+\pi^-p$ events were selected. These events were used to calculate the
invariant-mass and angular distributions.  The CLAS detector does not cover the full angular range. 
To reconstruct a $p\pi^+\pi^-$ event, only
two particles need to be detected, but some events may be lost. A dynamical model is required.  
For the {\it full data set} a model was used, developed by the JLab-MSU Collaboration~\cite{Mokeev:2012vsa,Mokeev:2008iw}, 
which was successfully applied to describe the cross section for photo- and electroproduction
(in the $Q^2$ range from 0.25\,GeV$^2$ to 1.5\,GeV$^2$) of two charged pions off protons. 
All ingredients of the model are the same when describing
reactions induced by real or virtual photons. Therefore, it was used for a model extrapolation
of the cross sections published in Ref.~\cite{CLAS:2018drk}. This data is called the {\it full data set}.   

The parameters of the resonances were optimized by minimizing the negative logarithm 
of the likelihood function:
\be
-\ln L=-\sum\limits_{i=1}^{N_{data}}\left (\ln \sigma_i-\ln \frac{1}{N_{MC}}{\sum\limits_{j=1}^{N_{MC}}\sigma_j}\right ),
\ee
where $\sigma_i$ is the differential cross section calculated for the reconstructed data events, 
normalized to the sum of the differential cross sections for the reconstructed  Monte Carlo events. 

The number of events is too large to be used in event-based likelihood fits. Therefore we used a subsample 
consisting of 75512 4C fit events ($C_0$), and 660019  events  with the $\pi^-$ missing ($C_1$), 616221  events
with the $\pi^+$ missing ($C_2$), and 490576  events with the proton missing ($C_3$). 
This is only a small sample compared to the full data sample. However, in an event-based likelihood fit,
all correlations within the three-particle dynamics are taken into account. As we shall see, the systematic uncertainties
of the physics results far exceed the statistical precision. Hence, the reduced statistics is not a real problem.
The acceptance of the {\it  event-based data sample} is determined by a comparison of reconstructed and 
generated Monte Carlo events weighted with the result of the BnGa coupled-channel analysis.  
\\[-2ex]

\paragraph{The four topologies:}

The four topologies, fully reconstructed events and events with a missing proton, $\pi^+$, or $\pi^-$,  
reflect one dynamical process only. Thus, the mass and angular distributions in all four topologies
should be described by one fit. However, the four topologies have different
acceptances. Thus, mass and angular distributions could show different shapes. Hence, acceptance has to be taken into account and a precise understanding of the detector is mandatory. The detector efficiency was
calculated using a detailed GEANT simulation of the CLAS detector called GSIM~\cite{gsim}
and an event generator based on the older JM05 reaction model~\cite{Aznauryan:2005tp, Ripani:2000va}. The use of an event generator simulating the reaction 
is necessary not to produce large Monte Carlo data sets in kinematic areas of low cross sections. The details of
the reaction model are irrelevant to the likelihood fit.

Figure~\ref{4top-full}  presents acceptance-corrected angular distributions of $p$, $\pi^+$, and $\pi^-$ for the four 
topologies at $W=2$\,GeV in bins of 25\,MeV or
0.04 in $\cos\Theta$ for the {\it full data set}. The consistency is very good. We conclude that the
acceptance is reasonably well accounted for by the Monte Carlo simulation.

In Fig.~\ref{4top-mass}, we present invariant mass distributions for the {\it  event-based data sample} 
and the four topologies at $W=2$\,GeV. The experimental distributions are shown by crosses, the histogram represents
the final fit. Although the distributions are not corrected for efficiency, they are strikingly similar. 
The
solid angle coverage of the CLAS detector for detecting two out of three particles is 
homogeneous.

In all four topologies,
the $\pi^+\pi^-$ mass distributions show the important role of $\rho^0(770)$ meson production in
photoproduction of two charged pions. The $p\pi^+$ mass distribution exhibits a strong peak due to
$\Delta(1232)^{++}$ production. The masses and widths of the $\Delta(1232)^{++}$ seem to be reasonably
consistent.  Small differences can be noticed in the $p\pi^+$ invariant mass distribution in the higher-mass 
region and in the  $p\pi^-$ invariant mass distribution in the lower-mass 
region.  The threshold enhancement
in the $p\pi^-$ mass distribution can be traced to the production of $\Delta(1232)^{0}$. It is much
weaker than the production of $\Delta(1232)^{++}$. The large enhancement 
in the $p\pi^-$ invariant mass distribution centered at $\sim 1.8$\,GeV
is a reflection of $\Delta(1232)^{++}$ and $p\rho^0(770)$ production.

A more detailed view is obtained by studying the angular distributions. These are shown in
Fig.~\ref{4top-angles}, again in four rows for the four topologies $C_i$.  Similar
conclusions can be drawn as for the mass distributions: the main features are well reproduced in the
fit, while some small but significant inconsistencies are seen. A remarkable effect is seen for the $b_i$ distributions:
$b_0$, $b_2$, $b_3$ show a pronounced dip at $\cos\Theta_{\pi^-}\lessapprox 0$ which is not seen in $b_1$.
In $b_1$ small discrepancies can be observed, but the dip is absent in both the experimental distribution and the fit.

Adding all four topologies will minimize the effects of small deviations of the Monte Carlo simulation and
the response of the CLAS detector. In any case, we have made fits to the individual topologies $i=0,1,2,3$.
The spread of the results obtained from these fits is included as part of the systematic uncertainty.

\section{\boldmath Results}
\label{SectionResults}
\subsection{Dalitz plots}

Figure~\ref{Dalitz1} shows the acceptance-corrected 
Dalitz plots from the {\it event-based data sample} for the reaction~(\ref{reaction}) in bins of 100\,MeV in
$W$ from 1600 to 2150\,MeV.  In the figures, 
$M^2 _{p\pi^-}$ is plotted versus
$M^2 _{p\pi^+}$, then  $M^2 _{p\pi^+}$ versus $M^2 _{\pi^+\pi^-}$. For the Dalitz plots, the four topologies $C_i$ are added. 

Production of neutral $\rho(770)$ mesons is one of the most significant features of the reaction: 
it dominates the reaction above energies of $W>1.80$ GeV.  
Also very important, in particular at low energies, is the production of $\pi^-\Delta(1232)^{++}$. It is much
larger than the production of $\pi^+ \Delta(1232)^{0}$.  The Clebsch--Gordan coefficients 
imply that high-mass $N^*$ resonances decaying to $\pi\Delta(1232)$
contribute to 
$\pi^-\Delta(1232)^{++}$ 9 times more than
to $\pi^+\Delta(1232)^{0}$, for $\Delta^*$ resonances, the ratio is 9:4.
The even higher yield of $\pi^-\Delta(1232)^{++}$ points to a production mode different from only
resonance contributions. This is due to the Kroll--Rudermann 
mechanism \cite{Kroll:1954}: the impinging photon
can dissociate the charges within the proton, creating a $\Delta(1232)^{++}$ and a charged pion.
This effect is particularly strong at low energies but persists up to the highest energies reported here.
Here, it is introduced through a formalism suggested in Ref.~\cite{Guidal:1997hy}.

\begin{figure}
\begin{tabular}{cc}
\phantom{z}\\
\hspace{-2mm}\includegraphics[width=0.24\textwidth]{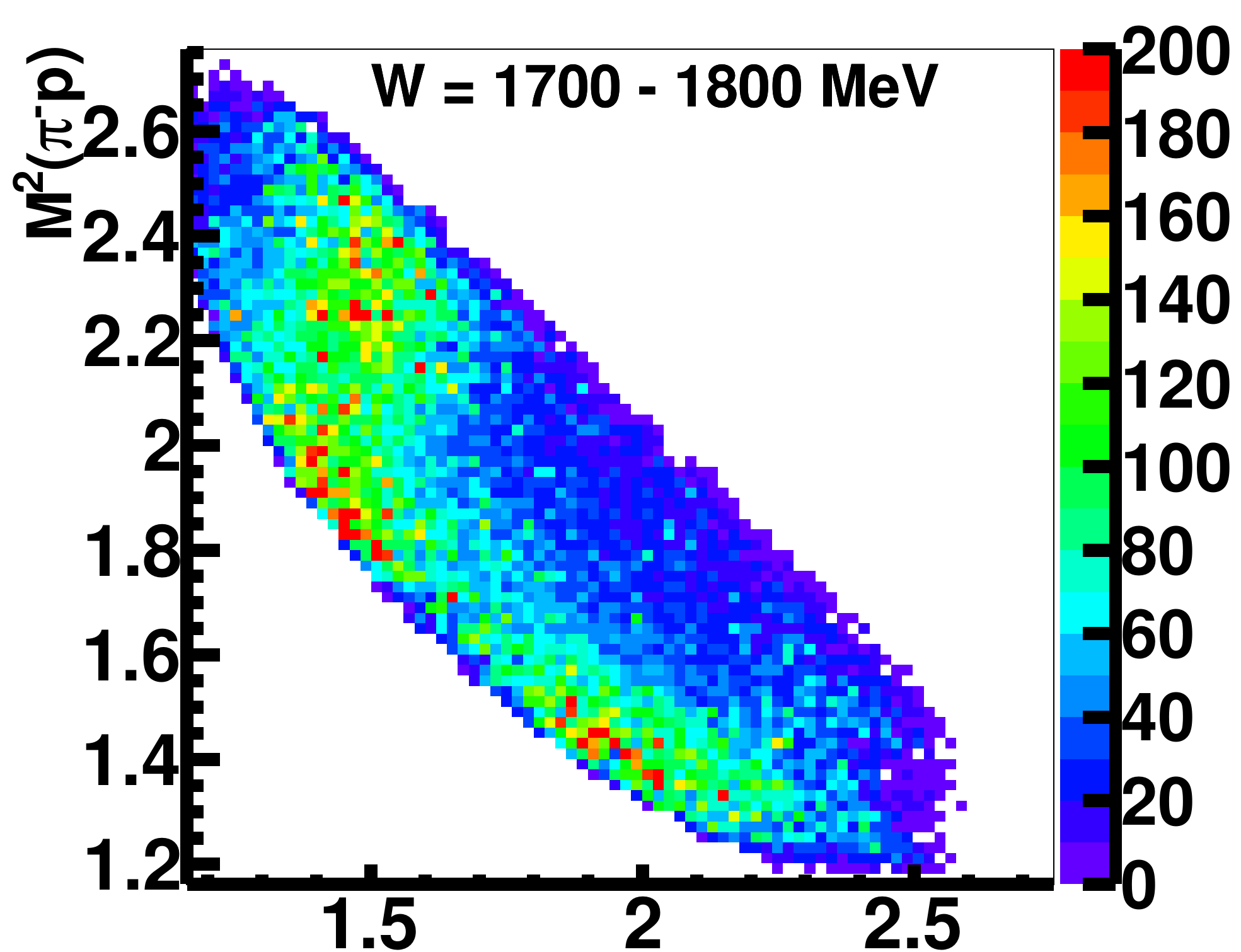}&
\hspace{-1mm}\includegraphics[width=0.24\textwidth]{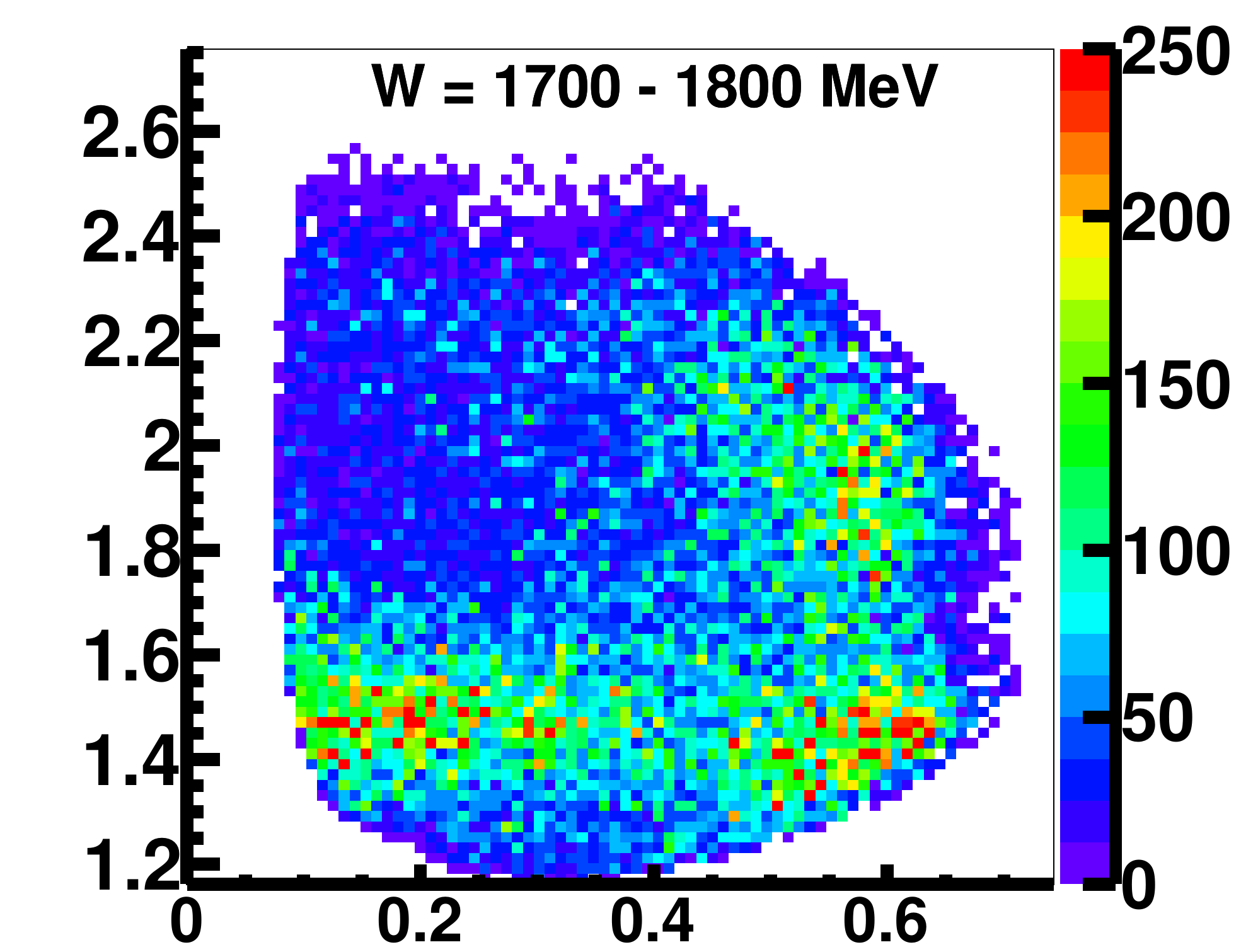}
\vspace{-1mm}\\
\hspace{-2mm}\includegraphics[width=0.24\textwidth]{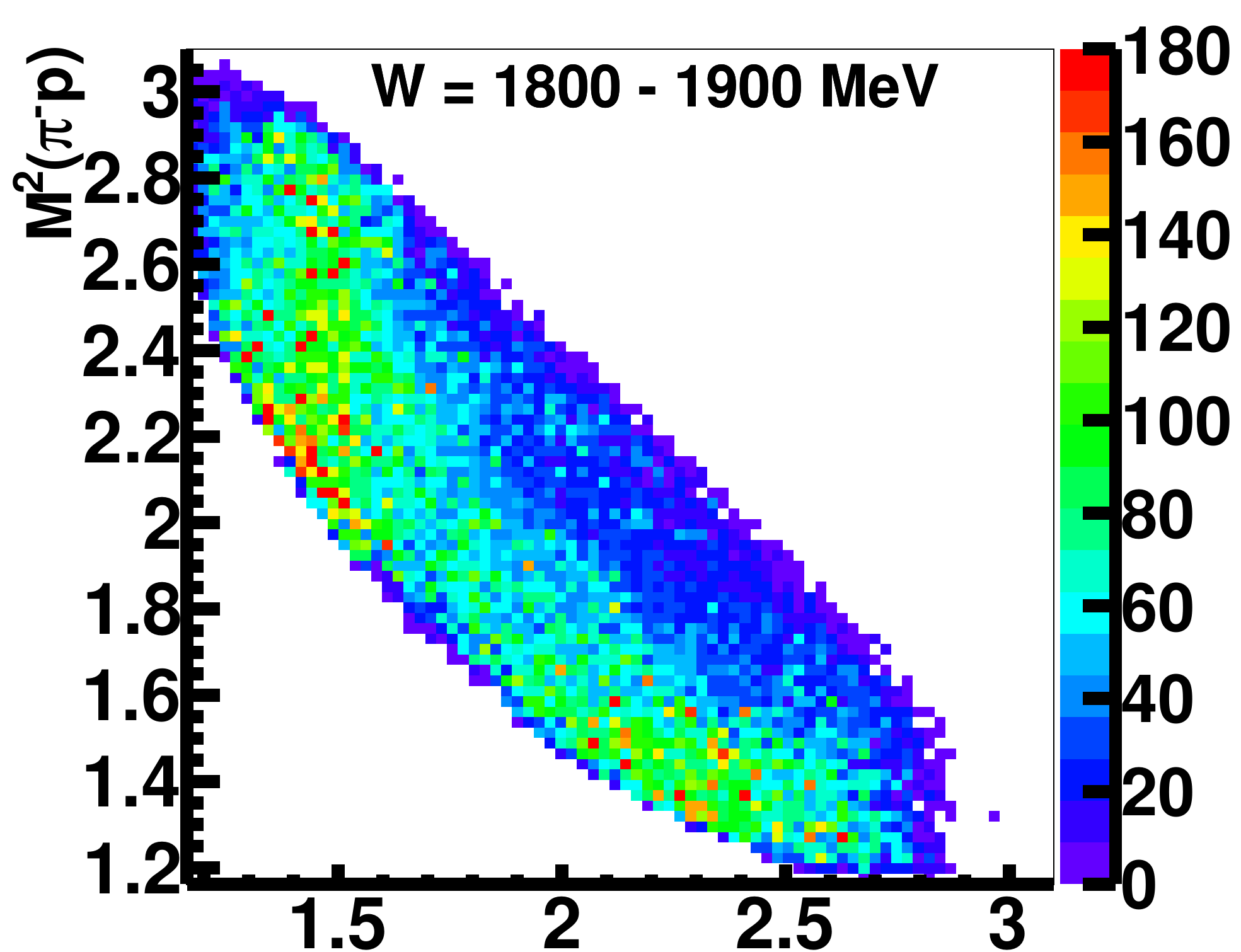}&
\hspace{-1mm}\includegraphics[width=0.24\textwidth]{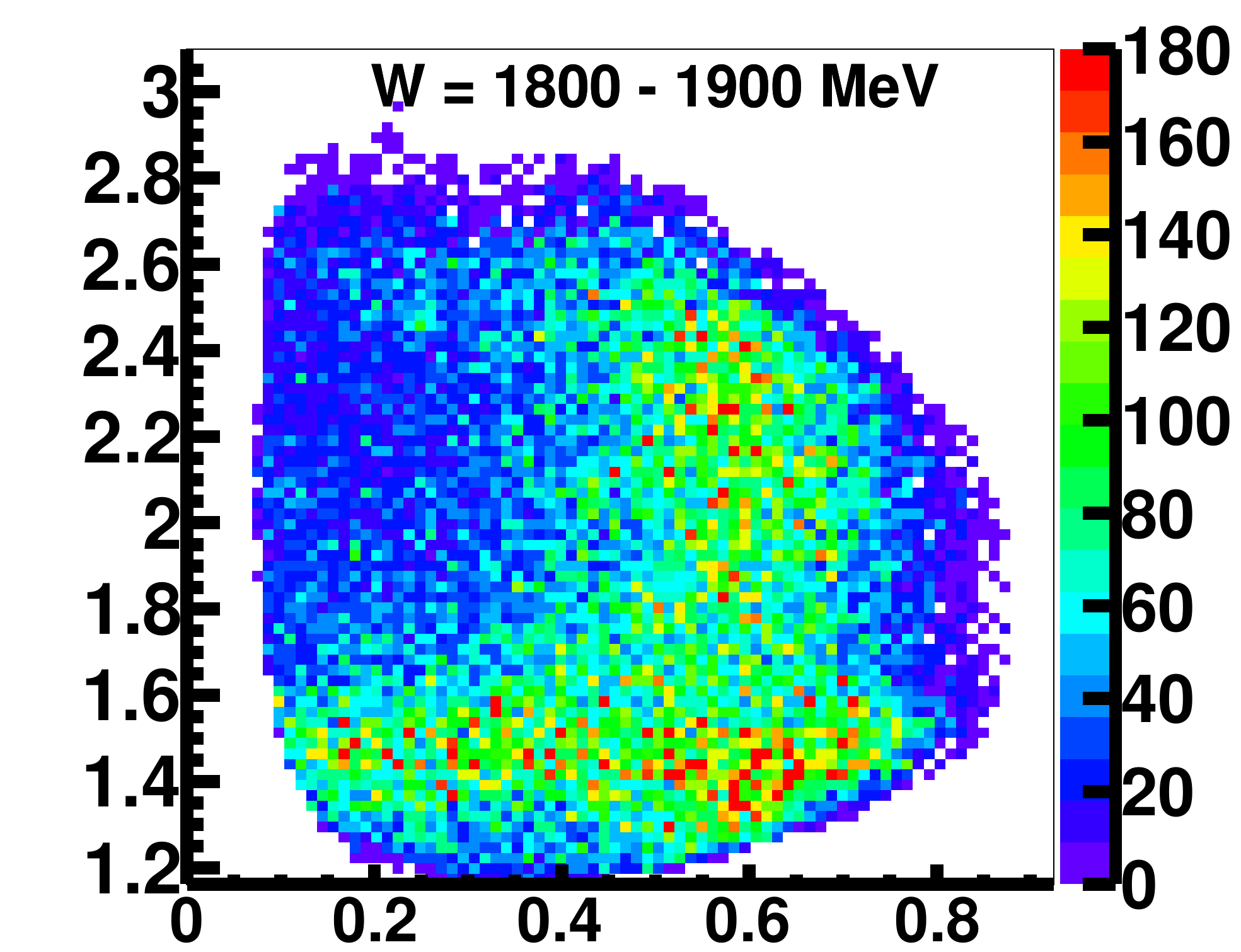}
\vspace{-1mm}\\
\hspace{-2mm}\includegraphics[width=0.24\textwidth]{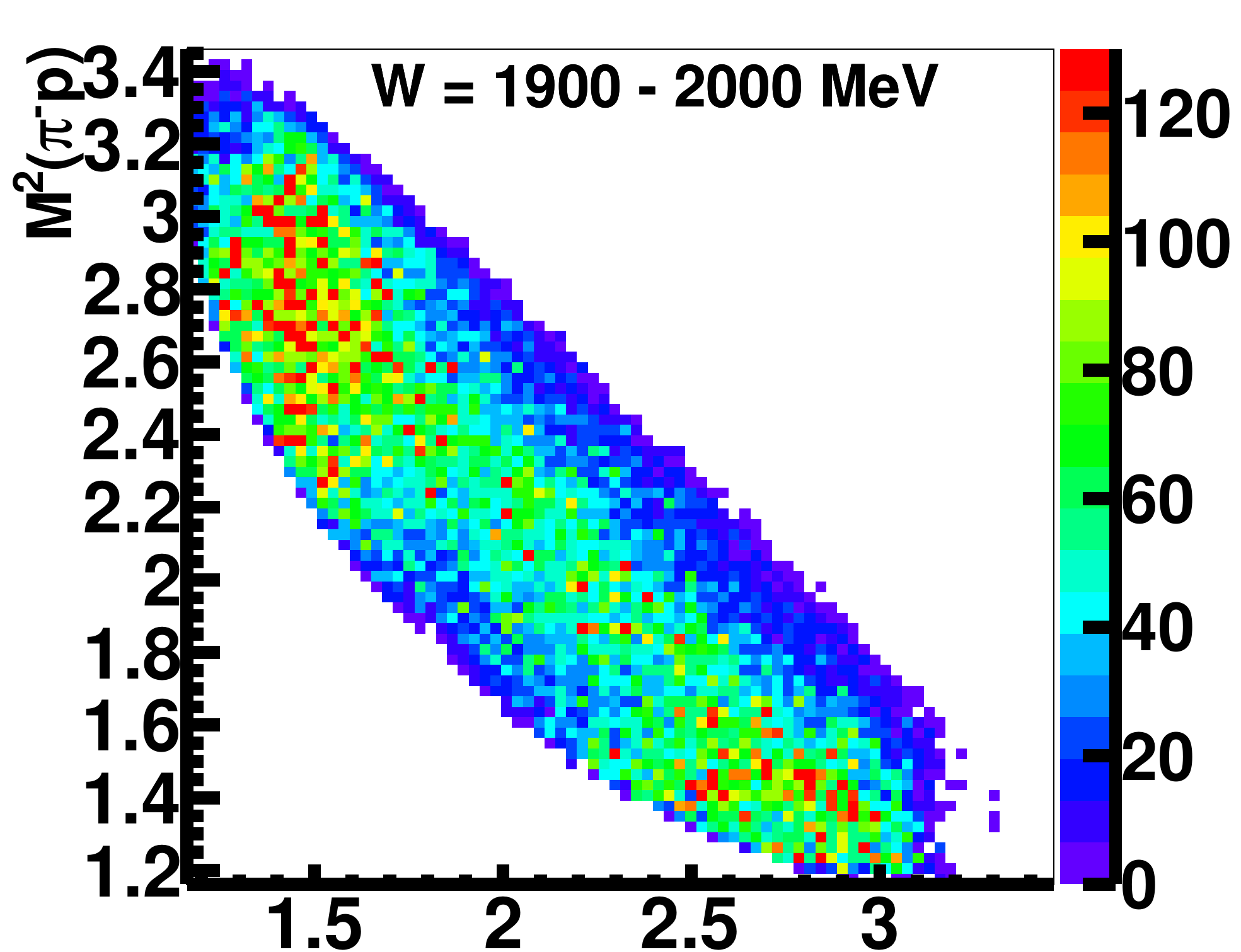}&
\hspace{-1mm}\includegraphics[width=0.24\textwidth]{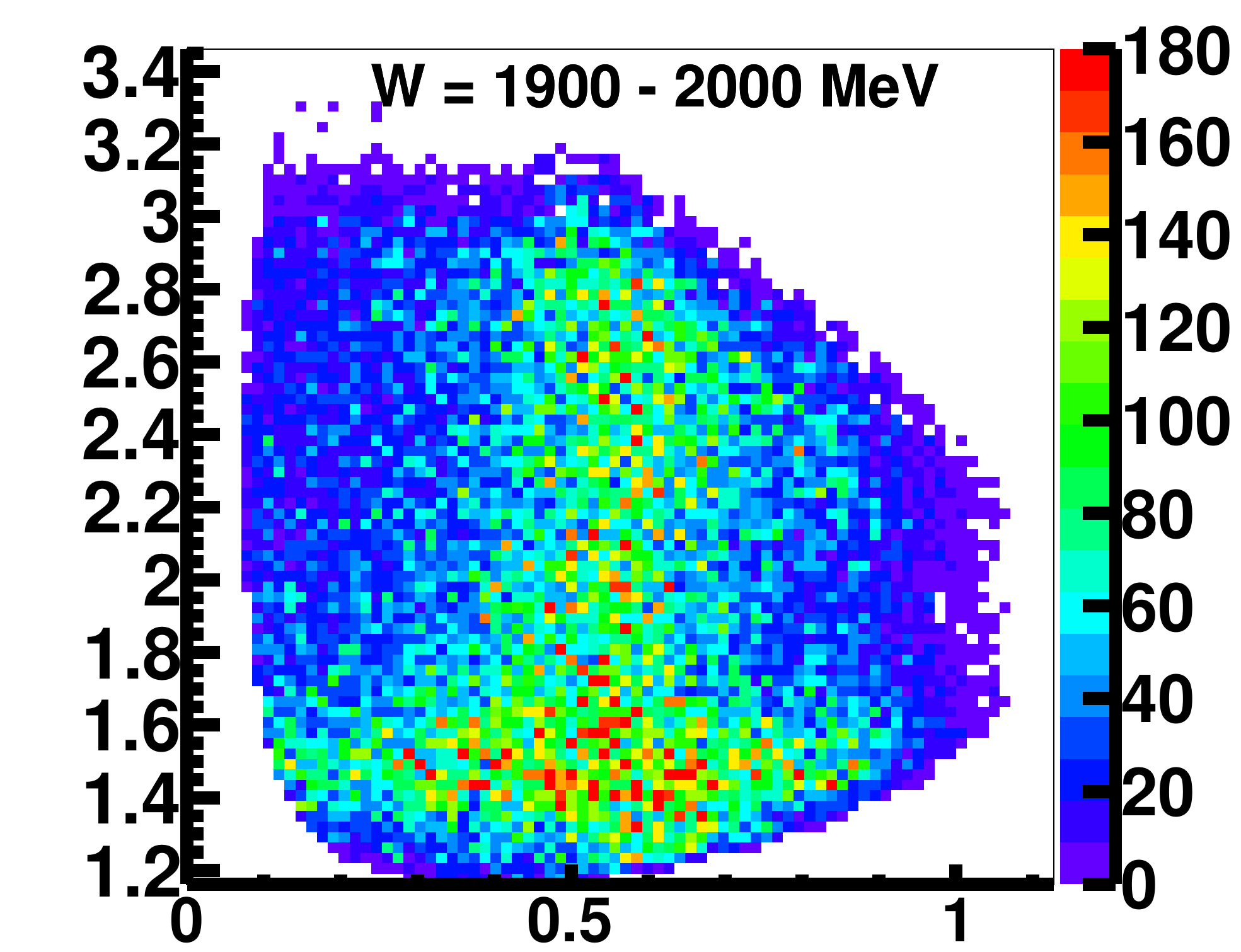}
\vspace{-1mm}\\
\hspace{-2mm}\includegraphics[width=0.24\textwidth]{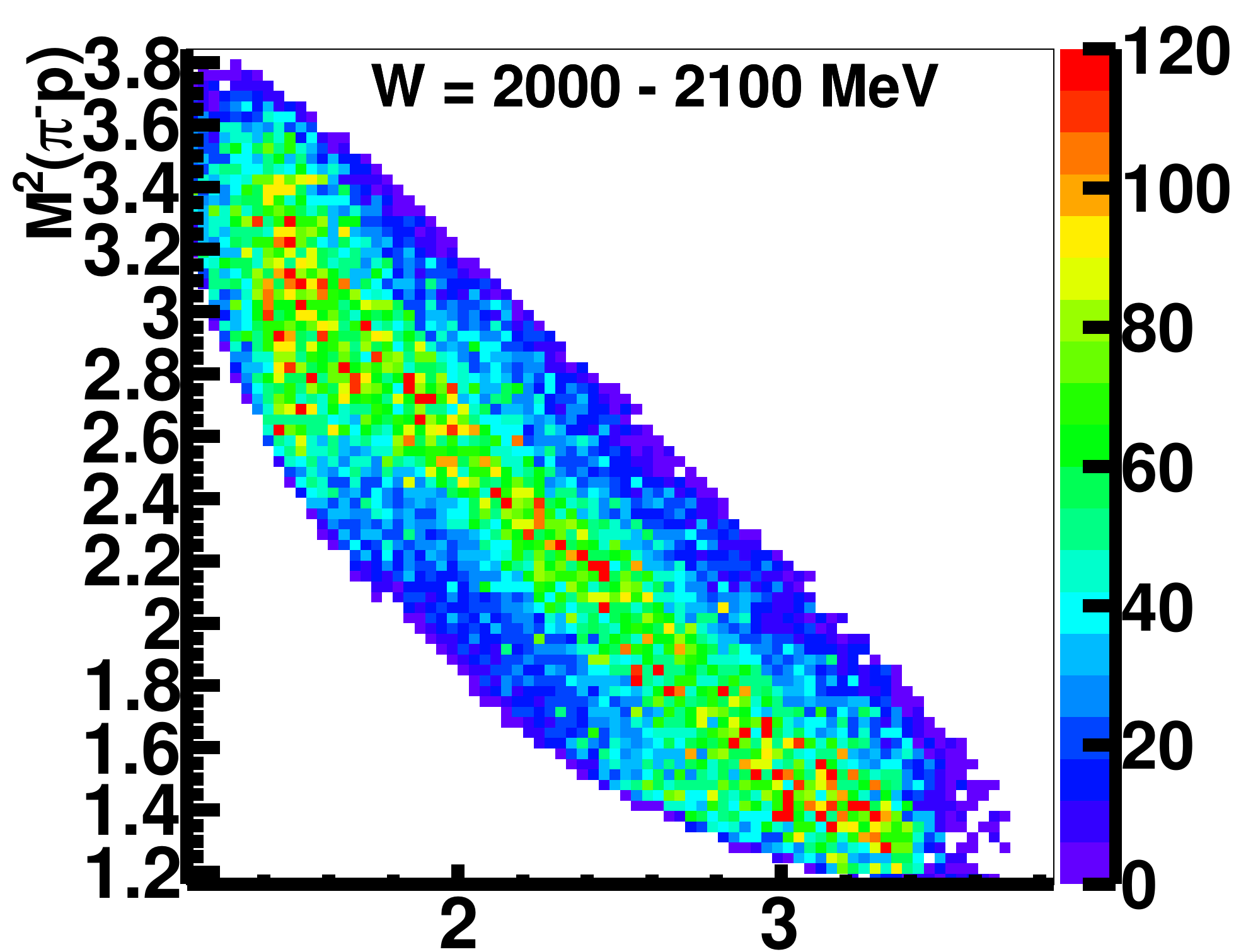}&
\hspace{-1mm}\includegraphics[width=0.24\textwidth]{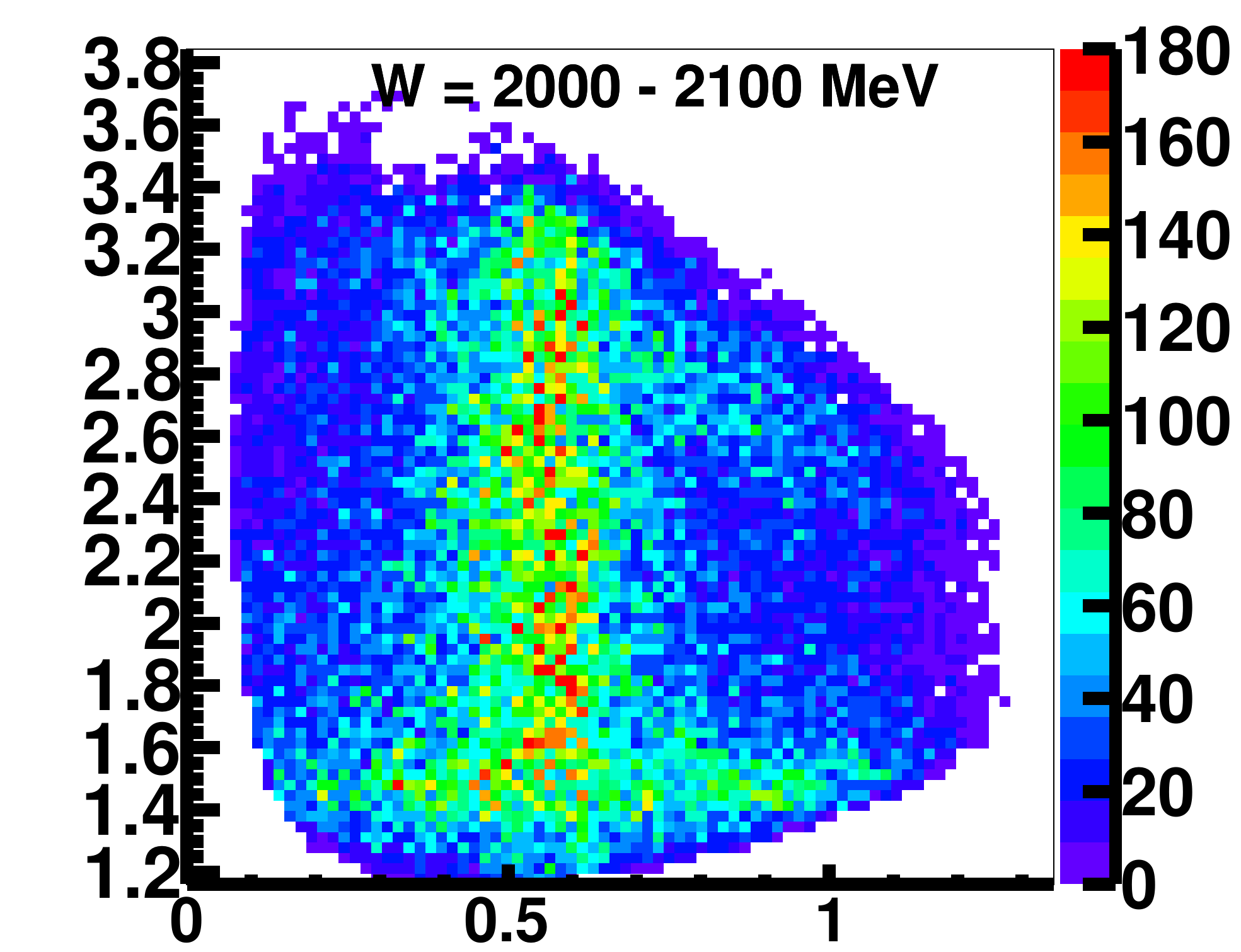}
\vspace{-1mm}\\
\hspace{-4mm}\includegraphics[width=0.24\textwidth]{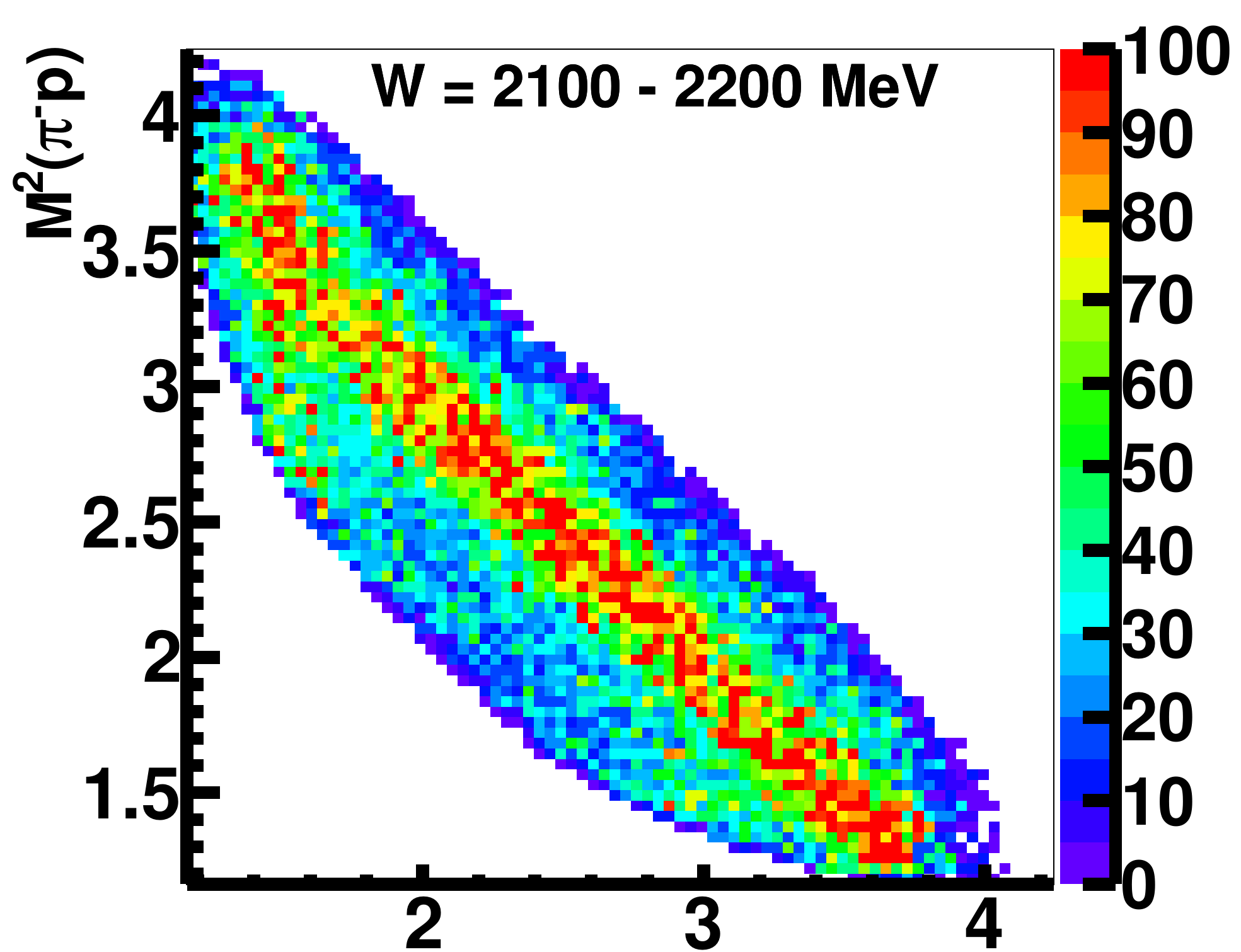}&
\hspace{-1mm}\includegraphics[width=0.24\textwidth]{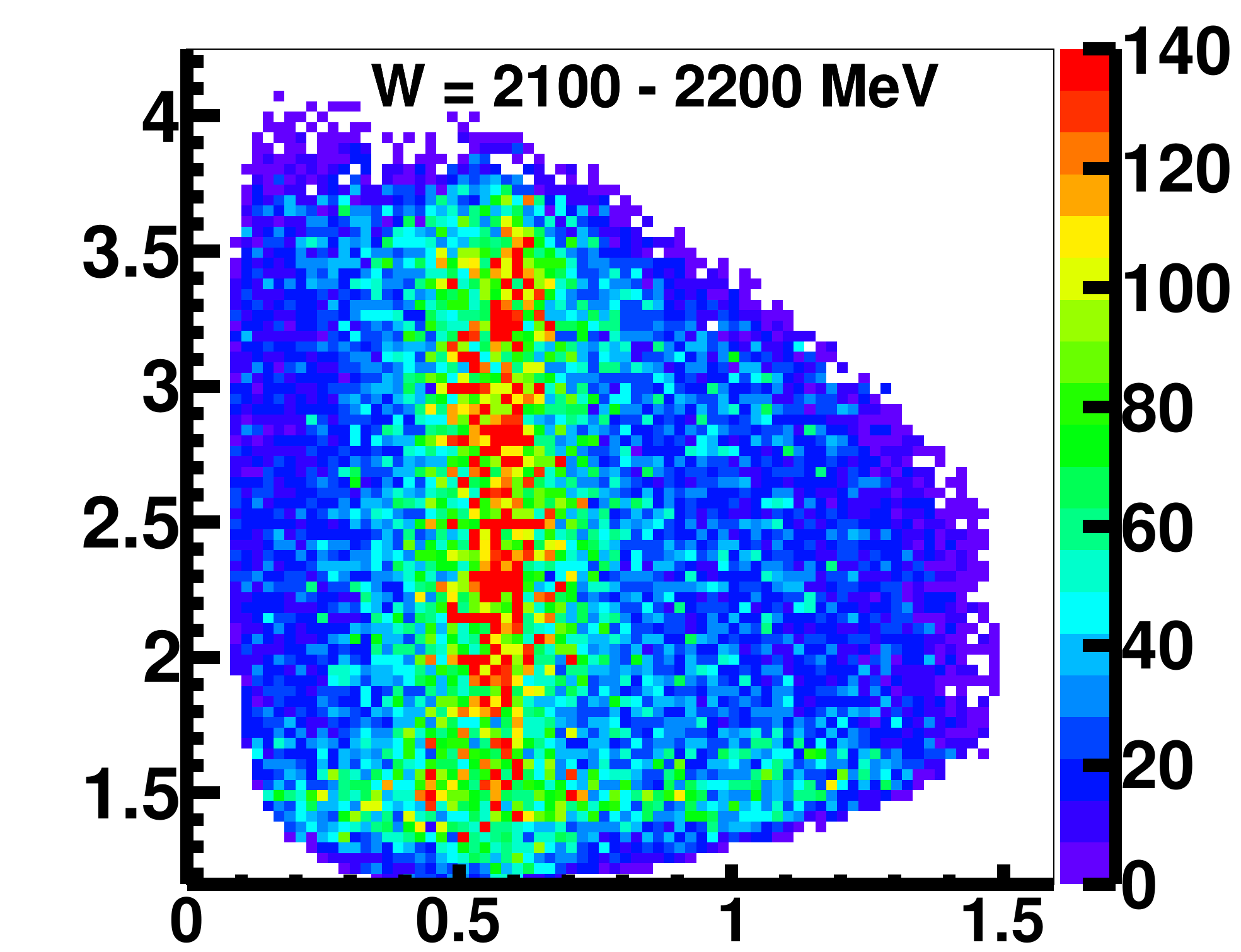}
\vspace{-1mm}\\
\hspace{-4mm}\includegraphics[width=0.24\textwidth]{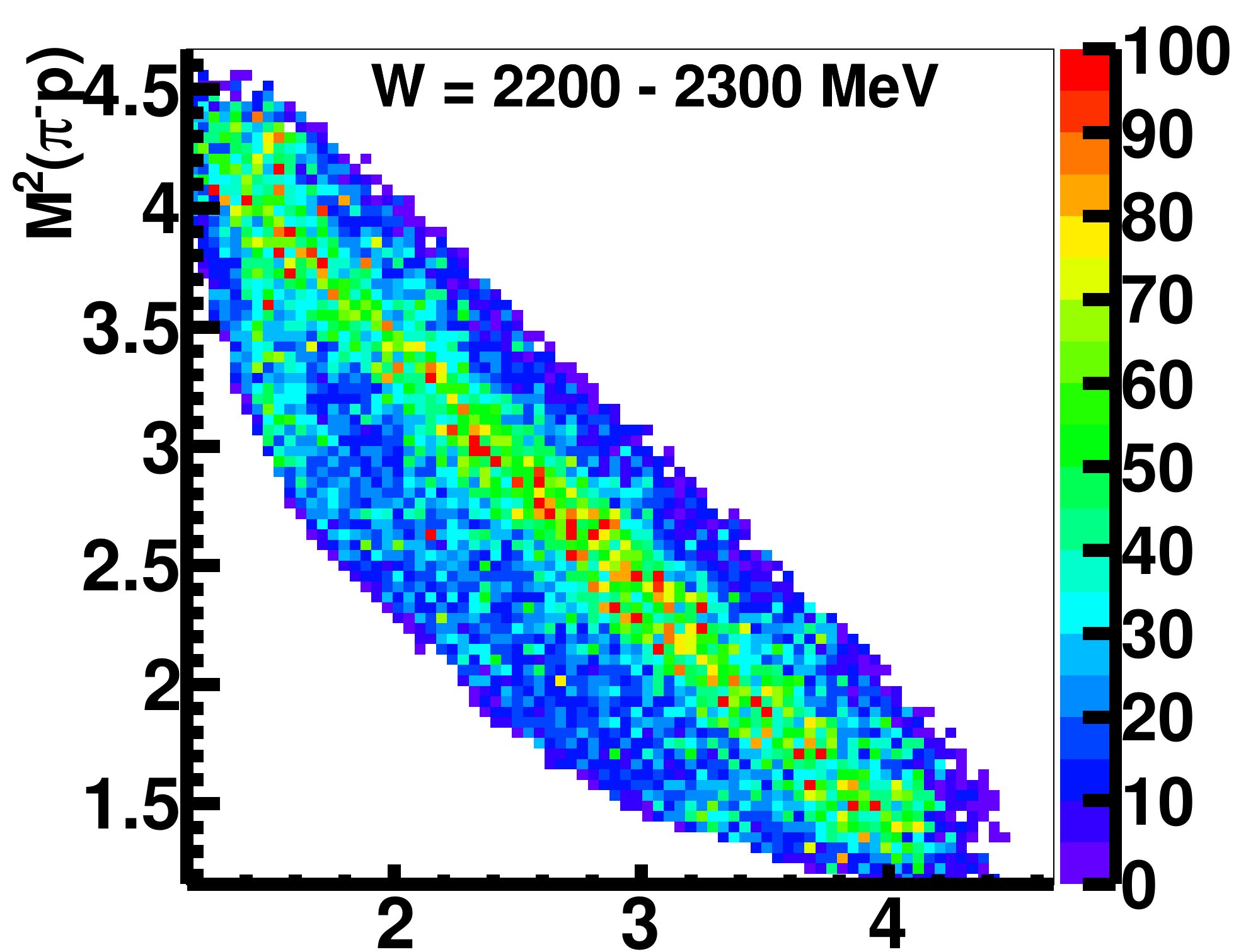}&
\hspace{-1mm}\includegraphics[width=0.24\textwidth]{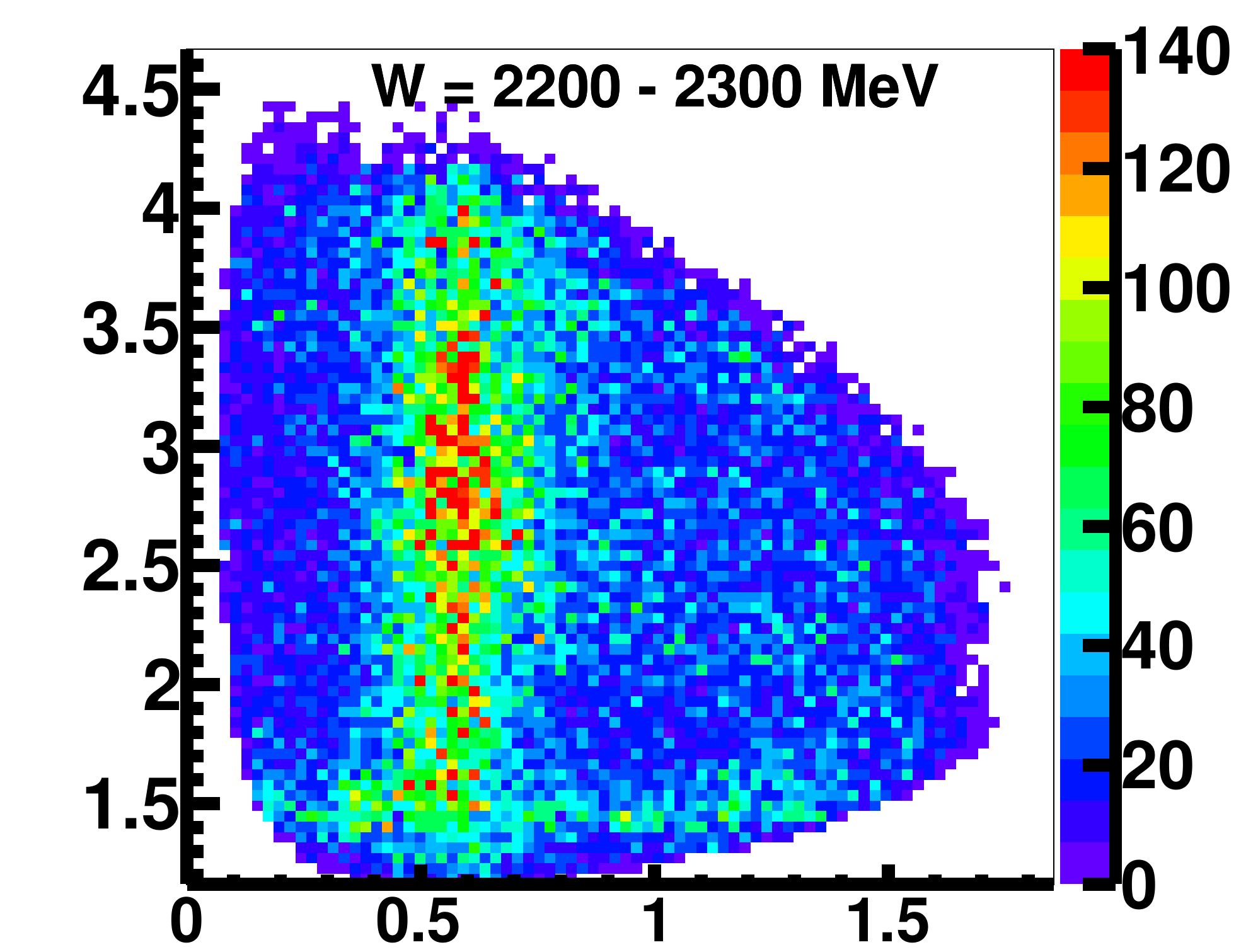}
\vspace{-1mm}\\
\hspace{-4mm}\includegraphics[width=0.24\textwidth]{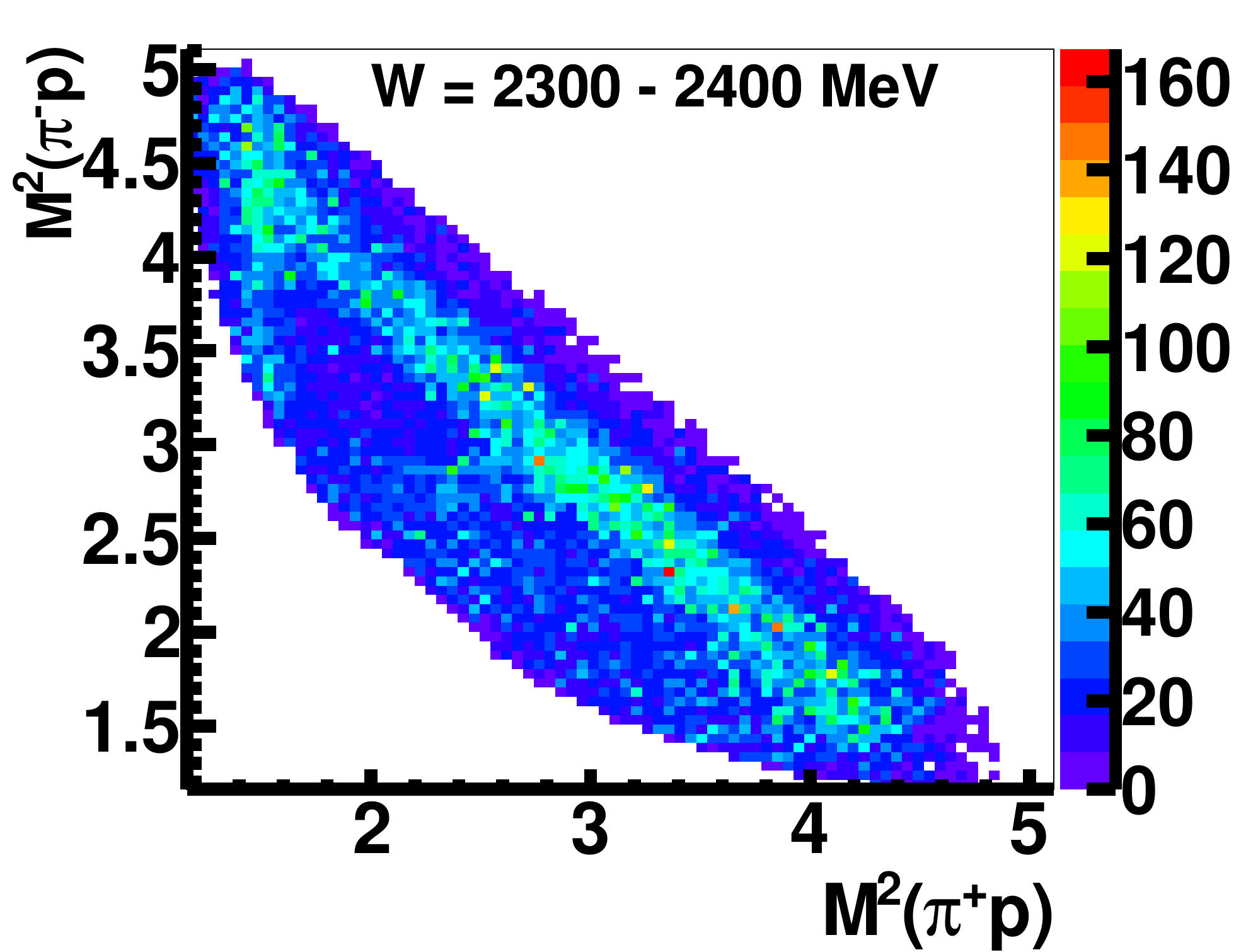}&
\hspace{-1mm}\includegraphics[width=0.24\textwidth]{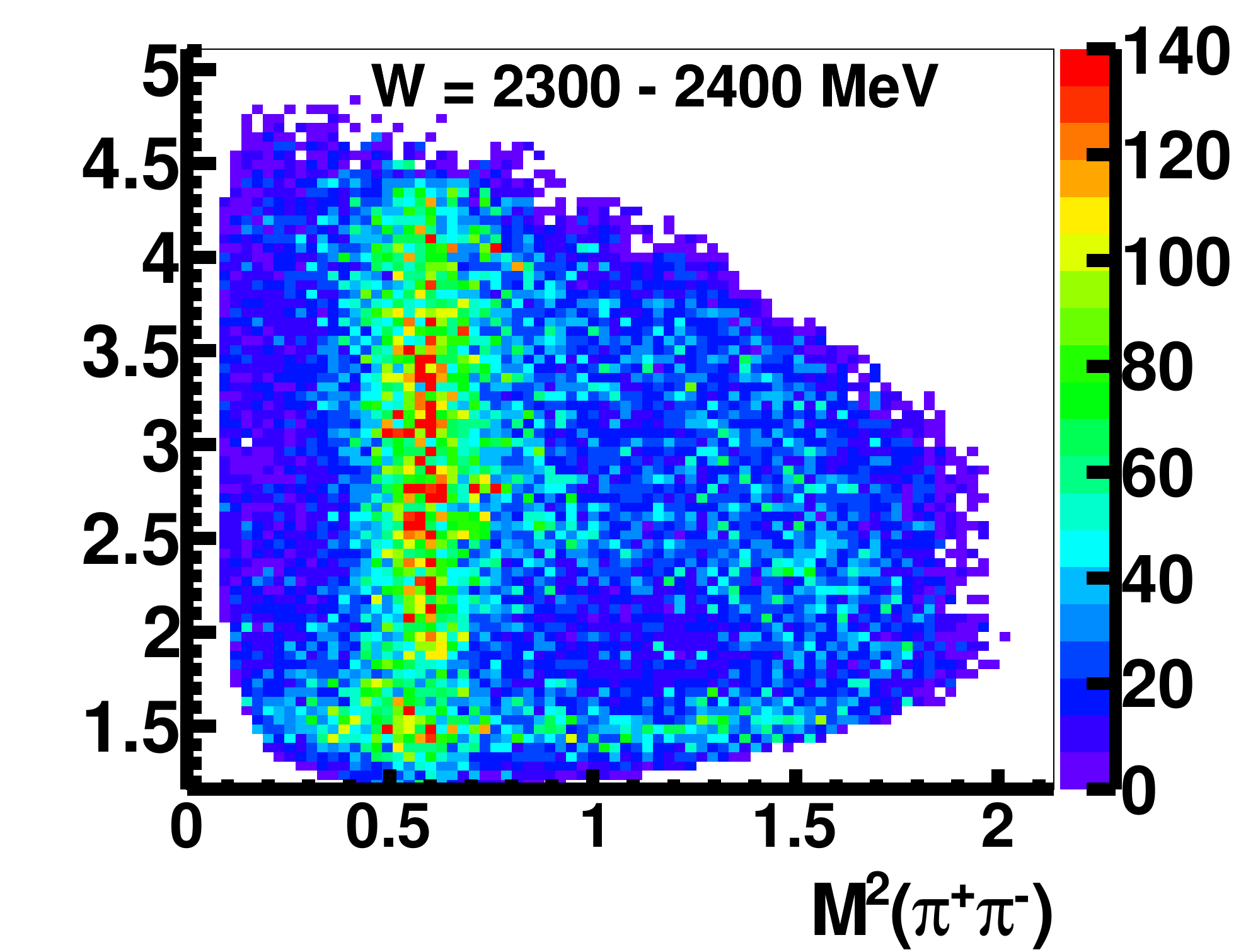}
\end{tabular}
\end{figure}
\subsection{Mass and angular distributions  of the \textit{full data set}}

\begin{figure*}
  \includegraphics[width=0.98\textwidth]{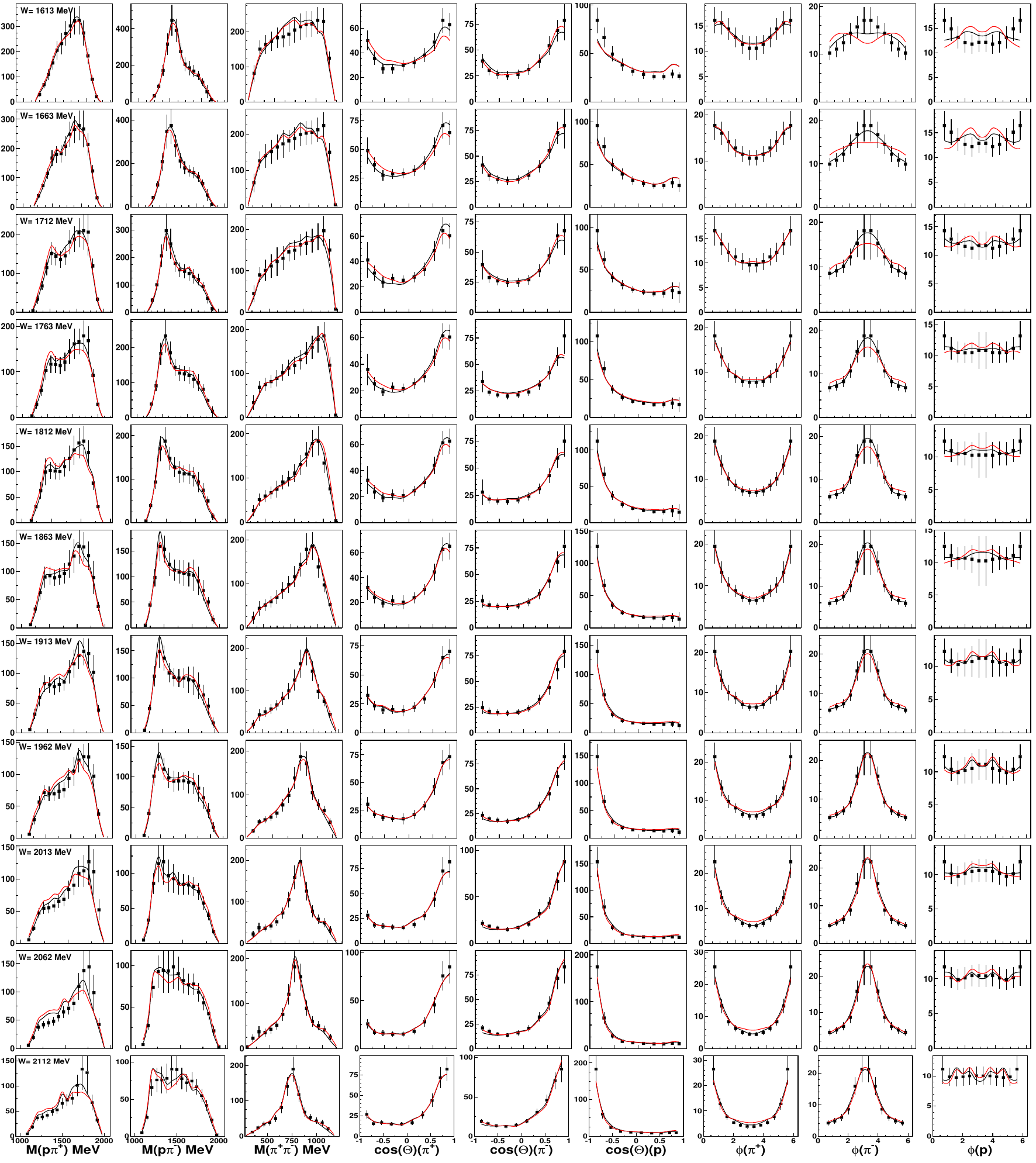}
  \caption{\label{mokeev}Invariant mass and angular distributions for the {\it  full data set}~\cite{CLAS:2018drk}. 
The black lines represent a fit to the {\it full data set} only, the red line a fit to the
{\it event-based data sample}. The central value of $M_{\gamma p}$ is shown in the left-most subfigure.
The bins cover a range of 25\,MeV and only every second bin is shown. 
}
\end{figure*}

Figure~\ref{mokeev} shows the mass and angular distributions of the {\it full data set}. The  black curves represent
a fit in which the {\it full data set} is included but not the {\it event-based data sample}, the red curve 
represents a fit in which the  {\it event-based data sample} is included but not the {\it full data set}. The data 
are mostly well  reproduced demonstrating good agreement between the {\it full data set} 
and the {\it event-based data sample}. At low energies, some discrepancies in the $\phi$ 
distributions are seen. 
Hence, the {\it full data set} and the {\it event-based data sample} are not fully consistent at low photon energies.
Including the {\it full data set} leads to a $\chi^2/N_{\rm data}$ of the fit to the polarization data of 1.33 for 2508 data
points and to  $\chi^2/N_{\rm data}=0.68$ when  the {\it event-based data sample} is included. 
The {\it event-based data sample} is perfectly consistent, and the {\it full data set} reasonably consistent
with the polarization data~\cite{Crede:2024tbd}. 
\begin{figure}[pb]
\caption{\label{Dalitz1}Acceptance-corrected Dalitz plots $M^2 _{p\pi^-}$ versus $M^2 _{p\pi^+}$ and $M^2 _{p\pi^+}$ versus $M^2 _{\pi^+\pi^-}$. All four topologies are added. In the low-energy region, the $\Delta(1232)^{++}$ provides the
most significant contribution, at high energies, $\rho(770)$ production dominates the reaction.
}
\end{figure}


\subsection{Total cross section and excitation functions}

\begin{figure}
\includegraphics[width=0.48\textwidth]{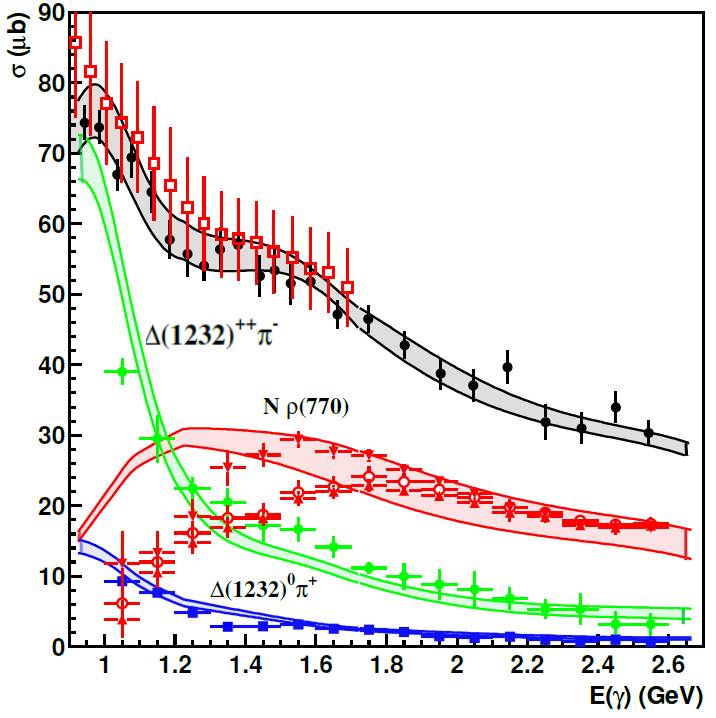}
\caption{\label{saphir-tot}The total cross section for the reaction $\gamma p\to p\pi^+\pi^-$ including different isobar contributions.  
Our total cross section is represented by the black-shaded band that is determined from the spread of different PWA solutions. 
Our result on the total cross section is compared to our earlier
results covering a smaller energy range ({\rd $\Box$} \cite{CLAS:2018drk}) and 
to results of the ABBHHM ($\bullet$ \cite{Aachen-Berlin-Bonn-Hamburg-Heidelberg-Munich:1968rzt}) 
that are consistent with those from the SAPHIR Collaboration (\cite{Wu:2005wf}). 
The red, green, and blue bands represent our result on the excitation
functions for $N\rho(770)$, $\Delta(1232)^{++}$ and $\Delta^0(1232)$ production, the
points with error bars are the SAPHIR results. 
The SAPHIR Collaboration used
three fit models for the $\rho(770)$ shape, a Breit--Wigner function ({\rd $\circ$}), the S\"oding model
({\rd $\blacktriangle$}), and the Ross--Stodolsky model ({\rd $\blacktriangledown$}). 
 }
\end{figure}

Figure \ref{saphir-tot} presents the total cross section for the
reaction $\gamma p\to p\pi^+\pi^-$. Our result is shown by bands representing the
total uncertainty.  
It is compared to the cross section as determined by the Aachen--Berlin--Bonn--Hamburg--Heidelberg--Munich (ABBHHM) Collaboration~\cite{Aachen-Berlin-Bonn-Hamburg-Heidelberg-Munich:1968rzt} 
and to an earlier CLAS analysis~\cite{CLAS:2018drk}. These earlier results are given by dots and 
error bars. The cross section obtained by the SAPHIR Collaboration~\cite{Wu:2005wf} is compatible. It is not shown here for the sake of clarity. 

The total uncertainty in the cross section measurement contains the contributions listed
in Table~\ref{uncertainties}. The PWA uncertainty is determined from the spread of a large number of fits.
In systematic studies, we changed the fit model by adding high-mass resonances in different partial waves in the 2.2 to 2.5\,GeV
mass range. Furthermore, we multiplied or divided the weight of data sets in $\gamma \vec p\to p\pi^+\pi^-$
by a factor of 2, performed fits of the {\it full $\gamma p\to p\pi^+\pi^-$ data set} and the {\it event-based data sample}
with all four topologies, or excluded the {\it full data set} or the {\it event-based data sample}, or used only one of the topologies. The largest part of the PWA uncertainty is due to the fit model; changing the
weights has a minimal effect on the fit results.

Figure \ref{saphir-tot} also shows excitation functions for $\gamma p\to p\rho, \Delta(1232)^{++}\pi^-$
and $\Delta(1232)^{0}\pi^-$ and compares these to those of
the SAPHIR Collaboration. The SAPHIR results are consistent but more precise than those
of the ABBHHM Collaboration that are not shown here. Both collaborations use three different
fit models to describe $N\rho^0(770)$ production. In a first model, the $\rho^0(770)$ meson is represented by a simple
non-relativistic Breit--Wigner function. Experimentally, the $\rho^0(770)$ meson in low-energy photoproduction has an asymmetric shape. This is taken into account in the S\"oding model \cite{Soding:1965nh}
of the Breit--Wigner amplitude with a Deck-type background~\cite{Ballam:1971yd}. In the
Ross--Stodolsky model \cite{Ross:1965qa}, the $\rho^0(770)$ mass distribution is multiplied by a factor
$(M_\rho/M_{\pi\pi})^4$. The two $\Delta(1232)$ isobars are described by Breit--Wigner
functions, their interference by an additional term. Non-resonant $N\pi\pi$ 
production is described by a phase-space distribution. 
In the BnGa approach, the three isobars shown in Fig.~\ref{saphir-tot} provide the largest contributions,
but other less important isobars are also admitted. Non-resonant production of $N\pi\pi$ is not needed.
Interference is allowed between all amplitudes and
not only for interference between $\Delta(1232)^{++}$ and $\Delta(1232)^{0}$.  
The asymmetric $\rho^0(770)$ shape is well described by a relativistic Breit--Wigner amplitude taking phase space and 
orbital-angular-momentum-barrier factors into account. 

\begin{table}[htbp]
\caption{\label{uncertainties}Uncertainties in the cross section measurements \cite{CLAS:2018drk}.
The uncertainty in the PWA depends on the kinematic range. The total uncertainty is represented by (red) curves
in Figs.~\ref{rho-diff}-\ref{Delta0_diff} and by a range in Table~\ref{BR}.}
\centering
\renewcommand{\arraystretch}{1.4}
\begin{tabular}{lc}
\hline\hline
Fiducial cuts & 4.0\%\\
Kinematic fitting & 3.0\%\\
Photon flux correction & 1.0\%\\
Particle detection efficiency correction & 1.5\%\\
Sum of constant uncertainties & 5.3\%\\\hline
Three-pion contamination& 0 -- 2\%\\\hline\hline
\end{tabular}
\end{table}

At low energies, the largest contribution to the cross section is provided by the Kroll--Ruderman mechanism.
The $\Delta(1232)^{++}\pi^-$ contribution continues to decrease. The SAPHIR analysis finds a weak
shoulder in the second and third resonance regions, which is not confirmed here. $\Delta^0(1232)\pi^+$
production is smaller by a factor 3 - 5, the reduction is much larger than expected if these two
isobars would be dominantly produced via intermediate $N^*$ or $\Delta^*$ resonances. We find a considerably larger cross section
for the $N\rho^0(770)$ production. Note that our uncertainty is mainly due to the different solutions
of the multichannel analysis, while SAPHIR gives the statistical uncertainty derived from one model.
The $N\rho^0(770)$ contribution is mostly in the (20--30)\,$\mu$b range. 
The $\gamma p\to N\rho^0(770)$ cross section is nearly constant above 1.3\,GeV in photon energy, 
pointing to diffractive scattering as the main production mechanism.
The small $N\rho^0(770)$ cross
section in the SAPHIR analysis, in particular at low energies, may originate from three reasons:
SAPHIR fits with a large phase-space contribution and  
the $N\rho(770) - \Delta(1232)^{++}\pi^-$ interference is neglected. In the BnGa solution a notable part 
comes from the $N(1520)3/2^-\to N\rho$ contribution which
produces a deeper slope for the $N\rho$ contribution at low energies than the $t-channel$ exchange amplitudes.

\subsection{\boldmath$\gamma p\to  N\rho^0(770)$}

\paragraph{Differential cross sections:}

\begin{figure}
\vspace{-5mm}
\centering
\includegraphics[width=0.5\textwidth]{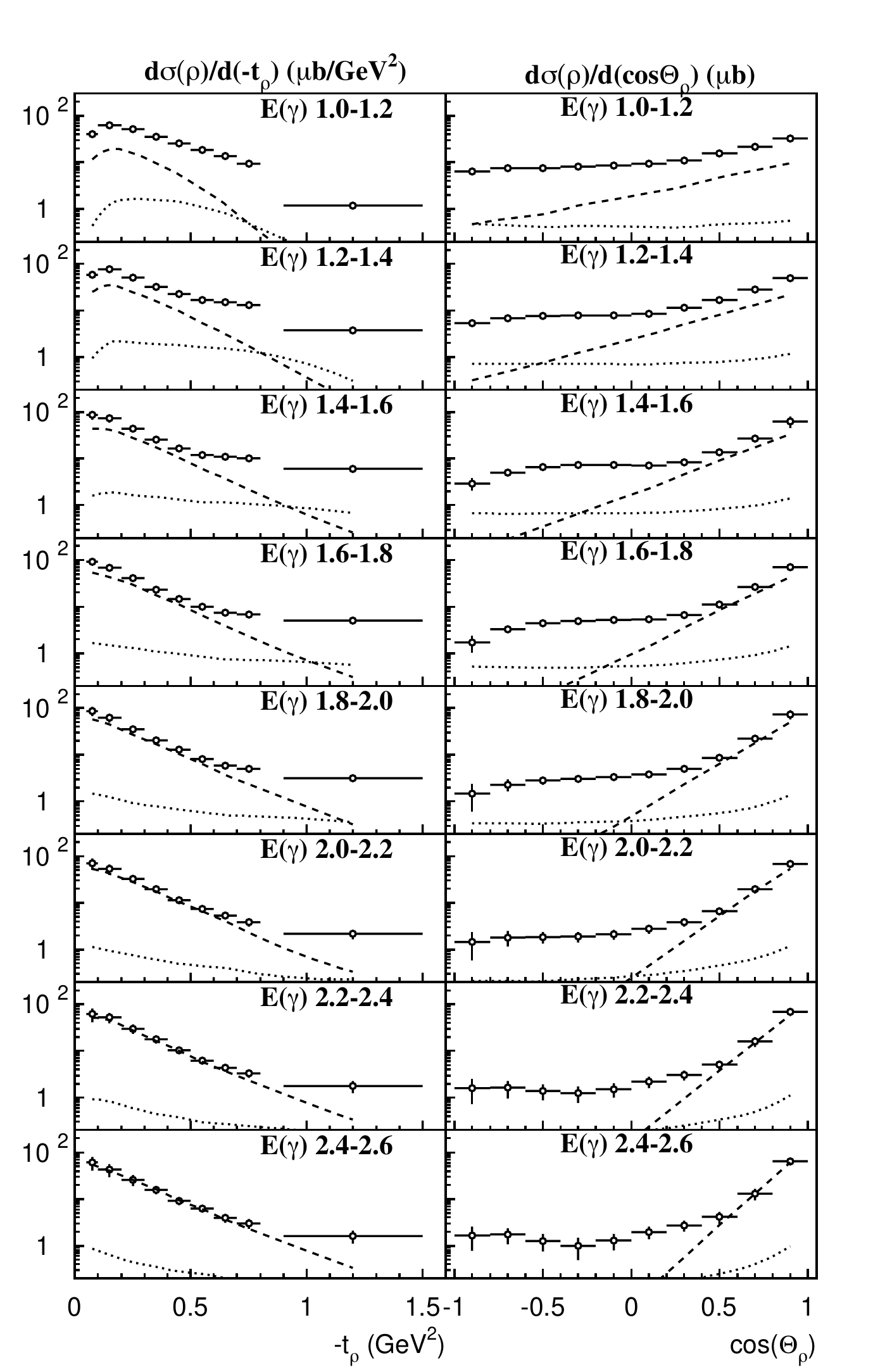}
\caption{\label{rho-diff}Differential cross sections for $\gamma p\to  N\rho^0(770)$ as a function
of $-t$ (left) and $\cos\Theta_\rho ^{cms}$ right. The vertical and horizontal lines define the
uncertainty and range of cross section measurement. The dashed curves give the contribution
due to exchange, and the dotted curves give the contribution due to pion exchange.}
\end{figure}

Figure~\ref{rho-diff} presents the differential cross sections as a function of the squared momentum
transfer $-t$  to the $\rho^0(770)$ meson and as a function of the cosine of the $\rho(770)$ scattering angle $\cos\Theta_\rho ^{cms}$.
The $t$-dependence shows a fast exponential
fall-off that is characteristic of diffractive scattering. The pion-exchange contribution is considerably
smaller. Over a wide range of energies, their sum almost corresponds to the full cross section,
leaving little room for $N^*$ or $\Delta^*$ to decay to $N\rho(770)$.
The $N\rho(770)$ differential cross sections in the forward direction are well described for $E_{\gamma}$ $>$ 1.4 GeV 
by Pomeron exchange only. This makes it difficult to see manifestations 
of the N$^*$ signals in this region. The additional intensity
at backward angles reveals additional $\rho(770)$-production modes beyond Pomeron or pion exchange that
suggests resonance contributions.
\begin{figure}
\centering
\vspace{-18mm}
\includegraphics[width=0.54\textwidth]{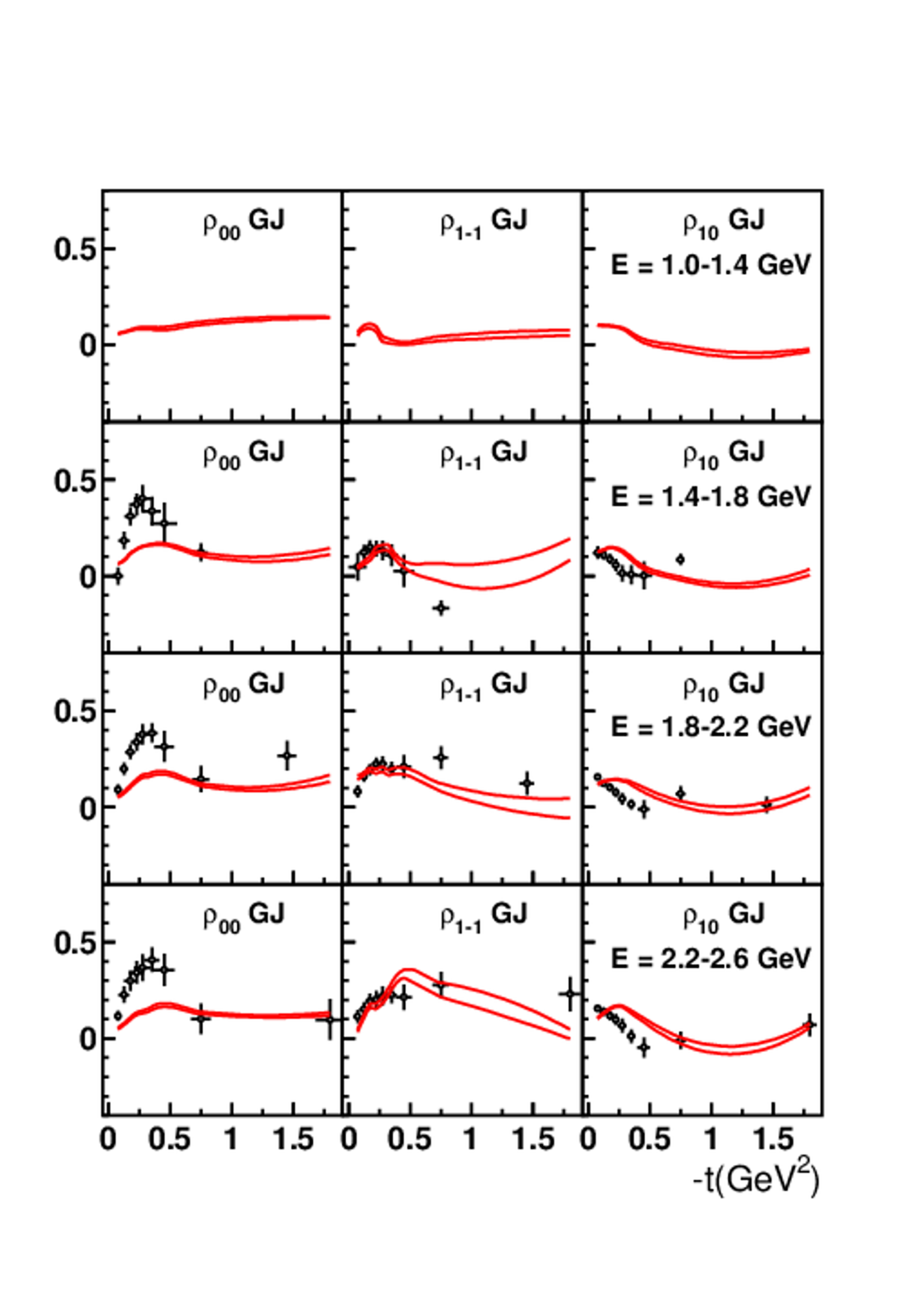}\vspace{-10mm}
\caption{\label{rho_gj}The spin-density matrix elements for the decay of the $\rho^0(770)$ meson as a function of the
(squared) transferred momentum calculated in the Gottfried--Jackson frame. 
The two red lines show the range of the spin-density matrix elements calculated from the set of the final
solutions. The data points with error bars show the result obtained by the SAPHIR Collaboration~\cite{Wu:2005wf}.}
\end{figure}
\begin{figure}
\centering
\vspace{-18mm}
\includegraphics[width=0.54\textwidth]{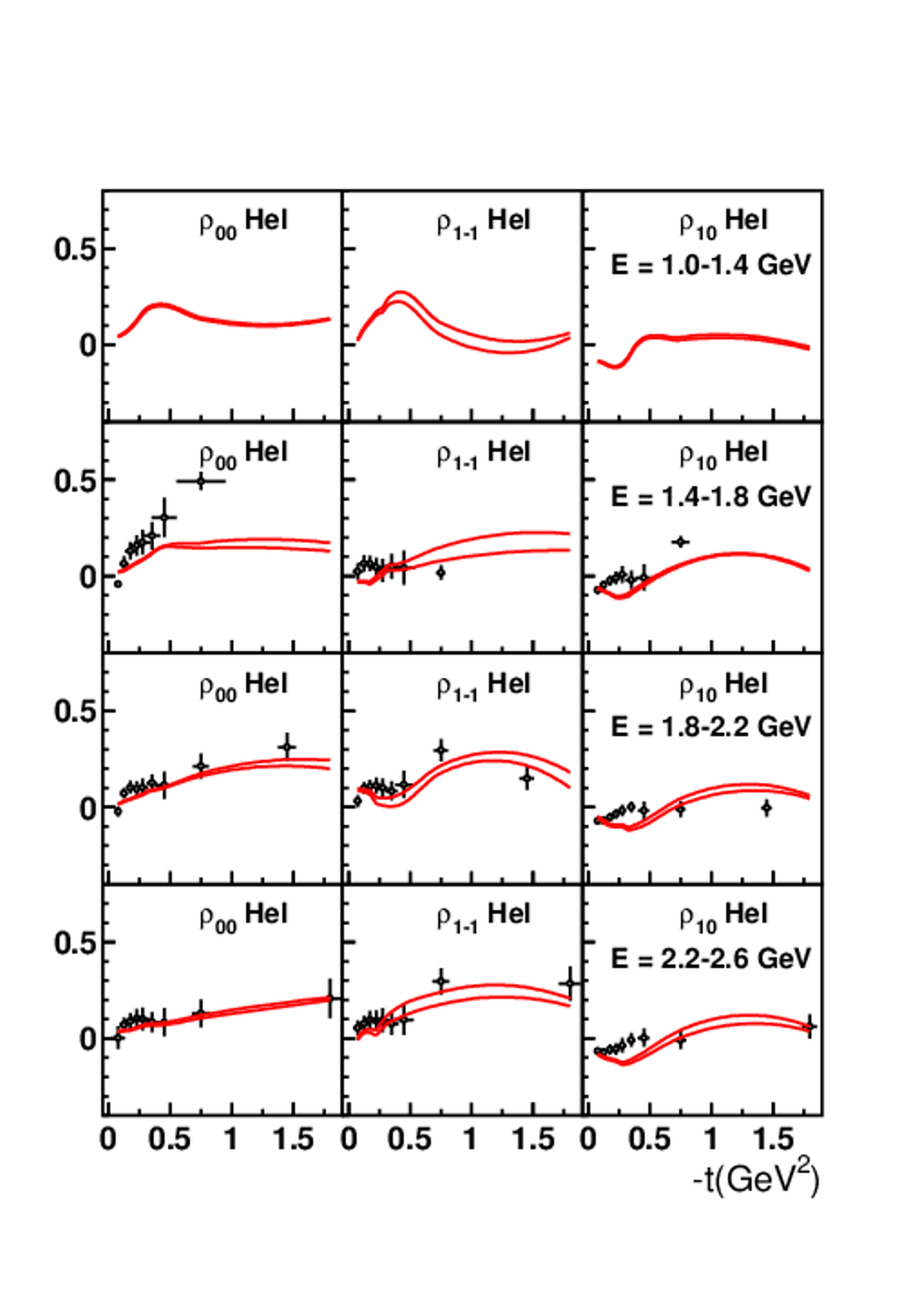}\vspace{-10mm}
\caption{\label{rho_hel} The spin-density matrix elements for the $\rho^0(770)$ meson decay in dependence 
on the  momentum transferred squared calculated in the helicity frame. 
The red lines show the range of the spin-density matrix elements calculated from the set of the final
solutions. The data points with error bars show the result obtained by the SAPHIR Collaboration~\cite{Wu:2005wf}.}
\end{figure}
\begin{table}[pb]
\caption{\label{slopes}Slope of the differential cross section for photoproduction
of $\rho^0(770)$ mesons from this analysis, and from the SAPHIR and ZEUS Collaborations.
\\[-2ex]}
\centering
\renewcommand{\arraystretch}{1.4}
\begin{tabular}{ccccc}
\hline\hline
                     &\multicolumn{2}{c}{This work}&\multicolumn{2}{c}{SAPHIR}\\[-1ex]
\phantom{rrrrr}$E_\gamma$\phantom{rrrrr} &\phantom{rrrrr}$a$\phantom{rrrrr}&\phantom{rrrrr} $b$\phantom{rrrrr}
&\phantom{rrrrr}$a$\phantom{rrrrr}&\phantom{rrrrr} $b$\phantom{rrrrr}\\[-1.5ex]
\scriptsize GeV\quad&\scriptsize$\mu$b/GeV & \scriptsize GeV$^{-2}$&\scriptsize$\mu$b/GeV & \scriptsize GeV$^{-2}$\\\hline
1.0 -- 1.2 &  115\er 8 &3.6\er 0.6 & &\\
1.2 -- 1.4 & 147\er 8 &4.3\er 0.5 & &\\
1.4 -- 1.6 & 154\er 5 &5.2\er 0.3 &170\er 6 & 6.69\er 0.17\\
1.6 -- 1.8 &  150\er 6 &5.3\er 0.3 &175\er 4 & 6.32\er 0.10\\
1.8 -- 2.0 & 138\er 6 &5.3\er 0.3 &164\er 4 & 5.96\er 0.09\\
2.0 -- 2.2 & 120\er 5 &5.2\er 0.3 &136\er 3 & 5.55\er 0.08\\
2.2 -- 2.4 & 114\er 6 &5.3\er 0.3 &121\er 2 & 5.42\er 0.08\\\cline{2-5}
3.2 -- 3.9  &\multicolumn{2}{c}{CLAS \cite{CLAS:2001zxv}} &  & 6.4 \er 0.3  \\
65 &\multicolumn{2}{c}{ZEUS \cite{ZEUS:1997rof}}&12.1\er1.2& 11.0\er1.0 \\
\hline\hline
\end{tabular}
\end{table}

We fitted the $t$-dependence below $-0.5$\,GeV$^2$ 
with an exponential function in the form $d\sigma/dt=a\exp{\{-b|t|\}}$. The fit values are given
in Table~\ref{slopes}. The SAPHIR Collaboration finds a steeper slope and a higher yield in the forward direction. Their uncertainties are much smaller, but do not include a systematic variation
of the PWA model. The ZEUS Collaboration
fitted $\rho^0(770)$ production in $e\,p$ scattering \cite{ZEUS:1997rof} at $W=55, 65$ and 84\,GeV
and found a mild $W$ dependence only. In Table~\ref{slopes}, we quote the central value.
At HERA energies, the cross section is considerably reduced and falls off faster with increasing
momentum transfer.  The $\cos\Theta_\rho ^{cms}$ distributions exhibit a forward rise. Note that forward $\rho^0(770)$ mesons are
produced at low values for $-t$.\\[-2ex]

\paragraph{Density matrix elements:}

The $\rho(770)$ meson is a vector particle. The $\rho(770)$-decay carries information
on its spin alignment relative to a given axis. This information is encoded in  
the spin-density matrix.  Its elements are extracted from 
angular distributions of the pions calculated in the rest system of
the $\rho^0(770)$ meson. The angular distribution of its decay into two pions can be parameterized as:\\[-4ex]


\bq &&W(\Theta_\pi,\phi_\pi)=\frac{3}{4\pi}\Big
[\frac12(1-\cos^2\Theta_\pi)+\frac12(3\cos^2\Theta_\pi-1)\rho_{00}
\nonumber \\
&&-\rho_{1-1}\sin^2\Theta\_\pi\cos2\phi_\pi-\sqrt2
Re\rho_{10}\sin2\Theta_\pi\cos\phi_\pi \Big ]. \eq
These angular distributions depend on the orientation of the axis in
the rest system of the two-pion system. For $\rho^0(770)$ meson
production two systems are considered: the helicity system,
where the $z$-axis is directed along the momentum of the final-state nucleon recoiling against
the $\rho^0(770)$ meson, and
the Gottfried--Jackson system, where the $z$-axis is directed along
the photon beam axis. In both systems the reaction plane is defined by the beam momentum and the proton momentum (see 
caption of Figure \ref{cms}). The elements of the density matrix reveal flips of the spin relative to the
direction of the $\rho^0(770)$ momentum in the helicity frame or to the direction of the 
photon in the Gottfried--Jackson frame. For helicity conservation in a given frame, 
the corresponding three spin-density matrix elements vanish.

\begin{figure}
\centering
\vspace{-5mm}
\includegraphics[width=0.5\textwidth]{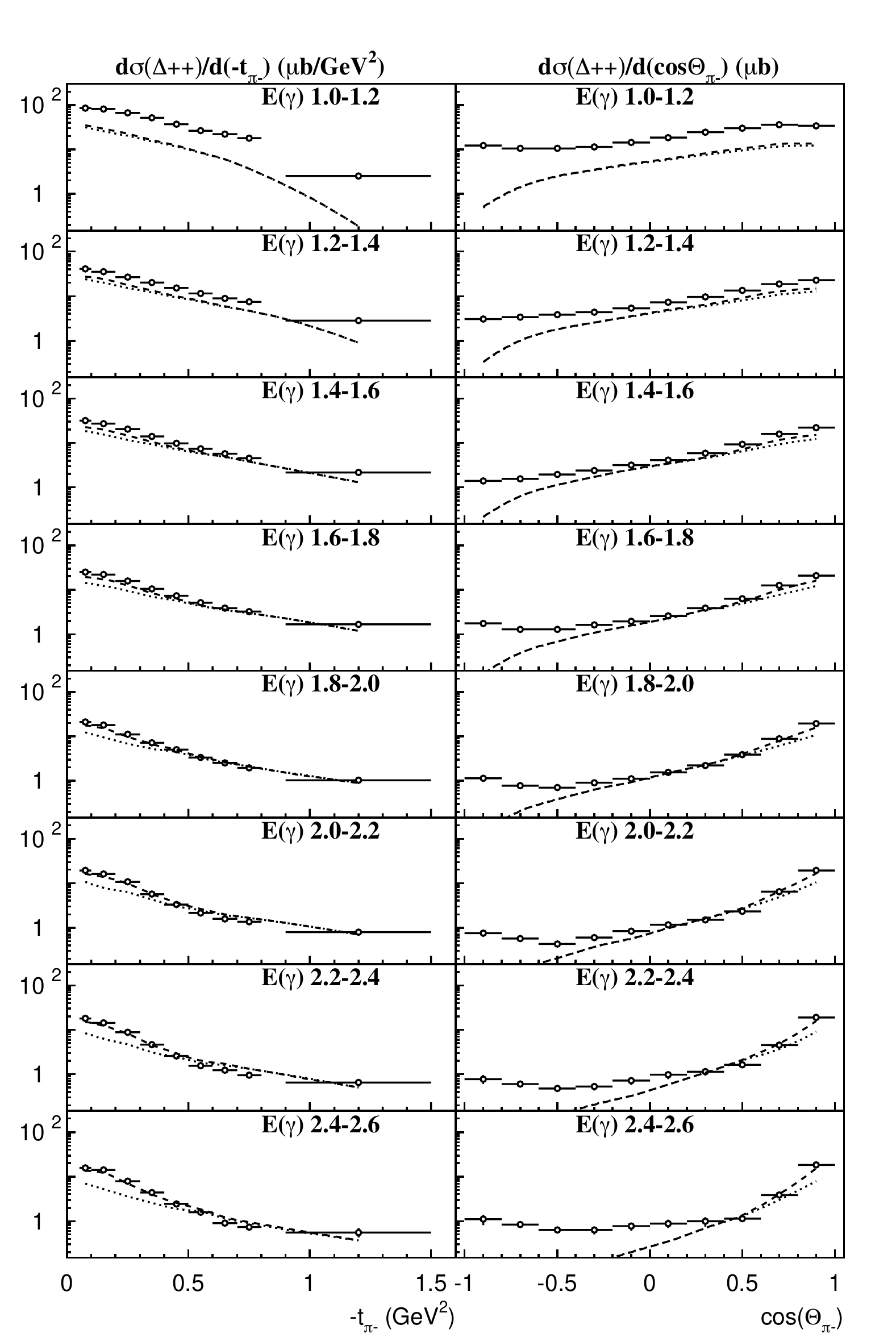}
\caption{\label{Deltapp_diff}Differential cross sections for $\gamma p\to \Delta(1232)^{++}\pi^-$ as a function
of $-t$ (left) and $\cos\Theta_\rho ^{cms}$ right. The vertical and horizoantal lines define
uncertainty and range of cross section measurement. The dotted curve represents the
Kroll--Rudermann contribution, the dashed curve the pion-exchange contribution. 
}
\end{figure}

\begin{figure}
\centering
\vspace{-14mm}
\includegraphics[width=0.54\textwidth]{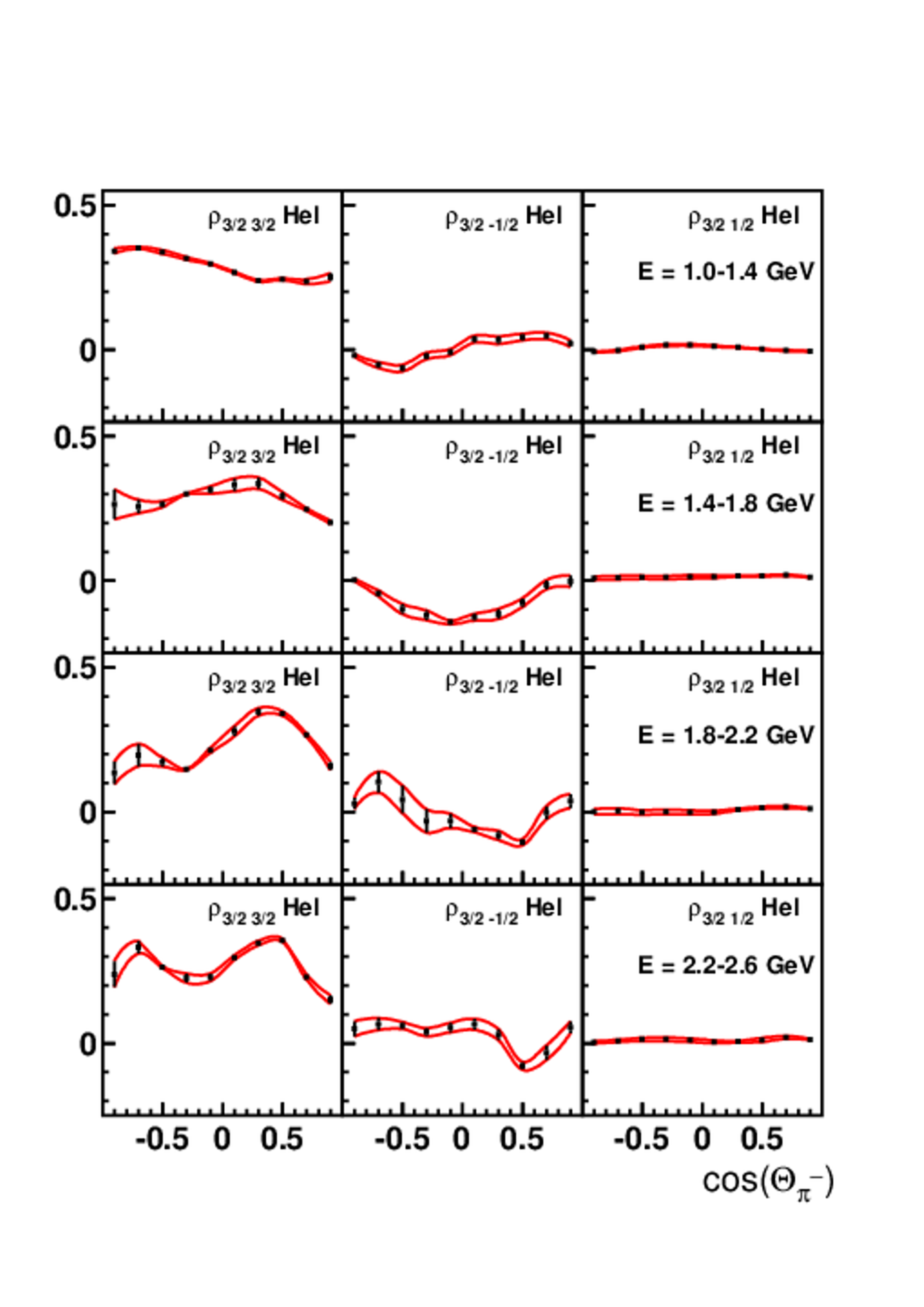}\vspace{-10mm}
\caption{\label{Deltapp_hel}The spin-density matrix elements for the $\Delta(1232)^{++}$ 
decay as a function of $\cos\Theta_{\pi^-}$ calculated in the helicity frame. 
The (red) curves show the range of the spin-density matrix elements calculated from the set of the final
solutions. 
\vspace{2mm}
}
\end{figure}
Figure~\ref{rho_gj} shows the spin-density matrix elements for photoproduction
of $\rho^0(770)$ mesons as a function of $-t$ in the Gottfried--Jackson frame; Fig.~\ref{rho_hel} is the same
in the helicity frame. In the helicity frame, vanishing spin-density matrix elements would be expected if photoproduction of $\rho(770)$ mesons were fully dominated
by diffractive scattering where the incoming photons virtually convert into a vector
meson that scatters off the proton via natural (e.g.\ Pomeron) or unnatural (e.g.\ pion) parity  exchange. 
Diffractive scattering is not the only process
contributing to $\rho(770)$ production.
With increasing energy, vanishing spin-density matrix elements in the helicity frame
lead to a $\sin 2\Theta$ distribution in the Gottfried--Jackson frame that can be
recognized in the experimental distributions. This behavior is expected for
diffractive scattering: the photon virtually converts to a $\rho^0(770)$ mesons that
scatters off the proton by Pomeron or pion exchange keeping the orientation
of its spin.

\subsection{\boldmath$\gamma p\to  \Delta(1232)^{++}\pi^-$}

\paragraph{Differential cross sections:}

Figure~\ref{Deltapp_diff} shows the differential cross section for the $\Delta(1232)^{++}$ 
as a function of $\cos\Theta_{\pi^-}$ where $\Theta_{\pi^-}$ is the c.m.s. angle.
The direct production of $\Delta(1232)^{++}$ plays a very significant role,
pion exchange is less important. Significant room is left for $N^*$ and $\Delta^*$ resonances
decaying into $\Delta(1232)\pi$. \\[-2ex]

\paragraph{Spin-density matrix elements:}
\begin{figure}
\vspace{-2mm}
\centering
\includegraphics[width=0.5\textwidth]{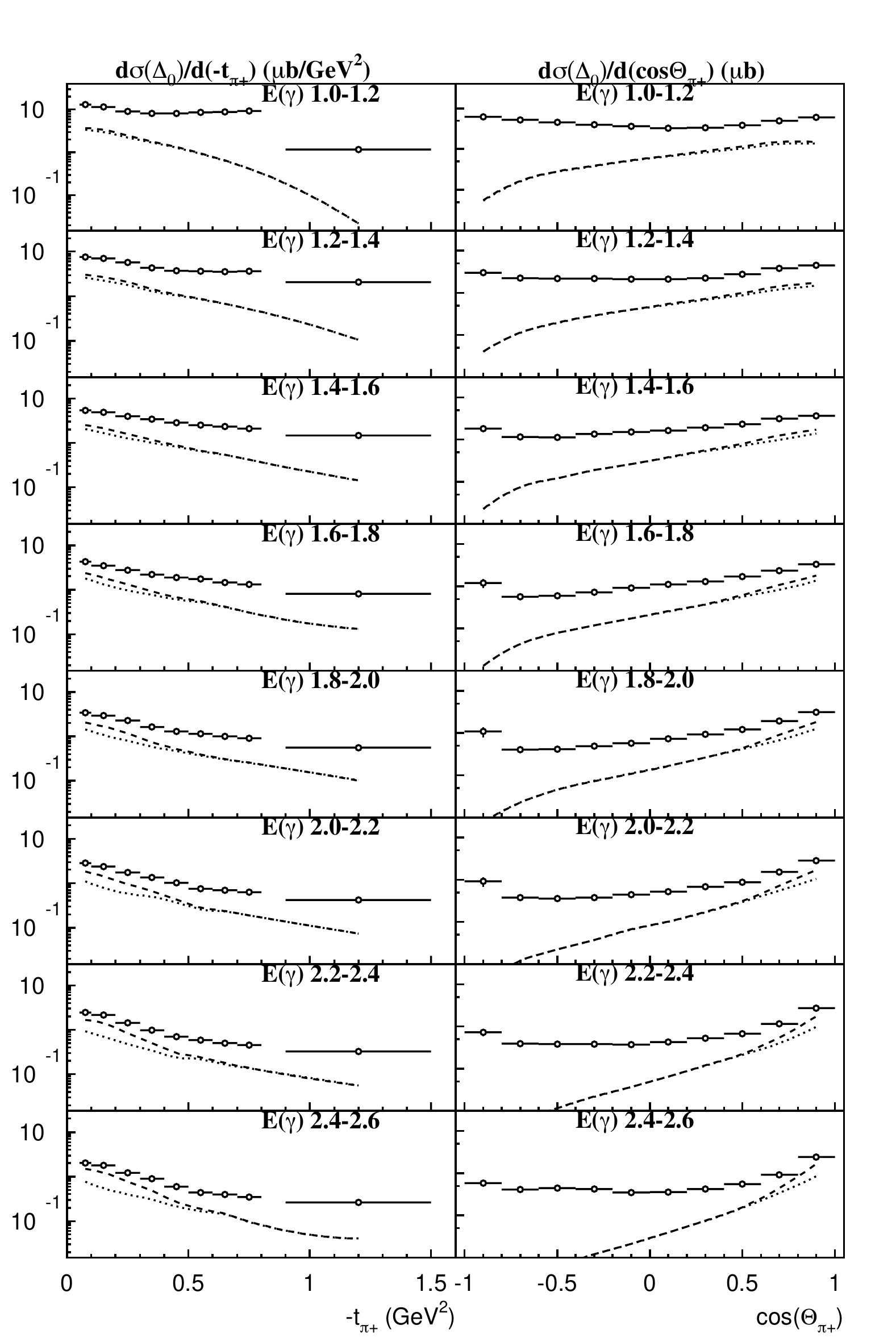}
\caption{\label{Delta0_diff}Differential cross sections for $\gamma p\to \Delta(1232)^{0}\pi^+$ as a function
of $-t$ (left) and $\cos\Theta_\rho ^{cms}$ (right). The uncertainties are smaller than the open circles; the horizontal lines define the
range of cross section measurement. The dashed curves gives the pion-exchange contribution and the dotted curves shows the Born term.
\vspace{-2mm}
}
\end{figure}
\begin{figure}[htbp]
\centering
\vspace{-14mm}
\includegraphics[width=0.54\textwidth]{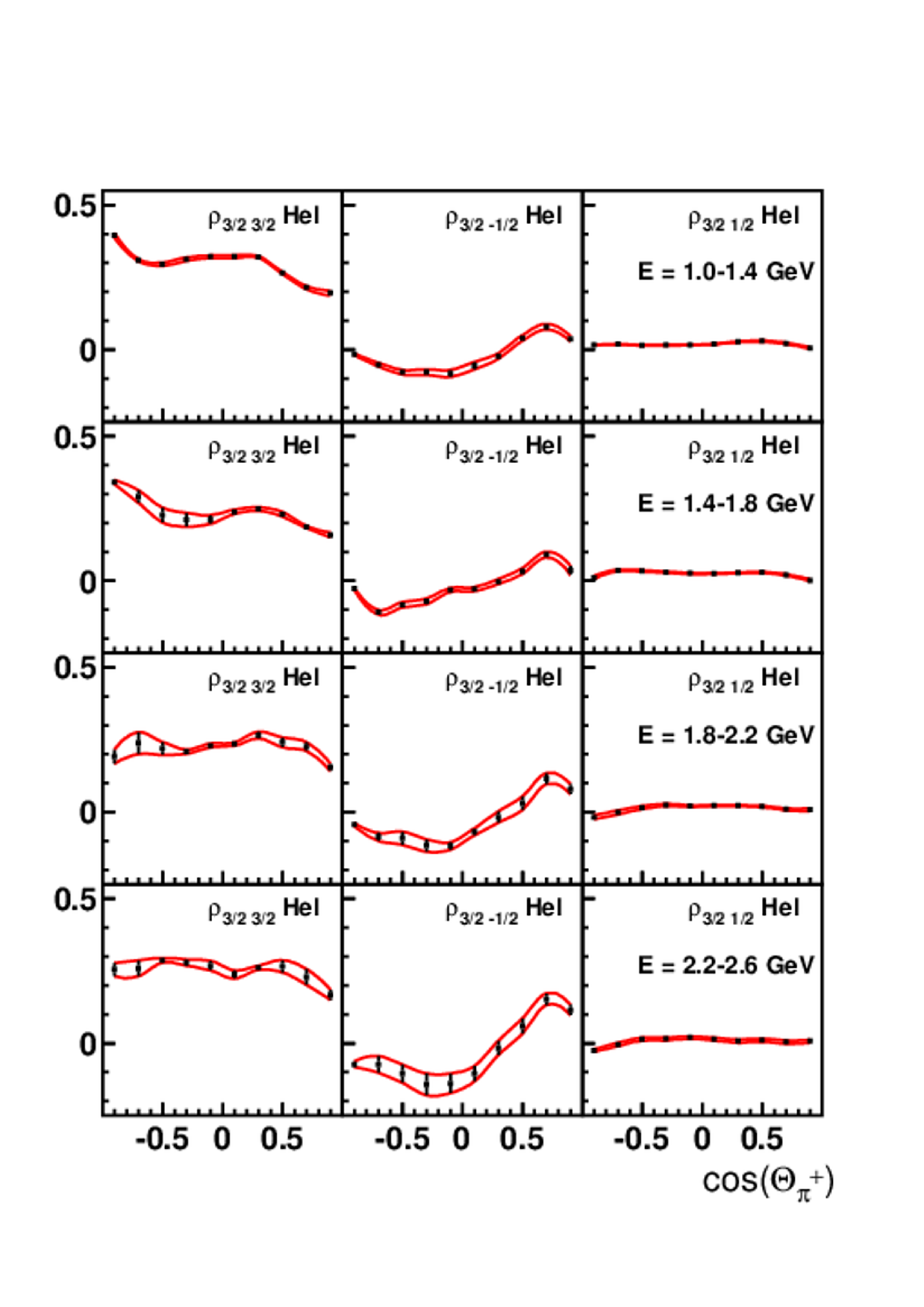}\vspace{-12mm}
\caption{\label{Delta0_hel}The spin-density matrix elements for the $\Delta(1232)^{0}$ 
decay as a function of $\cos\Theta_{\pi^+}$ calculated in the helicity frame. 
The two red lines show the range of the spin-density matrix elements calculated from the set of the final
solutions.\vspace{-3mm}}
\end{figure}

In the case of the production of a baryon with spin 3/2$^+$, the
elements of the density matrix can be extracted from the angular dependence of $\pi^+$ 
calculated in the rest frame of the resonance system:
\bq 
  &&W(\Theta_\pi,\phi_\pi)=\frac{1}{4\pi}\Big
  [\frac12(1+3\cos^2\Theta_\pi)+2(1-
  3\cos^2\Theta_\pi)\times
  \nonumber \\
  &&\rho_{\frac32\frac32}-2\sqrt3\,Re\rho_{\frac32-\frac12}\sin^2\Theta_\pi\cos2\phi_\pi-\nonumber \\ &&2\sqrt3
  Re\rho_{\frac32\frac12}\sin2\Theta_\pi\cos\phi_\pi \Big ].
  \eq
The elements of the spin-density matrix were extracted in the helicity system where the reaction plane is defined by the beam and $\pi^-$ momenta and the $z$-axis is directed along the $\pi^-$ momentum.
The elements show very significant structures
that support the evidence for the production of decays $N^*$ and $\Delta^*\to \Delta(1232)\pi$.  

\subsection{\boldmath $\gamma p\to  \Delta(1232)^{0}\pi^+$}

\paragraph{Differential cross sections:}
The differential cross section for $\gamma p\to \Delta^0(1232)\pi^+$ (see Fig.~\ref{Delta0_diff}) is considerably smaller
than the one for $\Delta(1232)^{++}$ production. At low photon energies, there is little backward--forward
asymmetry. Only at higher energies does pion exchange play a significant role leading to fast forward 
$\pi^+$ mesons.\\[-2ex]
\paragraph{Spin-density matrix elements:}
The spin-density matrix elements extracted in the helicity system for $\gamma p\to  \Delta(1232)^{0}\pi^+$
in Fig.~\ref{Delta0_hel} also show significant structures, again likely due to 
production of $N^*$ and $\Delta^*\to \Delta(1232)\pi$ decays.

\begin{table}[htbp]
\caption{\label{BR}
Branching ratios in \% for decays of $N^*$ and $\Delta^*$ resonances into $N\rho$. 
The results are obtained by integration over the $\rho$ width and over the width of
the resonance. The small numbers are estimates of the Particle Data Group~\cite{ParticleDataGroup:2022pth}.
These branching ratios are mostly determined with no integration. }
   \footnotesize
\renewcommand{\arraystretch}{1.47}
\begin{center}
\begin{tabular}{|rl|c|c|cc|}
\hline\hline
                &&\multicolumn{1}{c|}{\phantom{z} $(N\rho)_{\rm tot}$\phantom{z} } &\multicolumn{3}{c|}{\phantom{zzz}$N\rho$\phantom{zzz}}  \\\hline
                        &&\multicolumn{1}{c|}{ \phantom{zz} } &\multicolumn{1}{c|}{\phantom{zz}$S=\frac12$\phantom{zz}}  &\multicolumn{2}{c|}{$S=\frac32$}       \\[-1.8ex]
                        &&\multicolumn{1}{c|}{\tiny \phantom{z} \phantom{z}}&& \multicolumn{1}{c}{\tiny \phantom{z}${L<J}$ \phantom{z}}      & \multicolumn{1}{c|}{\phantom{zz}\tiny $L>J$\phantom{zz}}     \\[-0.5ex]\hline
$N(1440)$&$\frac12^+$                  &18\er6&9\er4&-&9\er4   \\[-2ex]
&&\tiny{$17-50$}&\tiny{$$}&&\tiny{$6-27$}  \\[-1ex]
$N(1520)$&$\frac32^-$                   &28\er4&4\er3&24\er3&$<1$  \\[-2ex]
& &\tiny{$10-16$}& \tiny{$0.2-0.4$} & \tiny{$10-16$} & \\[-1ex]
$N(1535)$&$\frac12^-$                    &9\er4&7\er3&-&2\er2 \\[-2ex]
&  &\tiny{$2-17$} &\tiny{$2-16$}&& \tiny{$<1$}  \\[-1ex]
$N(1650)$&$\frac12^-$                    &17\er6&12\er5&-&5\er3\\[-2ex]
&&\tiny{$12-22$}&\tiny{$<4$}&\tiny{$$}&\tiny{$12-18$} 
\\[-1ex]
$N(1675)$&$\frac52^-$                     &30\er8& 20\er7&10\er5& $<1$     \\[-2ex]
&&\tiny{$0.1-0.9$}&\tiny{$<0.2$}&\tiny{$0.1-0.7$}&    
\\[-1ex]
$N(1680)$&$\frac52^+$                    &10\er5&$<1$&9\er4&$<1$     \\[-2ex]
&&\tiny{$8-11$}&&\tiny{$6-8$}&\tiny{$2-3$}   
\\[-1ex]
$N(1700)$&$\frac32^-$                   & 21\er9& 5\er3& $<1$& 16\er8      \\[-2ex]
&&\tiny{$32-44$}&&\tiny{$32-44$}&   
\\[-1ex]
$N(1710)$&$\frac12^+$                     &17\er4& 6\er2&-& 11\er3    \\[-2ex]
&&\tiny{$11-23$}&\tiny{$11-23$}&\tiny{$$}&    
\\[-1ex]
$N(1720)$&$\frac32^+$                     &29\er8&12\er5&17\er6&$<1$     \\[-2ex]
&&\tiny{$1-2$}&\tiny{$1-2$}&& 
\\[-1ex]
$N(1860)$&$\frac52^+$                    &54\er16&16\er9&13\er5&25\er12      \\[-2ex]
&&\tiny{$<8.6$}&&\tiny{$<8.5$}&\tiny{$<0.1$}       
\\[-1ex]
$N(1875)$&$\frac32^-$                      &12\er4&4\er2&8\er3&$<1$\\[-2ex]
&&&&\tiny{$36-56$}&    
\\[-1ex]
$N(1880)$&$\frac12^+$                     &28\er7&20\er6&-&8\er3\\[-2ex]
&&\tiny{$19-45$}&\tiny{$19-45$}&&    
\\[-1ex] 
$N(1895)$&$\frac12^-$                     &43\er25&18\er5&&25\er14     \\[-2ex]
&&\tiny{$14-50$}&\tiny{$<18$}&-&\tiny{$14-32$}   
\\[-1ex]
$N(1900)$&$\frac32^+$                      &46\er17&7\er4&9\er3&30\er12           \\[-2ex]
&&\tiny{25-40}&\tiny{25-40}&&   
\\[-1ex]
$N(1990)$&$\frac72^+$                    &10\er5&8\er4&2\er2&$<1$      \\[-2ex]
&&&&&  
\\[-1ex]
$N(2000)$&$\frac52^+$                    &15\er4&8\er3&7\er3&$<1$      \\[-2ex]
&&&&&   
\\[-1ex]
$N(2060)$&$\frac52^-$                    &28\er11&24\er10&4\er4& $<1$     \\[-2ex]
&&\tiny 5-33&\tiny $<10$&\tiny 5-23&   
\\[-1ex]
$N(2100)$&$\frac12^+$                    &17\er7&12\er6&-&5\er3      \\[-2ex]
&&\tiny35-70&\tiny35-70&&   
\\[-1ex] 
$N(2120)$&$\frac32^-$                     &28\er6&4\er2&19\er5&5\er 3      \\[-2ex]
 &&&&\tiny{$<3$}&  
\\[-1ex]
$N(2190)$&$\frac72^-$                     &9\er7&$<1$&8\er7&$<1$      \\[-2ex]
&&&&\tiny{$<11$}&    
\\\hline\hline
$\Delta(1600)$& $\frac32^+$                  &7\er4&2\er2&5\er3&$<1$      \\[-2ex]
&&&&&   \\[-1ex]
$\Delta(1620)$&$\frac12^-$                   &52\er17&30\er12&-&22\er12      \\[-2ex]
&&\tiny23-32&\tiny 23-32&&\tiny$<0.04$ \\[-1ex] 
$\Delta(1700)$&$\frac32^-$                   &14\er4&$<1$&13\er4&$<1$     \\[-2ex]
&&&&\tiny 22-32&     \\[-1ex] 
$\Delta(1750)$&$\frac12^+$                   &27\er13&17\er10&-&10\er8      \\[-2ex]
&&\tiny&\tiny &&     \\[-1ex] 
$\Delta(1900)$&$\frac12^-$                    &38\er13&18\er8&-&20\er10     \\[-2ex]
&&\tiny22-60&\tiny11-35&&\tiny11-25     \\[-1ex] 
$\Delta(1905)$&$\frac52^+$                     &26\er10&$<1$&25\er10&$<1$      \\[-2ex]
&&&&\tiny17-35&    \\[-1ex] 
$\Delta(1910)$&$\frac12^+$                       &10\er4&5\er3&-&5\er3       \\[-2ex]
&&&&& \\[-1ex]
$\Delta(1920)$&$\frac32^+$                    &57\er9&8\er4&14\er5&35\er6      \\[-2ex]
&&&&&        \\[-1ex] 
$\Delta(1930)$&$\frac52^-$                    &33\er8&3\er2&$<1$&30\er8      \\[-2ex]
&&& &&  \\[-1ex] 
$\Delta(1950)$&$\frac72^+$                    &10\er5&10\er5&$<1$&$<1$      \\[-2ex]
&&&&&  \\[-1ex] 
$\Delta(2200)$&$\frac72^-$                     &36\er14&21\er12&$<1$&15\er9      \\[-4ex]
&&&&&   \\
\hline\hline
\end{tabular}
\end{center}
\end{table}
\section{Branching ratios for the decays $\mathbf{N^*}$ and $\mathbf{\Delta^*\to N\rho(770)}$}

The differential cross sections for $\rho^0(770)$ meson production as a function of the squared momentum
transfer $-t$ (see Fig.~\ref{rho-diff}) show some additional intensity
at backward angles suggesting resonance contributions. In fact, coupled-channel analysis requires
significant contributions from $N^*$ and $\Delta^*\to N\rho(770)$. The resulting branching ratios are
presented in Table~\ref{BR}. 
The branching ratios are calculated by integration over the $\rho$ line shape
and the width of the resonance under study, see Ref.~\cite{Burkert:2022bqo}.

In most cases, the values are compatible with those given in the Review of Particle
Physics (RPP)~\cite{ParticleDataGroup:2022pth} even though some branching ratios differ significantly. 
This may be not too surprising: the range of values listed in the RPP often represents just the statistical uncertainty
in an analysis. An extreme example is the decay $N(1720)\frac32^+\to N\rho$:
Shrestha and Manley~\cite{Shrestha:2012ep} give a branching ratio $N(1720)\frac32^+\to N\rho, S=\frac12, P$-wave of (1.4\er0.5)\%
that defines the RPP entry. The measurement of Vrana {\it et al.}~\cite{Vrana:1999nt}, giving (91\er1)\%, is not used to define a range. Contributions from $N\rho$ decays with $N$ and $\rho$ spins aligned to $S=3/2$ were not tested in either analysis. 
For $N(1875)\frac32^-\to N\rho$, Hunt and Manley~\cite{Hunt:2018wqz} report a branching ratio $N(1875)\frac32^-\to N\rho, S=\frac32, S$-wave 
of (46\er10)\% that defines the RPP entry; the values from Shrestha and Manley~\cite{Shrestha:2012ep}
and Vrana {\it et al.}~\cite{Vrana:1999nt}, who find $<5$\% and (6\er6)\%, are not used. The branching ratios $N(1675)\frac52^-$ and
$N(2120)\frac32^-\to N\rho$, where large discrepancies are seen, have so far been 
determined only by Hunt and Manley~\cite{Hunt:2018wqz}, in the analyses~\cite{Shrestha:2012ep,Vrana:1999nt}
there was no need to introduce these contributions.
The large discrepancies evidence the limitations of the
database that have existed so far and the difficulty of the analyses.

\section{Summary}
\label{SectionSum}
We studied the reaction $\gamma p\to p\pi^+\pi^-$ for photon energies
from 0.7 to 2.4\,GeV with the CLAS detector at Jefferson Lab. The full data
set of 400 million reconstructed events is included in the form of invariant mass and angular distributions, and
a subsample of nearly 2 million events is used in an event-based likelihood fit. This is the first analysis
in which an event-based likelihood fit is made on the reaction $\gamma p\to p\pi^+\pi^-$. 

Events with the three particles reconstructed and events
with a missing $\pi^+$, $\pi^-$, or proton are analyzed. 
The full data set and the subsample are interpreted within the BnGa  coupled-channel analysis. 
The analysis reveals strong $p\rho^0(770)$ and $\Delta(1232)^{++}\pi^-$ production and weaker 
production of $\Delta(1232)^{0}\pi^+$.
Differential cross section and spin-density matrix elements are presented. The production 
of $\rho^0(770)$ mesons
can be assigned largely  
to diffractive scattering via exchange of Pomeron and pion trajectories. $\Delta(1232)^{++}$ 
production is strong due to the Kroll--Rudermann mechanism. Deviations of the behavior expected 
for these dominant contributions point at a production via intermediate resonances decaying 
into $N\rho(770)$ and $\Delta(1232)\pi$. 

The BnGa  coupled-channel analysis provides masses, widths, and branching ratios for decays of the contributing
resonances into a large number of final states, including intermediate resonances. Here, we have reported 
the branching ratios for $N\rho$ decays that are uniquely 
determined by the new CLAS data on $\gamma p\to p\pi^+\pi^-$. In particular, the {\it event-based data sample}
plays a decisive role for the results presented here.

\begin{acknowledgments}
The authors thank the technical staff at Jefferson Lab and at all the participating institutions for their 
invaluable contributions to the success of the experiment. This material is based on work supported by
the U.S. Department of Energy, Office of Science, Office of Nuclear Physics, under Contract No.
DE-AC05-06OR23177. The work was supported by the \textit{Deutsche Forschungsgemeinschaft} (SFB/TR110),
the \textit{U.S. Department of Energy} (DE-AC05-06OR23177, DE-FG02-92ER40735),
and the \textit{Russian Science Foundation} (RSF 24-22-00322), 
the US National Science Foundation, the State Committee of Science of 
Republic of Armenia, the Chilean Comisi\'{o}n Nacional de Investigaci\'{o}n Cientifica y Tecnol\'{o}gica 
(CONICYT), the Italian Istituto Nazionale di Fisica Nucleare, the French Centre National de la Recherche 
Scientifique, the French Commissariat a l'Energie Atomique, the Scottish Universities Physics Alliance 
(SUPA), the United Kingdom's Science and Technology Facilities Council, and the National Research Foundation 
of Korea.
\end{acknowledgments}

\bibliographystyle{apsrev4-2}

\end{document}